\documentclass{article}
\pdfoutput=1
\usepackage{jheppub}
\usepackage{epic}
\usepackage{latexsym}
\usepackage{epsf}
\usepackage{epsfig}
\usepackage{amssymb,amsmath}
\usepackage{amsfonts}
\usepackage{dsfont}
\usepackage{graphicx}
\usepackage{epstopdf}
\usepackage{gensymb}
\usepackage{graphics}
\usepackage{subfigure}
\usepackage{multirow}
\usepackage{amssymb} 
\usepackage{ulem}
\usepackage{amsmath}
\usepackage{bbm} 
\usepackage{mathrsfs}   
\usepackage{xspace}
\usepackage{listings}

 \usepackage[utf8]{inputenc}

\newcommand{\beq}{\begin{equation}} 
\newcommand{\eeq}{\end{equation}} 
\newcommand{\ba}{\begin{array}}  
\newcommand{\ea}{\end{array}} 
\newcommand{\bea}{\begin{eqnarray}}  
\newcommand{\eea}{\end{eqnarray} }  
\newcommand{\be}{\begin{eqnarray}}  
\newcommand{\ee}{\end{eqnarray} }  
\newcommand{\bal}{\begin{align}}
\newcommand{\eal}{\end{align}}   
\newcommand{\bi}{\begin{itemize}}  
\newcommand{\ei}{\end{itemize}}  
\newcommand{\ben}{\begin{enumerate}}  
\newcommand{\een}{\end{enumerate}}  
\newcommand{\bc}{\begin{center}}
\newcommand{\ec}{\end{center}} 
\newcommand{\bt}{\begin{table}}
\newcommand{\et}{\end{table}}  
\newcommand{\btb}{\begin{tabular}}
\newcommand{\etb}{\end{tabular}}

\newcommand{\eVdist}{\kern-0.06em}

\newcommand{\SARAH}{{\tt SARAH}\xspace}
\newcommand{\Vevacious}{{\tt Vevacious}\xspace}
\newcommand{\SPheno}{{\tt SPheno}\xspace}

\newcommand{\HB}{{\tt HiggsBounds}\xspace}

\newcommand{\MS}{$\overline{\text{MS}}$\xspace}
\begin{document}


 \vspace*{-3.0cm}
 \begin{flushright}
 BONN-TH-2017-08\\
 KA-TP-30-2017
\end{flushright}

\title{\huge Perturbativity Constraints in BSM Models}

\author[a]{Manuel E. Krauss,} 
\affiliation[a]{Bethe Center for Theoretical Physics \& Physikalisches Institut der 
Universit\"at Bonn,\\ Nu{\ss}allee 12, 53115 Bonn, Germany}
\emailAdd{mkrauss@th.physik.uni-bonn.de}

\author[b,c]{Florian Staub}
\affiliation[b]{Institute for Theoretical Physics (ITP), Karlsruhe Institute of Technology, Engesserstra{\ss}e 7, D-76128 Karlsruhe, Germany}
\affiliation[c]{Institute for Nuclear Physics (IKP), Karlsruhe Institute of Technology, Hermann-von-Helmholtz-Platz 1, D-76344 Eggenstein-Leopoldshafen, Germany}
\emailAdd{florian.staub@kit.edu}

\abstract{
Phenomenological studies  performed for non-supersymmetric extensions of the Standard Model usually use  tree-level parameters 
as input to define the scalar sector of the model. This  implicitly assumes that a full on-shell calculation of the scalar sector is possible -- and meaningful.
However, this doesn't have to be the case as we show explicitly at the example of the Georgi-Machacek model.  This  model comes with an appealing custodial symmetry to explain the smallness 
of the $\rho$ parameter. However, the model cannot be renormalised on-shell without breaking the custodial symmetry. Moreover, we find that it can often happen that the radiative corrections are so large that any consideration  based on a perturbative expansion
appears to be meaningless: counter-terms to quartic couplings can become much larger than $4\pi$ and/or two-loop mass corrections can become larger than the one-loop ones. Therefore, conditions
are necessary to single out parameter regions which cannot be treated perturbatively. We propose and discuss different sets of such perturbativity conditions and show their impact on the parameter space of 
the Georgi-Machacek model. Moreover, the proposed conditions are general enough that they can be applied to other models as well. 
We also point out that the vacuum stability constraints in the Georgi-Machacek model, which have so far only been applied at the tree level, receive crucial radiative corrections.
We show that large regions of the parameter space which feature a stable
electroweak vacuum at the loop level would have been --wrongly-- ruled out by the tree-level conditions. 
}

\maketitle

\section{Introduction}
\label{sec:intro}
Already five years have past since the discovery of a new scalar at the LHC \cite{Aad:2012tfa,Chatrchyan:2012xdj}. In the meantime, the properties of this particle 
have been measured with an astonishing precision. All coupling measurements agree very well with the expectations for the standard model (SM)
Higgs boson. Thus, this particle is at least a SM-like Higgs. Maybe, it is {\it the} Higgs, i.e. the only fundamental scalar which exists in nature 
and which is involved in the breaking of the electroweak (ew) gauge symmetry. However, it is much too early to draw this conclusion and there is still
plenty of space where new Higgs-like particles as predicted by models beyond the SM (BSM) might show up. The motivation to introduce new scalars in BSM
models can be quite different and either stem from a top-down approach like a new proposed symmetry, or from a bottom-up approach where new scalars are needed 
to get some specific feature in the model. One can count supersymmetric models, models with a grand unified theory or models with an extended gauge sector to
the first category, models like those with only additional scalar doublets or triplets to the second one. The interesting feature of these models is that they 
can offer very interesting, phenomenological effects like changes in the couplings of the SM-like Higgs to fermions or vector bosons, charged or even doubly-charged scalars, 
or signatures of additional light or heavy  neutral scalars. Models with additional Higgs triplets are constrained by the $\rho$-parameter which relates $M_W$, $M_Z$ and the 
weak mixing angle. This parameter is measured to be very close to one. While additional doublets only contribute loop corrections to this parameter, vacuum expectation values (VEVs) 
of triplets, $v_T$, usually enter already at the tree-level. This imposes an upper limit on $v_T$ of just a few GeV \cite{Gunion:1990dt,Arhrib:2011uy,Kanemura:2012rs}. However,
this constraint can be circumvented if several triplets are arranged in such a way that they obey a new (custodial) symmetry:
in this case, the tree-level contributions to $\delta\rho$ cancel and the triplet VEVs can be much larger. 
This was the original idea proposed in Refs.~\cite{Georgi:1985nv,Chanowitz:1985ug}. The model known today as the `Georgi-Machacek' (GM) model comes with one complex and 
one real triplet.\\
Many aspects of the GM model were studied in detail in the last years like the properties of the couplings  of scalars
\cite{Haber:1999zh,Logan:2010en,Aoki:2012jj,Kanemura:2013mc,Kikuchi:2013gba,Godunov:2015lea,Logan:2015fka,Chang:2017niy,Blasi:2017zel,Chiang:2017vvo,Degrande:2017naf}
or vector bosons \cite{Arroyo-Urena:2016gjt}, or the potential signatures at colliders
\cite{Godfrey:2010qb,Chiang:2012cn,Godunov:2014waa,Chiang:2015kka,Chiang:2015gva,Chiang:2015rva,Chiang:2015amq,Yu:2016skv,Zhang:2017och}. 
Not only the phenomenological consequences of the new states were explored, but also the theoretical properties of the extended Higgs sector were studied. This was used to impose constraints 
on the parameters of the model. For instance, limits for the quartic couplings were derived from the unitarity of the tree-level $2 \to 2$ scattering \cite{Aoki:2007ah,Hartling:2014zca}. Also the stability 
of the ew vacuum was checked, and dangerous regions in which the potential is unbounded-from-below (UFB) were singled out \cite{Chiang:2014bia,Hartling:2014aga}. However, all of these constraints have up to now only been studied
at lowest order in perturbation theory and it has not been checked whether the conclusions still hold once loop corrections are included. 
Moreover, it was so far not clear if higher-order corrections could even be implemented in a sensible way or if some regions in parameter space cannot be treated perturbatively.
If this were the case, Born-level results would not be good approximations of the `full' result, but might be completely 
meaningless. 
This is actually an issue which has so far hardly been discussed in any non-supersymmetric BSM model, which might sound surprising as it was already shown in Ref.~\cite{Nierste:1995zx} for the SM that a naive limit like 
$\lambda < 4\pi$ is too weak and that perturbation theory can break down at much smaller coupling values. 
Among the reasons why this breakdown of perturbation theory has not been checked exhaustively for many BSM models are the missing but necessary 
higher order results: the only rigorous way to claim a breakdown of perturbation theory is to calculate observables at different loop levels and compare the residual scale dependence which, if a perturbative treatment is possible, must shrink from every order to the next. This 
would make it necessary to calculate decays or scattering processes at least up to the two-loop level, which would cause a tremendous amount of work. However, some results can be obtained in an easier way which should
give more sophisticated hints as to when perturbation theory is in trouble than the simple rules of thumb which say that a quartic coupling must be smaller than some factor times $\pi$. A first idea in that direction was shown in Ref.~\cite{Braathen:2017izn} where the one- and two-loop 
corrections to the scalar masses in the \MS scheme were compared. If the two-loop corrections become larger than the one-loop corrections, this already points already towards a problem. Indeed, it has been 
shown in Ref.~\cite{Braathen:2017izn} at the example of the GM model that the two-loop corrections can be several times as large as the one-loop corrections. \\
We are going to investigate this potential breakdown of perturbation theory in the GM model in more detail in this work. We propose different conditions which could be used to check for dangerous regions in 
parameters space. These are not only the size of the one- and two-loop corrections in the \MS scheme, but also the values of the counter-terms (CTs) when performing an on-shell (OS) normalisation of the scalar sector 
at the one-loop level. Since these CTs will enter at all higher loop levels, they cannot be too large without disturbing the convergence of the loop series. The methodology which we develop to apply perturbativity 
constraints is not restricted to the GM model, but can be applied similarly to other BSM models. 
\\
A second main aim of this work is the promotion of the vacuum stability constraints to the loop level: as we will see, the loop corrections in the GM model are always sizeable. Even if perturbation theory might 
still be under control, tree-level results can be very misleading. For instance, it was shown in Ref.~\cite{Staub:2017ktc} that UFB directions in two Higgs doublet Models (THDM) usually disappear once loop 
corrections are included. We will find similar results for the GM model. We will show that many parameter regions which seem to pass all tree-level constraints are often in conflict with the constraints from perturbativity. On the 
other side, parameter regions which are unstable at tree-level become stable once the loop corrections are included. We will present some indications where these effects are most likely to happen in the parameter space 
of the GM model. \\

This work is organised as follows: in sec.~\ref{sec:model}, we explain the basic ingredients of the GM model with and without the custodial symmetry, as well as the two different normalisation schemes, \MS and OS, which we use in this paper. In sec.~\ref{sec:constraints}, the existing tree-level constraints 
on the model are summarised, and our new constraints are explained. In sec.~\ref{sec:results}, the impact of our new constraints is discussed. We start with specific examples to discuss the impact of the perturbativity constraints as well as the loop-improved vacuum stability checks before we explore the consequences in the full parameter space. We conclude in sec.~\ref{sec:conclusions}. The appendix contains 
a long collection of additional material: all tree-level couplings of all scalars, the expressions for the one-loop corrections to self-energies and tadpoles as well as the necessary CTs to renormalise the scalar sector of the GM model on-shell.

\section{The Georgi-Machacek model}
\label{sec:model}
In the GM model, the SM is extended by one real scalar $SU(2)_L$-triplet $\eta$ with hypercharge $Y=0$ and one complex scalar triplet  $\chi$ with $Y=-1$ (using the notation $Q_{\rm em} = T_{3L}+Y$). Those can be written as
\begin{equation} \label{eq:fieldsGM}
 \eta  = \frac{1}{\sqrt{2}} \, \left( \begin{array}{cc}
\eta^0 & - \sqrt{2} \left(\eta^-\right)^*\\
- \sqrt{2} \, \eta^- & - \eta^0  \end{array} \right) \,, \quad 
 \chi  = \frac{1}{\sqrt{2}}  \, \left( \begin{array}{cc}
\chi^- & \sqrt{2} (\chi^0)^* \\
-\sqrt{2}\, \chi^{--} & -\chi^-  \end{array} \right) \,.
\end{equation}
with $(\eta^0)^* = \eta^0$. 

\subsection{Unbroken custodial symmetry}
After imposing a global $SU(2)_L \times SU(2)_R$ symmetry on the model, the most compact way to write the Lagrangian in a form which explicitly conserves this custodial symmetry is to re-express the Higgs doublet $\Phi$ as a bi-doublet under $SU(2)_L \times SU(2)_R$ and the two scalar triplets as a bi-triplet:
\begin{align}
\Phi =
\left(
\begin{array}{cc}
\phi^{0*}  & \phi^+     \\
 \phi^- & \phi^0      
\end{array}
\right)\,, \qquad
\quad \Delta =
\left(
\begin{array}{ccc}
 \chi^{0*} & \eta^+  & \chi^{++}   \\
 \chi^- & \eta^0  & \chi^+  \\
 \chi^{--} & \eta^{-}  & \chi^0   
\end{array}
\right)\,.
\end{align}
The scalar potential can then be written as
\begin{align}
V(\Phi, \Delta) & =  \frac{\mu_2^2}{2}  \mathrm{Tr} \Phi^\dagger \Phi + \frac{\mu_3^2}{2} \mathrm{Tr} \Delta^\dagger \Delta  + \lambda_1 \left[  \mathrm{Tr} \Phi^\dagger \Phi \right]^2 + \lambda_2  \mathrm{Tr} \Phi^\dagger \Phi \, \mathrm{Tr} \Delta^\dagger \Delta \nonumber \\
&  + \lambda_3 \mathrm{Tr} \Delta^\dagger \Delta \Delta^\dagger \Delta +   \lambda_4 \left[\mathrm{Tr} \Delta^\dagger \Delta \right]^2   - \lambda_5 \mathrm{Tr} \left( \Phi^\dagger \sigma^a \Phi \sigma^b  \right) \, \mathrm{Tr}  \left(\Delta^\dagger t^a \Delta t^b \right) \nonumber \\
&   - M_1  \mathrm{Tr} \left( \Phi^\dagger \tau^a \Phi \tau^b \right) (U \Delta U^\dagger)_{ab}\nonumber    - M_2 \mathrm{Tr} \left( \Delta^\dagger t^a \Delta t^b \right) (U \Delta U^\dagger)_{ab} \, ,
\end{align}
where $\tau^a$ and $t^a$ are the $SU(2)$ generators for the doublet and triplet representations respectively, and $U$ is for instance given  in Ref.~\cite{Hartling:2014zca}. \\
The vacuum expectation values (VEVs) of the triplets read
\begin{equation} \label{eq:GM:VEVs}
\langle \eta \rangle = \frac{1}{\sqrt{2}} \, \left( \begin{array}{cc}
v_\eta & 0\\
0 & -v_\eta  \end{array} \right) \,, \quad 
\langle \chi \rangle =  \, \left( \begin{array}{cc}
0 & v_\chi \\
0 & 0  \end{array} \right) \, ,
\end{equation}
while the doublet VEV is $\langle \phi^0 \rangle = v_\phi/\sqrt{2}$ as usual.
The gauge boson masses then read at tree level 
\begin{equation}
M_W^2 = \frac{1}{4}\,g_2^2\, \left( v_\phi^2 + 4(v_\eta^2 + v_\chi^2) \right)\,,\qquad M_Z^2 = \frac{1}{4} (g_1^2+g_2^2) (v_\phi^2 + 8 v_\chi^2)\,.
\label{eq:gaugebosonmasses}
\end{equation}
If the custodial symmetry is preserved, the triplet VEVs are identical, $v_\eta=v_\chi$, and there are no tree-level contributions to electroweak precision observables as a consequence.
The electroweak VEV $v$ can then be written as 
\begin{equation}
v^2 = v_\phi^2 + 8 v_\chi^2 \ \simeq 246~\text{GeV}\,,
\end{equation}
so that the massive gauge bosons end up with the known SM tree-level masses, 
and it is common to define the angle 
\begin{equation}
s_H \equiv \sin \theta_H = 2 \sqrt{2} \, \frac{v_\chi}{v}\,,\qquad c_H \equiv \cos \theta_H = \frac{v_\phi}{v}\,.
\end{equation}

The minimisation conditions for the scalar potential read
\begin{eqnarray}
	0 = \frac{\partial V}{\partial v_{\phi}} &=& 
	v_{\phi} \left(\mu_2^2 + 4 \lambda_1 v_{\phi}^2 + 6 \lambda_2 v_\chi^2 - 3 \lambda_5 v_\chi^2  - \frac{3}{2} M_1 v_\chi \right)\,, 
		\label{eq:phimincond} \\
	0 = \frac{\partial V}{\partial v_{\chi}} &=& 
	3 v_\chi \left( \mu_3^2 + 2 \lambda_2 v_\phi^2 - \lambda_5 v_\phi^2 - 6 M_2 v_\chi + 4 \lambda_3 v_\chi^2 + 12 \lambda_4 v_\chi^2 \right) -\frac{3}{4} M_1 v_\phi^2\,,
	\label{eq:chimincond}
\end{eqnarray}
which we solve for $\mu_2^2$ and $\mu_3^2$ to eliminate these parameters from the scalar potential. Note that we have assumed all parameters to be real, resulting in CP conservation in the scalar sector.

The mass eigenstates of the scalar fields can be divided into singlets, triplets and fiveplets under the custodial symmetry.
At tree level, the masses within each triplet and fiveplet are degenerate, and, using the above equations, are given by 
\begin{eqnarray}
	m_5^2 &=& \frac{M_1}{4 v_{\chi}} v_\phi^2 + 12 M_2 v_{\chi} 
	+ \frac{3}{2} \lambda_5 v_{\phi}^2 + 8 \lambda_3 v_{\chi}^2
    = v\left( s_H (3 \sqrt{2} M_2 + s_H \lambda_3 v) + c_H^2 (\frac{M_1}{\sqrt{2} s_H} + \frac{3}{2} \lambda_5 v) \right)\,, \label{eq:mass_fiveplet} \\
	m_3^2 &=& 
	\frac{v_\phi^2 + 8 v_\chi^2}{4 v_\chi} \left(M_1 + 2 v_\chi \lambda_5 \right)
	= \frac{v M_1}{\sqrt{2}\, s_H} + \frac{1}{2} \lambda_5 v^2\,. \label{eq:mass_triplet}
\end{eqnarray}

The mass matrix of the CP-even scalars reads
\begin{eqnarray}
m_{h^0}^2 = \begin{pmatrix}
8 c_H^2 \lambda_1 v^2  & -\frac{v}{2} c_H \tilde M & -\frac{v}{\sqrt{2}} c_H \tilde M \\
\cdot
& \frac{v \left( \sqrt{2} (c_H^2 M_1 + 3 s_H^2 M_2) + 2 s_H v (s_H^2 (\lambda_3+\lambda_4) + c_H^2 \lambda_5)\right)}{2 s_H}  
&  \frac{v^2 \left( 2 s_H^2 \lambda_4-c_H^2 \lambda_5 \right)}{\sqrt{2}} - 3 s_H M_2 v \\
\cdot &
\cdot &
\frac{v\left( 2 s_H^3 v (\lambda_3+2 \lambda_4) + c_H^2 (\sqrt{2} M_1+s_H \lambda_5 v)  \right)}{2 s_H}
\end{pmatrix}\,,
\end{eqnarray}
where
\begin{equation}
\tilde M  = M_1 + \sqrt{2} s_H v (\lambda_5 - 2 \lambda_2)\,.
\end{equation}
After the diagonalisation, one mass eigenstate corresponds to the fiveplet mass, Eq.~(\ref{eq:mass_fiveplet}), whereas the other two mix to form the mass eigenstates $h$ and $H$.

They are given in the gauge basis by \cite{Hartling:2014zca}
\begin{eqnarray}
	h &=& \cos \alpha \, H_1^0 - \sin \alpha \, H_1^{0\prime}\,,   \\ \nonumber 
	H &=& \sin \alpha \, H_1^0 + \cos \alpha \, H_1^{0\prime}\,,
\end{eqnarray}
where 
\begin{equation}
	H_1^0 = {\mathcal Re}[\phi^{0}]\,, \qquad 
	H_1^{0 \prime} = \sqrt{\frac{1}{3}}\, \eta^0 + \sqrt{\frac{2}{3}}\, {\mathcal Re}[\chi^{0}]\,.
\end{equation}

\subsection{Broken custodial symmetry}
Without the custodial symmetry, the most general gauge-invariant and CP-conserving form of the Higgs potential is given by \cite{Blasi:2017xmc}
\begin{align}
\label{eq:broken_custodial_literature}
V&=m_\phi^2(\phi^\dagger \phi)+m_\chi^2\text{Tr}(\chi^\dagger\chi)+m_\eta^2\text{Tr}(\eta^2)\notag\\
&+\mu_1\phi^\dagger \eta\phi +\mu_2 (\phi^T(i\tau_2) \chi \phi+\text{h.c.}) +\mu_3\text{Tr}(\chi \chi^\dagger \eta)+\lambda (\phi^\dagger \phi)^2 \notag\\
&
+\rho_1\left(\text{Tr}(\chi^\dagger\chi)\right)^2+\rho_2\text{Tr}(\chi^\dagger \chi\chi^\dagger \chi)
+\rho_3\text{Tr}(\eta^4)
+\rho_4 \text{Tr}(\chi^\dagger\chi)\text{Tr}(\eta^2)
+\rho_5\text{Tr}(\chi^\dagger \eta)\text{Tr}(\eta \chi)\notag\\
&+\sigma_1\text{Tr}(\chi^\dagger \chi)\phi^\dagger \phi+\sigma_2 \phi^\dagger \chi^\dagger \chi \phi
+\sigma_3\text{Tr}(\eta^2)\phi^\dagger \phi 
+\sigma_4 (\phi^\dagger \chi^\dagger \eta \phi^c + \text{h.c.})\,. 
\end{align}
Note that the above equation differs from Ref.~\cite{Blasi:2017xmc} in $\chi \leftrightarrow \chi^\dagger$ because of the differing definitions of $\chi$. Here, $\phi^c = i\tau \phi^*$.
For easier comparison with the limit of conserved custodial symmetry, we re-write the potential as 
\begin{align}
\label{eq:broken_custodial_ours}
V = & \mu^2_2 (\phi^\dagger \phi) + \mu^2_\chi \mathrm{Tr}(\chi^\dagger \chi) + \frac12 \mu^2_\eta   \mathrm{Tr}(\eta^\dagger \eta) + 4 \lambda_1 (\phi^\dagger \phi)^2  +  2 (\phi^\dagger \phi) \left(2\lambda_{2a} \mathrm{Tr}(\chi^\dagger \chi) + \lambda_{2b} \mathrm{Tr}(\eta^\dagger \eta)\right)   \nonumber \\
&
+ 2 \lambda_{3a} \mathrm{Tr}\left((\eta^\dagger \eta)^2\right)+ 2 \lambda_{3b} \left(\mathrm{Tr} \left( ( \chi^\dagger \chi)^2\right) + 3 \mathrm{Tr}(\chi^\dagger \chi \chi \chi^\dagger) \right) 
+ 4 \lambda_{3c} \mathrm{Tr}(\chi^\dagger \chi \eta^\dagger \eta + \chi^\dagger \eta^\dagger  \chi \eta) +   \nonumber \\
& + \lambda_{4a} \left(\mathrm{Tr}(\eta^\dagger \eta)\right)^2 + 4 \lambda_{4b} \left(\mathrm{Tr}(\chi^\dagger \chi)\right)^2 + 4\lambda_{4c} \mathrm{Tr}(\eta^\dagger \eta) \mathrm{Tr}(\chi^\dagger \chi) 
- \lambda_{5a} (\phi^\dagger \chi^\dagger \chi \phi - \phi^\dagger \chi \chi^\dagger   \phi)  \\
& + \sqrt{2} \lambda_{5b} (\phi^\dagger \eta \chi^\dagger \phi^c +{\rm h.c.}) \nonumber 
 +\frac{1}{\sqrt{2}}M_{1a} \phi^\dagger \eta \phi + \frac12 M_{1b} \left( \phi^T (i \tau_2)\chi \phi + \text{h.c.} \right) + 3 \sqrt{2} M_2 \left(\mathrm{Tr}( \chi^\dagger \chi \eta) + {\rm h.c.} \right)\,.
\end{align}
At the tree-level where custodial symmetry is conserved, we have 
\begin{align}
\lambda_{Na}=\lambda_{Nb}\equiv \ & \ \lambda_2\,, \hspace{1cm} N=\{2,5\}\,,\\
\lambda_{Na}=\lambda_{Nb}=\lambda_{Nc}\equiv \ & \ \lambda_3\,, \hspace{1cm} N=\{3,4\}\,,
\end{align}
as well as 
\begin{align}
M_{1a}=M_{1b}\equiv\ & \ M_1 \,,\\
\mu_\chi^2 = \mu_\eta^2 \equiv \ &  \ \mu_3^2 \,.
\end{align}

The parameters of Eqs.~(\ref{eq:broken_custodial_literature}) and (\ref{eq:broken_custodial_ours}) are related via
\begin{align}
\label{eq:parameter_translation_begin}
\lambda_1 &= \frac{1}{4} \lambda\,, \qquad 
\lambda_{2a} = \frac{1}{4} \left(\sigma_1 + \frac{1}{2} \sigma_2\right)\,, \qquad
\lambda_{2b} = \frac{1}{2} \sigma_3\,, \\
\lambda_{3a} + \lambda_{4a} = \frac{1}{2} \rho_3\,, \qquad
\lambda_{3b} &= -\frac{1}{4} \rho_2 \,, \qquad
\lambda_{3c} = \frac{1}{4} \rho_5\,,\qquad
\lambda_{4b} = \frac{1}{4} \left(\rho_1 + \frac{3}{2} \rho_2\right) \,,\qquad
\lambda_{4c} = \frac{1}{4} \rho_4 \,, \\
&~~~~~~~\lambda_{5a} = -\frac{1}{2} \sigma_2\,, \qquad
\lambda_{5b} = -\frac{1}{\sqrt{2}} \sigma_4\,, \\
\mu_2^2 = m_\phi^2\,, \qquad  \mu_\chi^2 = m_\chi^2\,,\qquad &\mu_\eta^2 = 2\, m_\eta^2\,, \qquad M_{1a} = \sqrt{2}\, \mu_1\,,\qquad M_{1b} = 2 \mu_2\,,\qquad M_2 = -\frac{\mu_3}{6 \sqrt{2}}\,.
\label{eq:parameter_translation_end}
\end{align}
The tadpole equations in the case of broken custodial symmetry read
\begin{eqnarray}
0&=&\frac{\partial V}{\partial v_\phi} = 
\frac{v_\phi}{2} (2 \mu_2^2 + 8 \lambda_1 v_\phi^2 - M_{1a} v_\eta + 4 \lambda_{2b} v_\eta^2 - 
   2 M_{1b} v_\chi - 4 \lambda_{5b} v_\eta v_\chi + 8 \lambda_{2a} v_\eta^2 - 
   2 \lambda_{5a} v_\chi^2)\,,\\
 0 &=&  \frac{\partial V}{\partial v_\chi} = 
 -\frac{M_{1b} v_\phi^2}{2} - \lambda_{5b} v_\phi^2 v_\eta  \\ \nonumber
 & +& v_\chi (2 \mu_{\chi}^2 + 4 \lambda_{2a} v_\phi^2 - \lambda_{5a} v_\phi^2 - 12 M_2 v_\eta + 
    8 \lambda_{4c} v_\eta^2 + 8 \lambda_{3b} v_\chi^2 + 16 \lambda_{4b} v_\chi^2)\,, \\
 0 &=&   \frac{\partial V}{\partial v_\eta} =
 -\frac{M_{1a} v_\phi^2}{4} + \mu_\eta^2 v_\eta + 2 \lambda_{2b} v_\phi^2 v_\eta + 
 4 \lambda_{3a} v_\eta^3 + 4 \lambda_{4a} v_\eta^3 \\ \nonumber
 &-& \lambda_{5b} v_\phi^2 v_\chi - 
 6 M_2 v_\chi^2 + 8 \lambda_{4c} v_\eta v_\chi^2\,.
\end{eqnarray}
%

After solving these equations for $\mu_2^2\,,\,\mu_\chi^2$ and $\mu_\eta^2$, the CP-even scalar mass matrix is at tree level given by
\begin{align}
m_{h^0}^2 = \left(
\begin{array}{ccc}
 8 \lambda_1 v_\phi^2 & -\frac{1}{2} v_\phi \left(4 v_\chi \lambda
   _{5 b}-8 v_\eta \lambda_{2 b}+M_{1a}\right) & -\frac{v_\phi
   \left(-8 v_\chi \lambda_{2 a}+2 v_\chi \lambda_{5 a}+2 v_\eta
   \lambda_{5 b}+M_{1b}\right)}{\sqrt{2}} \\
 \cdot
   & \frac{32 v_\eta^3 \lambda_{3 a}+32 v_\eta^3
   \lambda_{4 a}+4 v_\chi v_\phi^2 \lambda_{5 b}+M_{1a}
   v_\phi^2+24 M_{2} v_\chi^2}{4 v_\eta} & -\frac{v_\phi^2
   \lambda_{5 b}+4 v_\chi \left(3 M_{2}-4 v_\eta \lambda_{4
   c}\right)}{\sqrt{2}} \\
 \cdot
   &
 \cdot
   & 
   \frac{v_\phi^2 \left(2 v_\eta \lambda
   _{5 b}+M_{1b}\right)+32 v_\chi^3 \lambda_{3 b}+64 v_\chi^3 \lambda
   _{4 b}}{4 v_\chi} \\
\end{array}
\right)\,, \label{eq:mass_matrix_h}
\end{align}
and the mass of the physical pseudo-scalar state is 
\begin{equation}
m_{A}^2 = \left(8 v_\chi^2+v_\phi^2\right) \left(\frac{v_\eta \lambda_{5 b}}{2
   v_\chi}+\frac{M_{1b}}{4 v_\chi}\right)\,.
\end{equation}
The mass matrix for the singly-charged scalars reads, in Landau gauge,
\begin{align}
m_{H^\pm}^2 = 
\left(
\begin{array}{ccc}
 \tilde M_b v_\chi+\tilde M_a v_\eta & -\frac{\tilde M_a
   v_\phi}{2} & -\frac{\tilde M_b v_\phi}{2} \\
 \cdot
 & \frac{4 \tilde M_2 v_\chi^2+2
   v_\phi^2 \lambda_{5 b} v_\chi+\tilde M_a v_\phi^2}{4
   v_\eta} & -\frac{1}{2} \lambda_{5 b} v_\phi^2-\tilde M_2
   v_\chi \\
 \cdot
  & \cdot 
   & \frac{\tilde M_b v_\phi^2+2
   v_\eta \lambda_{5 b} v_\phi^2+4 \tilde M_2 v_\chi
   v_\eta}{4 v_\chi} \\
\end{array}
\right)\,, \label{eq:mass_matrix_hp}
\end{align}
where
\begin{equation}
\tilde M_a = M_{1a} + 2 v_\chi \lambda_{5b} \,,\qquad \tilde M_b = M_{1b} + 2 v\chi \lambda_{5a}\,, \qquad \tilde M_2 = 6 M_2 + 4 v_\eta \lambda_{3c}\,,
\end{equation}
and the mass of the doubly-charged scalar is given by
\begin{equation}
m^2_{H^{\pm\pm}} = v_\phi^2 \lambda_{5 a}+8 v_\chi^2 \lambda
   _{3 b}+\frac{v_\eta v_\phi^2 \lambda_{5
   b}}{2 v_\chi}+\frac{M_{1b} v_\phi^2}{4
   v_\chi}+12 M_2 v_\eta\,.
\end{equation}

\subsection{Renormalisation of the scalar sector of the Georgi-Machacek model}
So far, the scalar sector of the GM model has only been studied at tree level in the literature. The only exception is Ref.~\cite{Braathen:2017izn} which has pointed out that an on-shell renormalisation of the model is not possible without breaking the custodial symmetry. Moreover, it was found that the loop corrections to the scalar mass matrices can become huge. Here we are going to apply two different renormalisation schemes to this model. We start with the discussion of a \MS scheme, before we turn to an on-shell scheme.

\subsubsection{\MS renormalisation}
\label{sec:MS}
The easiest option to renormalise the scalar sector of the GM model is to use the \MS scheme which is applicable to any model. Another advantage of this scheme, beside its general applicability, is that it makes large loops corrections immediately visible without hiding them in counter-terms. The renormalisation procedure is as follows:
\begin{enumerate}
 \item We match the measured SM parameters ($\alpha_s,\, \alpha_{ew},\,  m_f,\,  G_F$) to the running \MS parameters at the scale $Q=M_Z$. The matching procedure is explained in Ref.~\cite{Staub:2017jnp}.
 \item We run the SM parameters using three-loop SM renormalisation group equations (RGEs) to the scale of new physics, which we set to either $m_5$ or $m_H$. At this scale, the model-specific input parameters ($\lambda_1$,\dots$\lambda_5$,\, $M_1$,\, $M_2$,\, $s_H$) are taken as input. In addition, we solve the tadpole equations to obtain the numerical values of the quadratic mass terms. 
 \item All tree-level masses are calculated and the shifts to the SM gauge and Yukawa couplings are included. This is done  by imposing that the 
 eigenvalues of the one-loop corrected mass matrix of the fermions
 \begin{equation}
m_f^{(1L)}(p^2_i) =  m_f^{(T)} - \tilde{\Sigma}_S(p^2_i) - \tilde{\Sigma}_R(p^2_i) m_f^{(T)} - m_f^{(T)} \tilde{\Sigma}_L(p^2_i)   
 \end{equation}
 agree with the running \MS masses. In the gauge sector, the electroweak coupling is shifted as
 \begin{equation}
\alpha^{\rm GM}_{ew}(Q) =  \frac{\alpha^{SM}_{ew}(Q)}{1-\frac{\alpha_{ew}}{2\pi}\Delta_{ew}} \hspace{1cm} \text{with} \hspace{1cm} \Delta_{ew} = \frac43 \log(m_{H^{\pm\pm}}/Q) + \frac13 \sum_{i=1}^2 \log(m_{H_i^{\pm}}/Q)\,.
 \end{equation}
The changes in the 
weak mixing angle are calculated from the $Z$ and $W$ self-energies, see Ref.~\cite{Staub:2017jnp} for more details of the matching procedure at the example of the MSSM.
 \item In order to ensure that we are operating at the bottom of the potential at each order of perturbation theory, we demand that the loop-corrected tadpole equations fulfil 
 \begin{equation}
  T_i + \delta^{(n)} t_i  = 0\,.
 \end{equation}
Here, $T_i$ are the tree-level tadpoles, and $\delta^{(n)} t_i$ are the shifts at the $n$-loop level. These conditions introduce the only three finite CTs which we need:
\begin{align}
\label{eq:tad1}
\delta^{(n)} \mu_2^2 = & - \delta^{(n)} t_v \,,\\
\delta^{(n)} \mu_\eta^2 = & - \delta^{(n)} t_\eta \,,\\
\label{eq:tad2}
\delta^{(n)} \mu_\chi^2 = & - \delta^{(n)} t_\chi \,.
\end{align}
Note that the breaking of the custodial symmetry at the loop level already becomes visible at this stage due to $\delta^{(n)} t_\eta \neq \delta^{(n)} t_\chi$ and therefore $\mu_\eta \neq \mu_\chi$.
 We will make use of the functionality of the tools \SARAH/\SPheno which are able to calculate $\delta^{(n)} t_i$ up to two-loop. 
 \item We calculate the one-loop corrections $\Pi$ to the scalar mass matrices. At the one-loop level, the full dependence on the external momenta is included, while at two-loop, the approximation $p^2=0$ is used. 
 The pole masses are the eigenvalues of the loop-corrected mass matrix calculated as
 \begin{equation}
  M_{\phi}^{(2L)}(p^2) = \tilde{M}^{(2L)}_{\phi} - \Pi_{\phi}(p^2)^{(1L)} -   \Pi_{\phi}(0)^{(2L)}\,.
 \end{equation}
Here, $\tilde{M}_{\phi}$ is the tree-level mass matrix including the shifts eqs.~(\ref{eq:tad1})--(\ref{eq:tad2}). The two-loop self-energies are available for all real scalars in \SARAH/\SPheno. For charged scalars, the scalar masses are available at the one-loop level,
\begin{equation}
 M_{\phi}^{(1L)}(p^2) = \tilde{M}^{(1L)}_{\phi} - \Pi_{\phi}(p^2)^{(1L)} \,.
\end{equation}
The calculation of the one-loop self-energies in both cases is done iteratively for each eigenvalue $i$ until the on-shell condition
\begin{equation}
  \left[\text{eig} M_{\phi}^{(n)}(p^2=m^2_{\phi_i})\right]_i \equiv m_{\phi_i}^2
\end{equation}
is fulfilled. 

We present the explicit expressions for the one-loop corrections to the tadpoles and self-energies in appendix~\ref{app:loopcorrections}. The two-loop corrections are too long to be presented in this work. However, they are available in the {\tt Mathematica} format and can be sent on demand or generated automatically with \SARAH. \\
\end{enumerate}
In the \MS scheme, all masses receive finite corrections at the loop level. Thus, mass parameters used as input are only Lagrangian (\MS) parameters which are different from the values of the pole masses. 
This has the drawback that one can't use physical parameters as input. On the other side, as we already mentioned, it makes the presence of large loop corrections immediately visible. Moreover, if one wants to draw the connection to a more fundamental theory which predicts the Lagrangian parameter at a higher scale, one must start with the running \MS parameters at the given scale and include the higher order corrections to all masses.

\subsubsection{On-shell renormalisation}
\label{sec:OS}
In an OS scheme, the tree-level masses and rotation angles are taken to be equivalent to the loop-corrected ones. Therefore, an OS scheme has the advantage that physical parameters can directly be used as input. This scheme is often the preferred option if a sufficient number of free parameters exists to absorb all finite corrections. However, one needs to be careful as this is not always possible. The best known exception are supersymmetric models: the SUSY relations among the terms in the potential reduce the number of free parameters and make a full OS calculation of the Higgs sector even in the simplest models impossible. Also the custodial symmetry of the GM model reduces the number of free parameters and there are not sufficient CTs to renormalise the scalar sector on-shell. Therefore, we need to give up this symmetry at the loop level and introduce CTs for all potential parameters
\begin{equation}
\label{eq:CTshift}
x \to x + \delta_x\,, \hspace{1cm} x \in \{\lambda_1,\lambda_{2a}, \lambda_{2b}, \lambda_{3a},\lambda_{3b},\lambda_{3c},\lambda_{4a},\lambda_{4b},\lambda_{4c},\lambda_{5a},\lambda_{5b},M_{1a},M_{1b},M_2, \mu_2^2, m_\eta^2, m_\chi^2\} 
\end{equation}
With this extended set of CTs, it is now possible to renormalise the scalar sector completely on-shell. We are doing this at the one-loop level using a similar ansatz as in Ref.~\cite{Basler:2016obg} for the THDM. The CTs
are fixed by the following renormalisation conditions:
\begin{align}
\label{eq:RenCond1}
\delta T_i + \delta^{(1L)} t_i \equiv & 0 \,, \\
\delta M^h + \Pi^h \equiv & 0  \,,\\
\delta M^A + \Pi^A \equiv & 0  \,,\\
\delta M^{H^+} + \Pi^{H^+} \equiv & 0  \,,\\
\label{eq:RenCond2}
\delta M^{H^{++}} + \Pi^{H^{++}} \equiv & 0 \,.  
\end{align}
Here, $\delta T_i$ and $\delta M^\Phi$ are the counter-term contributions to the tadpoles and mass matrices. $\delta t_i$ are the one-loop corrections to the tadpoles, and $\Pi^\Phi$ are the one-loop self-energies. For simplicity we assume that $\Pi^\Phi$ are calculated with vanishing external momentum, i.e. $p^2=0$. This approximation is justified because we are only interested in the overall size of the CTs and their impact on the vacuum stability constraints. The explicit expressions for the CTs stemming from these conditions are given in appendix~\ref{app:CTs}. \\
In order to obtain the finite values for the CTs, we perform the following steps:
\begin{enumerate}
 \item[1--3.] These steps to get the parameters at the renormalisation scale are identical to the \MS calculation.
 \item[4.] The one-loop tadpoles and self-energies are calculated.
 \item[5.] Eqs.~(\ref{eq:CT1})--(\ref{eq:CT2}) are used to obtain the finite values for the CTs. 
\end{enumerate}

As explained, the custodial symmetry gets broken by the one-loop shifts in the parameters $\lambda_i$. This effect is triggered by the hypercharge and one might expect that it is rather mild. This was 
for instance also found in Ref.~\cite{Blasi:2017xmc} where the effects of custodial symmetry breaking through RGE evolution have been studied. However, if we compare the dominant contributions to the CTs of $\delta\lambda_{3b}$ and $\delta\lambda_{3c}$, see eqs.~(\ref{eq:app:couterterm_lam3b}) and (\ref{eq:app:couterterm_lam3c}), we find for small $s_H$
\begin{equation}
\delta\lambda_{3b}- \delta\lambda_{3c} \simeq \frac{1}{s_H^2 v^2}\left(\Pi^{H^{++}} + \Pi^A_{22} -  2\, \Pi^{H^+}_{33}\right)\,.
\end{equation}
Including only the contributions from the ew gauge couplings and assuming the new scalars to be degenerate, we get for $m_5 \gg v$
\begin{equation}
\delta\lambda_{3b}- \delta\lambda_{3c} \simeq \frac{g_1^2}{16\pi^2}\frac{m_5^2}{s_H^2 v^2} \,,
\end{equation}
i.e. the effects are enhanced by a factor $m^2_5/(s_H v)^2$. 

\section{Theoretical constraints}
\label{sec:constraints}
\subsection{Tree-level unitarity constraints}
The first, and already at tree level rather severe, constraint on the parameter space of the GM model is perturbative unitarity of the 
 $2 \to 2$ scalar field scattering amplitudes. This means that the 0th partial wave amplitude $a_0$ must satisfy either $|a_0| \leq 1$ or $|{\mathcal Re} [ a_0]| \leq \frac{1}{2}$.  
The scattering matrix element $\mathcal{M}$  is given by 
\begin{equation}
\mathcal{M} = 16\pi\sum_J (2J+1) a_J P_J(\cos\theta),
\end{equation}
where $J$ is the angular momentum and $P_J(\cos\theta)$ are the Legendre polynomials.
At the tree level, 
 the $2 \to 2$ amplitudes are real, which is why one usually uses the more severe constraint  $|{\mathcal Re} [ a_0]| \leq \frac{1}{2}$, which leads to $|\mathcal{M}| < 8 \pi$. 
For analysing whether perturbative unitarity is given or not, it is common to work in the high energy limit, i.e. the dominant tree-level diagrams contributing to $|\mathcal M|$ involve only quartic interactions. All other diagrams with propagators are suppressed by the collision energy squared and can be neglected.  Moreover, effects of electroweak symmetry breaking (EWSB) are usually ignored, i.e. Goldstone bosons are considered as physical fields. 

The condition $|\mathcal{M}| < 8 \pi$ must be satisfied by all of the eigenvalues $\tilde{x}_i$ of the scattering matrix $\mathcal{M}$. $\mathcal{M}$ must be derived by  including each possible combination of two scalar fields in the initial and final states. The explicit expressions for the eigenvalues $\tilde{x}$ for conserved custodial symmetry are for instance given in Refs.~\cite{Aoki:2007ah,Hartling:2014zca}. They can be translated into the following tree-level unitarity conditions:
\begin{eqnarray}
\sqrt{ \left( 6 \lambda_1 - 7 \lambda_3 - 11 \lambda_4 \right)^2 + 36 \lambda_2^2}
+ \left| 6 \lambda_1 + 7 \lambda_3 + 11 \lambda_4 \right| &<& 4 \pi \,, \nonumber \\
\sqrt{ \left( 2 \lambda_1 + \lambda_3 - 2 \lambda_4 \right)^2 + \lambda_5^2}
+ \left| 2 \lambda_1 - \lambda_3 + 2 \lambda_4 \right| &<& 4 \pi \,, \nonumber \\
\left| 2 \lambda_3 + \lambda_4 \right| &<& \pi \,, \nonumber \\
\left| \lambda_2 - \lambda_5 \right| &<& 2 \pi \,.
\label{eq:uni}
\end{eqnarray}

Without the custodial symmetry, the eigenvalues of the scattering matrix have not been calculated before, but are given here for the first time. Still most eigenvalues of the more complicated scattering matrix have simple, analytical expressions:
\begin{align}
\pm 8 (2 {\lambda}_{3b}  + {\lambda}_{4b})\,,&  \\
\pm 4 (2 {\lambda}_{4b}  + {\lambda}_{3b})\,,&  \\
\pm 2 \left(2 {\lambda}_{3a}  + 2 ({\lambda}_{4a} + {\lambda}_{4b}) + 3 {\lambda}_{3b}  \pm \sqrt{(2 \lambda_{3a}-3 \lambda_{3b}+2 \lambda_{4a}-2 \lambda_{4b})^2+8 \lambda_{3c}^2} \right)\,,&  \\
\pm 8 (2 {\lambda}_{3c}  + {\lambda}_{4c})\,,&  \\
\pm 4 (2 {\lambda}_{4c}  + {\lambda}_{3c})\,,&  \\
\pm (4 {\lambda}_{2a}   - {\lambda}_{5a}) \,,&  \\
\pm 2 (2 {\lambda}_{2a}  + {\lambda}_{5a})\,,&  \\
\pm 2 \left(2 {\lambda}_{1}  + 2 {\lambda}_{4b}  - {\lambda}_{3b}  \pm \sqrt{(2 \lambda_1+\lambda_{3b}-2 \lambda_{4b})^2+\lambda_{5a}^2}\right)\,,&  \\
\pm 2 \left(2 {\lambda}_{1}  + 2 {\lambda}_{4c}   - {\lambda}_{3c}  \pm \sqrt{(2 \lambda_1+\lambda_{3c}-2 \lambda_{4c})^2+\lambda_{5b}^2}\right)\,,&  \\
\pm \frac{1}{2} \left(-4 {\lambda}_{2a}   -4 {\lambda}_{2b}   - {\lambda}_{5a}   \pm \sqrt{(4 \lambda_{2a}-4 \lambda_{2b}+\lambda_{5a})^2+8 \lambda_{5b}^2} \right)\,,&  \\
\pm \left(2 {\lambda}_{2a}  + 2 {\lambda}_{2b}   - {\lambda}_{5a}   \pm \sqrt{(-2 \lambda_{2a}+2 \lambda_{2b}+\lambda_{5a})^2+8 \lambda_{5b}^2}\right) \,.&  
\end{align}
Three other eigenvalues are the solutions $x_{1,2,3}$ of the polynomial
\begin{align}
 0 = & 384 \Big[2 {\lambda}_{1} \left(5 ({\lambda}_{3a} + {\lambda}_{4a})(3 {\lambda}_{3b}  + 4 {\lambda}_{4b} ) + (3 {\lambda}_{4c}  + {\lambda}_{3c})^{2}\right) \nonumber \\
  & \hspace{1cm}- 2 {\lambda}_{2a} {\lambda}_{2b} (3 {\lambda}_{4c}  + {\lambda}_{3c}) +5 {\lambda}_{2a}^{2} ({\lambda}_{3a} + {\lambda}_{4a}) + {\lambda}_{2b}^{2} (3 {\lambda}_{3b}  + 4 {\lambda}_{4b} )\Big] \nonumber \\
+ &x \left[16 (2 (3 {\lambda}_{1} (5 {\lambda}_{3a}  + 5 {\lambda}_{4a}  + 6 {\lambda}_{3b}  + 8 {\lambda}_{4b} ) - (3 {\lambda}_{4c}  + {\lambda}_{3c})^{2}  + 5 ({\lambda}_{3a} + {\lambda}_{4a})(3 {\lambda}_{3b}  + 4 {\lambda}_{4b} )) -3 {\lambda}_{2b}^{2}  -6 {\lambda}_{2a}^{2} )\right] \nonumber \\
\label{eq:UniPoly}
+ &x^2 \left[4 (5 {\lambda}_{3a}  + 5 {\lambda}_{4a}  + 6 {\lambda}_{1}  + 6 {\lambda}_{3b}  + 8 {\lambda}_{4b} ) \right] + x^3\,.
\end{align}
These expressions were extracted from \SARAH as described in appendix~\ref{app:scattering}.\\

From these eigenvalues, we can derive the following tree-level unitarity conditions in the case of broken custodial symmetry
\begin{align}
\label{eq:uni1}
|2 {\lambda}_{3b}  + {\lambda}_{4b}| & < \pi   \,, \\
|2 {\lambda}_{3c}  + {\lambda}_{4c}| & < \pi   \,,\\
|2 {\lambda}_{3a}  + 2 ({\lambda}_{4a} + {\lambda}_{4b}) + 3 {\lambda}_{3b}|  + \sqrt{(2 \lambda_{3a}-3 \lambda_{3b}+2 \lambda_{4a}-2 \lambda_{4b})^2+8 \lambda_{3c}^2} ) & < 4\pi  \,,\\
|2 {\lambda}_{4b}  + {\lambda}_{3b}| & < 2\pi   \,,\\
|2 {\lambda}_{4c}  + {\lambda}_{3c}| & < 2\pi  \,,\\
|4 {\lambda}_{2a}  - {\lambda}_{5a}| & < 8 \pi \,, \\
|2 {\lambda}_{2a}  + {\lambda}_{5a}|& < 4\pi  \,,\\
|4 {\lambda}_{2a} +4 {\lambda}_{2b}   + {\lambda}_{5a}| +  \sqrt{(4 \lambda_{2a}-4 \lambda_{2b}+\lambda_{5a})^2+8 \lambda_{5b}^2} & < 16\pi  \,,\\
|2 {\lambda}_{2a}  + 2 {\lambda}_{2b}   - {\lambda}_{5a} |  + \sqrt{(-2 \lambda_{2a}+2 \lambda_{2b}+\lambda_{5a})^2+8 \lambda_{5b}^2} & < 8 \pi  \,,\\ 
|2 {\lambda}_{1}  + 2 {\lambda}_{4b}  - {\lambda}_{3b}| + \sqrt{(2 \lambda_1+\lambda_{3b}-2 \lambda_{4b})^2+\lambda_{5a}^2}& < 4\pi  \,,\\
|2 {\lambda}_{1}  + 2 {\lambda}_{4c}  - {\lambda}_{3c}|  + \sqrt{(2 \lambda_1+\lambda_{3c}-2 \lambda_{4c})^2+\lambda_{5b}^2}& < 4\pi  \,,\\
\label{eq:uni2}
\text{max}(x_{1,2,3}) & < 8\pi \,.
\end{align}
Even though we will always assume that the custodial symmetry is conserved at tree-level, we will make use of these `generalised' unitarity constraints in combination with 
new perturbativity constraints as explained in sec.~\ref{sec:constraints_per}.

\subsection{Vacuum stability  constraints}
\subsubsection{Tree-level considerations}
\label{sec:constraints_ufb}
Another theoretical constraint which has already been studied in the context of the GM model is the vacuum stability constraint at tree-level. In general, there are two possible situations which can cause an instability of the 
vacuum with correct EWSB: either directions in the scalar potential exist in which the potential is unbounded from below, or other local minima exist in the scalar potential which are deeper than the ew one. 

\paragraph{Unbounded from below}
UFB directions exist if quartic couplings fulfil specific conditions. The simplest condition is $\lambda_1<0$ since in this case the potential approaches $-\infty$ for $v_\phi\to \infty$.
The full list of tree-level conditions to avoid unboundedness from below in the case of conserved custodial symmetry was derived in Ref.~\cite{Hartling:2014zca}. It reads:
\begin{eqnarray}
\lambda_1 &>& 0\,,  \\
\nonumber \\
\lambda_4 &>& \left\{ \begin{array}{l l}
- \frac{1}{3} \lambda_3\,, &\quad\ \lambda_3 \geq 0\,, \\
- \lambda_3\,, &\quad  \ \lambda_3 < 0\,, \end{array} \right. \\
\nonumber \\
\lambda_2 &>& \left\{ \begin{array}{l l}
\frac{1}{2} \lambda_5 - 2 \sqrt{\lambda_1 \left( \frac{1}{3} \lambda_3 + \lambda_4 \right)}\,, &
\quad 
 \quad \ \lambda_5 \geq 0 \hspace{.2cm} \land \hspace{.2cm} \lambda_3 \geq 0\,, \\
\omega_+(\zeta) \lambda_5 - 2 \sqrt{\lambda_1 ( \zeta \lambda_3 + \lambda_4)}\,, &
\quad 
 \quad  \ \lambda_5 \geq 0 \hspace{.2cm} \land \hspace{.2cm} \lambda_3 < 0\,, \\
\omega_-(\zeta) \lambda_5 - 2 \sqrt{\lambda_1 (\zeta \lambda_3 + \lambda_4)}\,, &
\quad  \quad  \ \lambda_5 < 0\,, 
\end{array} \right.
\label{eq:bfbcond2}
\end{eqnarray}
with 
\begin{equation}
\omega_{\pm}(\zeta) = \frac{1}{6}(1 - B(\zeta)) \pm \frac{\sqrt{2}}{3} \left[ (1 - B(\zeta)) \left(\frac{1}{2} + B(\zeta)\right)\right]^{1/2}\,, \qquad B(\zeta) = \sqrt{\frac{3}{2}\left(\zeta - \frac{1}{3}\right)} \,.
\end{equation}
Eq.~(\ref{eq:bfbcond2}) must be satisfied for all values of $\zeta \in \left[ \frac{1}{3}, 1 \right]$. \\

If the custodial symmetry is broken, more conditions need to be checked. These conditions were derived in Ref.~\cite{Blasi:2017xmc} assuming two simultaneously non-vanishing field directions. 
We have re-derived these conditions using our parametrisation, shown below. Not derived in Ref.~\cite{Blasi:2017xmc} were UFB conditions on the ``custodial'' direction with $\langle \eta^0\rangle = \langle \chi^0 \rangle \neq 0$ and $\langle \phi^0 \rangle \neq 0$ which we also present here. The set of UFB conditions reads
\begin{align}
\lambda_1 >& 0     \,, \\
\lambda_{3a} + \lambda_{4a} > &0  \,, \\
\lambda_{3b} + 2 \lambda_{4b} >& 0 \,, \\
\lambda_{3b} + \lambda_{4b} > &0   \,, \\
\lambda_{3c} + \sqrt{2} \sqrt{(\lambda_{3a} + \lambda_{4a}) (\lambda_{3b} + 2 \lambda_{4b})} +   2 \lambda_{4c} >& 0   \,,  \\
\sqrt{2} \sqrt{(\lambda_{3a} + \lambda_{4a}) (\lambda_{3b} + 2 \lambda_{4b})} +   2 \lambda_{4c} >& 0 \,, \\
\sqrt{(\lambda_{3a} + \lambda_{4a}) (\lambda_{3b} + \lambda_{4b})} +   \lambda_{4c} > &0  \,, \label{eq:UFB_diff1}\\
\lambda_{3c} + \sqrt{(\lambda_{3a} + \lambda_{4a}) (\lambda_{3b} + \lambda_{4b})} +   \lambda_{4c} > &0\,, \label{eq:UFB_diff2}\\ 
4 \lambda_{2a} + 4 \sqrt{2} \sqrt{\lambda_1 (\lambda_{3b} + 2 \lambda_{4b})} - \lambda_{5a}  >& 0 \,, \\
4 \lambda_{2a} + 4 \sqrt{2} \sqrt{\lambda_1 (\lambda_{3b} + 2 \lambda_{4b})} +   \lambda_{5a} >& 0 \,, \\
\lambda_{2a} + 2 \sqrt{ \lambda_1 (\lambda_{3b} +   \lambda_{4b})} >& 0 \,, \\
\lambda_{2b} + 2 \sqrt{   \lambda_1 (\lambda_{3a} +      \lambda_{4a})} >& 0\,, \\ 
\lambda_{3a} + 2 \lambda_{3b} + \lambda_{4a} + 4 \lambda_{4b} +   4 \lambda_{4c} >& 0 \,, \label{eq:UFB_secon_to_last_line}\\ 
2 \lambda_{2a} + \lambda_{2b} +  2 \sqrt{\lambda_1 (\lambda_{3a} + 2 \lambda_{3b} + \lambda_{4a} +    4 (\lambda_{4b} +     \lambda_{4c}))} - (\lambda_{5a}/2) - \lambda_{5b}  >&0\,.
\end{align}
Note that after translating the parameters according to eqs.~(\ref{eq:parameter_translation_begin})--(\ref{eq:parameter_translation_end}), eqs.~(\ref{eq:UFB_diff1}) and (\ref{eq:UFB_diff2}) differ w.r.t. the fourth and fifth line in eq.~(12) of Ref.~\cite{Blasi:2017xmc} in that we do not have a factor 2 in front of the square-root.
The last two conditions are shown here for the first time. \\

\paragraph{Other minima} 
Even if the potential is bounded from below in all possible field directions, additional minima are usually present. In these minima, the sum of all neutral VEVs $\sqrt{v_\eta^2+4(v_\chi^2+v_\phi^2)}$
usually doesn't agree 
with the measured value of 246~GeV, i.e. those minima are not viable. Moreover, also minima can occur at which charge is broken spontaneously by the VEV of a charged scalar. If any of those minima is the global minimum of the scalar tree-level potential, then the ew minimum is unstable. If tunnelling is assumed to be
instantaneous on cosmological scales -- as it is usually done -- this vacuum configuration is forbidden. All possible minima of the scalar potential at tree-level can be found by solving the minimisation conditions of the potential, i.e. in the most general case a set of seven coupled, cubic equations must be solved. Another method of finding all minima by using a re-parametrisation of the scalar potential is discussed in Ref.~\cite{Hartling:2014zca}.

\subsubsection{Loop effects}
\begin{figure}[tb]
\centering
\includegraphics[width=0.5\linewidth]{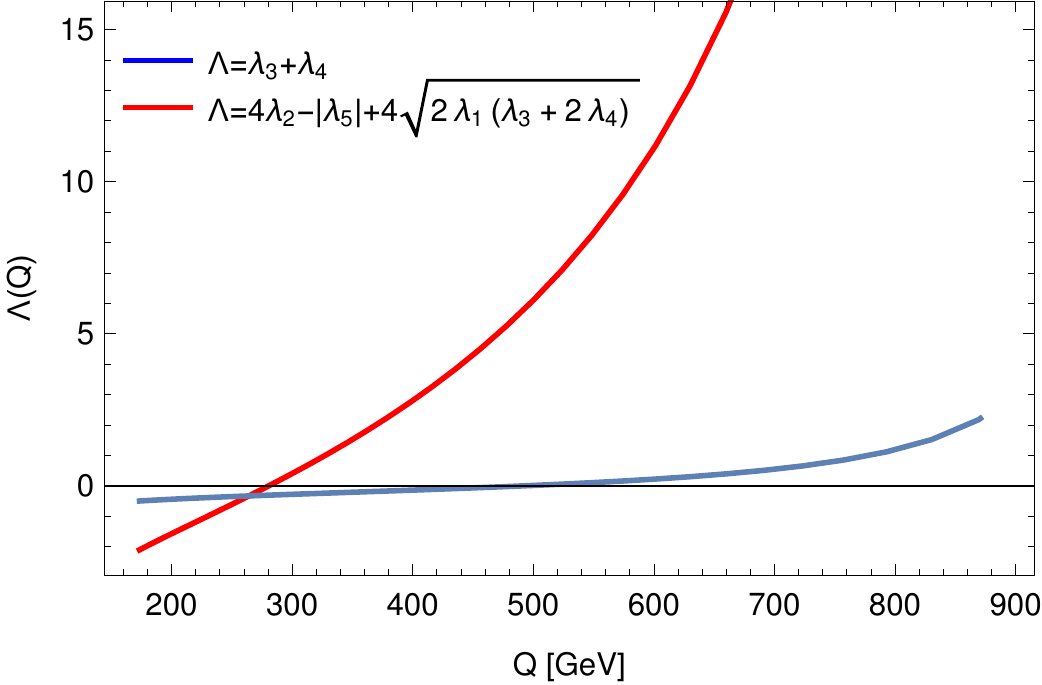} 
\caption{The scale dependence of two combinations of quartic couplings. The input parameters at $m_t$ have been $\lambda_1 = 0.05$, $\lambda_2 = 0.5$, $\lambda_3= -1.5$, $\lambda_4 = 1$, $\lambda_5=5$.  Negative 
values would point towards UFB directions, i.e. it is shown that these directions disappear at large energy scales. }
\label{fig:rge}
\end{figure}

Up to now, the vacuum stability in the GM model has  only been checked at tree level. However, using UFB conditions at tree level can be very misleading since these conditions involve very large field excursions -- which demand a proper treatment of radiative corrections. 
Usually, the best method to deal with very large field excursions is to use the RGE-improved potential. In the limit of very large VEVs, the potential can be approximated very well by 
the tree-level potential where the running quartic couplings are inserted. Thus, to single out UFB directions, it is necessary to check that the conditions hold in the limit
\begin{equation}
\lambda_N \to \lambda_N(Q)|_{Q\to \infty}\,, \qquad N=1,\dots,5 \,.
\end{equation}
It has already been pointed out in the context of the THDM \cite{Staub:2017ktc} that UFB directions usually disappear in the RGE-improved potential in the presence of large quartic couplings.  
This can be understood from the general form of the RGEs: bosonic contributions increase the size of the quartic couplings with increasing energy while fermionic contributions decrease them.
We can confirm that a similar observation holds in the GM model. If we forget for a moment the breaking of the custodial symmetry via hypercharge effects, the one-loop RGEs for the quartic couplings 
are given by \cite{Blasi:2017xmc}:
\begin{align}
16\pi^2 \beta_{\lambda_1} =&\frac{9 g_2^4}{32}-9 g_2^2 \lambda_1+\frac{3}{2} \left(64 \lambda_1^2+8 \lambda_1 y_t^2+12 \lambda_2^2+\lambda_5^2-y_t^4\right)\,,\\
16\pi^2 \beta_{\lambda_2} =&\frac{3 g_2^4}{2}-\frac{33 g_2^2 \lambda_2}{2}+2 \lambda_2 \left(24 \lambda_1+8 \lambda_2+28 \lambda_3+44 \lambda_4+3 y_t^2\right)+4 \lambda_5^2\,,\\
16\pi^2 \beta_{\lambda_3} =&\frac{3 g_2^4}{2}-24 g_2^2 \lambda_3+80 \lambda_3^2+96 \lambda_3 \lambda_4-\lambda_5^2\,,\\
16\pi^2 \beta_{\lambda_4} =&\frac{3 g_2^4}{2}-24 g_2^2 \lambda_4+8 \left(\lambda_2^2+3 \lambda_3^2+14 \lambda_3 \lambda_4+17 \lambda_4^2\right)+\lambda_5^2\,,\\
16\pi^2 \beta_{\lambda_5} =&\frac{1}{2} \lambda_5 \left(-33 g_2^2+8 (4 \lambda_1+8 \lambda_2-2 \lambda_3+4 \lambda_4+\lambda_5)+12 y_t^2\right)\,.
\end{align}
We show the running of the two coupling combinations $\lambda_3+\lambda_4$ and $4 \lambda_2-|\lambda_5| + 4 \sqrt{2} \sqrt{\lambda_1 (\lambda_3 + 2 \lambda_4)}$ in Fig.~\ref{fig:rge}.
The input values at $m_t$ were chosen to be $\lambda_1 = 0.05$, $\lambda_2 = 0.5$, $\lambda_3= -1.5$, $\lambda_4 = 1$, $\lambda_5=5$, i.e. both combinations of couplings are negative at $m_t$. This would give the 
impression of two UFB directions. However, we already see at scales which are below one TeV that both combinations of couplings turn positive. Thus, the UFB directions disappear once the loop effects are 
included. Since the running of the quartic couplings is usually very fast and since the scale at which the couplings (or combinations of them) change their sign is not far above the input scale, we can assume that
the dominant radiative effects are also covered by the effective potential without an RGE resummation. See also Ref.~\cite{Staub:2017ktc} for a similar discussion in the context of the THDM. \\

We will use in our numerical studies the one-loop effective potential $V_{EP}^{(1)}$ to check the vacuum stability. The different ingredients are
\begin{equation}
V_{EP}^{(1)} = V_{\rm Tree} + V^{(1)}_{\rm CT}  + V^{(1)}_{\rm CW} \,.
\label{eq:one_loop_potential}
\end{equation}
Here,  $V^{(1)}_{\rm CT}$ is the counter-term potential, and the sum $V_{\rm Tree} + V^{(1)}_{\rm CT}$ 
is given by eq.~(\ref{eq:broken_custodial_ours}) and replacing
 \begin{equation}
 \lambda_N ~\to~ \tilde{\lambda}_{Nx} \equiv \lambda_N + \delta\lambda_{Nx}\,.
 \end{equation}
Note, the derived CTs depend on the ew VEVs, i.e. they result in a cancellation between $V_{\rm CT}$  and $V_{\rm CW}$ {\it only} at the ew minimum, but not at other positions of the potential. Thus, the conditions eqs.~(\ref{eq:RenCond1})--(\ref{eq:RenCond2}) don't imply that the full one-loop potential is in general identical to the tree-level potential. 
%
The Coleman-Weinberg 
potential $V^{(1)}_{\rm CW}$ is given by \cite{Coleman:1973jx}
\begin{equation}
V^{(1)}_{\rm CW} = \frac{1}{16\pi^2} \sum_{i}^{\text{all fields}} r_i s_i C_i m_i^4 \left(\log\frac{m_i^2}{Q^2} - c_i \right)\,,
\end{equation}
with $r_i = 1$ for real bosons or Majorana fermions, otherwise 2; $C_i =3$ for quarks, otherwise 1; $\{s_i,c_i\}=\{-\frac{1}{2},\frac{3}{2}\}$ for fermions, $\{\frac{1}{4},\frac{3}{2}\}$ for scalars 
and $\{\frac{3}{4},\frac{5}{6}\}$ for vector bosons. It is important to stress that, for the check for spontaneous charge breaking via VEVs of the charged scalars at the loop level, the calculation of the physical masses must be adjusted. The reason is that the additional VEVs, which can be potentially present, also mix particles with different electric charge. If we assume no spontaneous CP violation, this results in a $7\times 7$  mass matrix for CP-even ($\Phi$) and a $6\times 6$ mass matrix for CP odd scalars ($\Sigma$). Analogously, both fermions and vector bosons of the same colour mix. Thus, in this case the more explicit expression for the CW potential is
\begin{align}
V^{(1)}_{\rm CW} =& \frac{1}{16\pi^2} \Bigg[ \frac{1}{4} \left( \sum_{i=1}^7 m_{\Phi_i}^4 \left(\log\frac{m_{\Phi_i}^2}{Q^2} - \frac{3}{2} \right) 
 + \sum_{i=1}^6 m_{\Sigma_i}^4 \left(\log\frac{m_{\Sigma_i}^2}{Q^2} - \frac{3}{2} \right) \right)\nonumber \\
& - 3 \sum_{i=1}^6 m_{Q_i}^4 \left(\log\frac{m_{Q_i}^2}{Q^2} - \frac{3}{2} \right) - \frac{1}{2} \sum_{i=1}^9 m_{L_i}^4 \left(\log\frac{m_{L_i}^2}{Q^2} - \frac{3}{2} \right) \nonumber \\
& +  \frac{3}{4} \sum_{i=1}^4 m_{V_i}^4 \left(\log\frac{m_{V_i}^2}{Q^2} - \frac{5}{6} \right) 
\Bigg]\,.
\end{align}
Because of the length of the mass matrices in the case of charge breaking VEVs, we don't give them explicitly in this paper. Instead, we provide on request the \SARAH model files for the charge breaking GM model to generate them.

\subsection{Perturbativity constraints}
\label{sec:constraints_per}
As we have seen, the GM model provides in principle a sufficient number of CTs to renormalise all masses on-shell if the custodial symmetry is given up at the loop level.  However, this does not yet ensure that such an on-shell calculation would also be trustworthy as one always needs to assume that the perturbative expansion is working. If this is not the case, then the calculation of the CTs and also all other loop calculations are not meaningful. Naively, one might expect that problems with perturbativity occur once quartic couplings $\mathcal O(4\,\pi)$ are involved, or that at least the tree-level unitarity constraints are strong enough to filter out points which violate perturbation theory. However, it was shown for the SM that problems can occur much earlier \cite{Nierste:1995zx} and that a better limit for the quartic coupling in the SM is $2\,\pi$. In the specific case of the GM model, it was pointed out in Ref.~\cite{Braathen:2017izn} that problems with perturbation theory can show up for even smaller coupling values. It was observed that, in sizeable regions of the parameter space of the GM model, the two-loop corrections to masses can become larger than the one-loop contributions. This was demonstrated at the example of the SM-like Higgs mass for which the one- and two-loop corrections in the $\overline{\text{MS}}$ scheme were compared. Of course, for a robust statement about whether perturbation theory is still working or has already broken down, it would be necessary to compare physical processes and their scale dependence at different loop levels. However, this is hardly possible in the GM model -- or any other BSM model. Therefore, we want to use information which is easier accessible to get some indication whether loop calculations for a given parameter point are trustworthy or not. For this, we are going to check the effects of four different conditions which might point towards the breakdown of perturbation theory. These conditions are:
\begin{enumerate}
 \item A parameter point is considered to violate perturbation theory if the two-loop corrections to at least one scalar mass are larger than the one-loop corrections, i.e.
 \begin{equation} 
  |(m_{\phi}^{2})_{\rm Tree} - (m_\phi^{2})_{\rm 1L}| < |(m_\phi^{2})_{\rm 2L} - (m_\phi^{2})_{ \rm 1L}| \,.
  \label{eq:size_2loop_vs_1loop}
 \end{equation}
 This is very close to the ansatz of Ref.~\cite{Braathen:2017izn}. However, we do not only consider the SM-like mass, but test all three neutral CP-even states, i.e. $\phi=h_{1,2,3}$. In addition, we impose a lower threshold on $|(m_{\phi}^{2})_{\rm Tree} - (m_\phi^{2})_{\rm 1L}|$ of $20^2~\text{GeV}^2$ for this test. 
The reason for this exception is that the one-loop corrections might be very small due to an accidental cancellation which is not any more present at two-loop level -- which would therefore otherwise lead to a constraint according to eq.~(\ref{eq:size_2loop_vs_1loop}) without actually violating perturbation theory.
 \item A parameter point is considered to violate perturbation theory if the CT to at least one parameter is larger than the tree-level value of this parameter times some constant, i.e. 
 \begin{equation}
  \left|\frac{\delta x}{x}\right| > v\,.
 \end{equation}
 For the most conservative choice, $v=1$, this forbids points with $|\delta x|>|x|$. We apply this constraint to all quartic couplings.
 \item A parameter point is considered to violate perturbation theory if the CT of at least one quartic coupling becomes larger than some fixed value, i.e.  
 \begin{equation}
  |\delta x| >  c \cdot \pi\,.
 \end{equation}
 We are going to test  $c$ within 1 and 4. Since $\delta x$ enters the two-loop corrections, a CT as large as $4\pi$ is for sure problematic. However, problems might occur even for smaller values as one has seen in the SM. 
 \item A parameter point  is considered to violate perturbation theory  if the generalised unitarity constraints eqs.~(\ref{eq:uni1})-(\ref{eq:uni2}) are violated when inserting the renormalised couplings, i.e.  
 \begin{equation}
 |\mathcal{M}(\lambda_{Nx} \to \lambda_N + \delta\lambda_{Nx})| > 8 \pi \,. 
 \label{eq:generalized_unitarity_bound} 
 \end{equation}
This condition is similar to the third condition but doesn't involve any (arbitrary) upper limit on the quartic couplings. In addition, it indicates the robustness of the unitarity constraints under radiative corrections. Of course, to be sure if the unitarity constraints are really violated or not, one would need to calculate in addition the virtual and real corrections to all possible $2\to 2$ scattering processes. 
\end{enumerate}
All four conditions are not rigorous in the sense that they can provide a definite answer if perturbation theory is still working or not. It is also in some sense a matter of taste which condition is considered as the most reasonable or reliable one. However, as we will see, one can get some very clear hints if problems with perturbation theory are present or not. In particular, if several conditions fail at the same time, one should be tempted to take results obtained from a calculation at Born- or even one-loop-level with care.

\section{Results}
\label{sec:results}
\subsection{Numerical setup}
For our numerical study we used the {\tt Mathematica} package \SARAH \cite{Staub:2008uz,Staub:2009bi,Staub:2010jh,Staub:2012pb,Staub:2013tta} with the implementation of the GM Model discussed in Ref.~\cite{Staub:2016dxq}. 
In a first step, we used the model files to generate a spectrum generator based on {\tt SPheno}~\cite{Porod:2003um,Porod:2011nf}. {\tt SPheno} calculates by default the mass spectrum 
at the full one-loop level  and includes all important two-loop corrections to the neutral scalar masses \cite{Goodsell:2014bna,Goodsell:2015ira,Braathen:2017izn}. Special care is needed at the two-loop level 
to avoid the so-called Goldstone boson catastrophe \cite{Braathen:2016cqe}. 
In addition, \SPheno calculates all decay modes of the particles at tree- and one-loop-level \cite{Goodsell:2017pdq}, including the modes discussed recently in Ref.~\cite{Degrande:2017naf}. This information is also used to write the files necessary to test a parameter point with {\tt HiggsBounds} \cite{Bechtle:2008jh,Bechtle:2011sb,Bechtle:2013wla}.  \SPheno performs in addition a calculation of flavour and precision observables like $\delta\rho$ or $g-2$ \cite{Porod:2014xia}. \\
For this project, we have extended the list of precision observables by the oblique parameters $S$, $T$ and $U$ \cite{Peskin:1991sw}. The main reason for this was mainly that $\delta\rho$, respectively the $T$ parameter can't be used to constrain the GM model: even if the tree-level contribution vanishes  for $v_\chi=v_\eta$, the one-loop correction is formally next-to-leading order. Thus, a fine-tuned CT can be added to cancel this contribution in principle. Therefore, we use the $S$ parameter as main constraint as proposed in Ref.~\cite{Hartling:2014aga}. We also compared our numerical values of a full one-loop calculation of the $S$ parameter with those obtained with {\tt gmcalc} and usually found  agreement within 5--10\%.  \\
We have modified one instance of \SPheno to include the 
CTs to keep the scalar sector on-shell as explained in sec.~\ref{sec:OS}. A second version was kept unmodified and used to get the size of the one- and two-loop corrections to the masses in the \MS scheme, see sec.~\ref{sec:MS} for more details. \\
We further used \Vevacious\cite{Camargo-Molina:2013qva} to test the stability of the one-loop effective potential. The necessary model files have been generated with \SARAH. Here, we use two two different implementations: the standard one with only the three standard VEVs for the neutral scalars, and one with the possibility that all seven scalars can obtain a VEV. Since the check of the vacuum stability with seven VEVs is quite time consuming, we only test points which have passed all other constraints.\footnote{We find that roughly 1\% of the otherwise valid points have a deeper global minimum 
where electric charge is broken spontaneously.} 
As input for \Vevacious, we used the spectrum files written by \SPheno. \Vevacious automatically adjusts the counter-term potential based on CTs which are present in the \SPheno spectrum file.\footnote{In practice, \Vevacious checks for SLHA blocks starting with {\tt TREE} and {\tt LOOP} in the spectrum files.} \\ 
We made use of {\tt gmcalc} \cite{Hartling:2014xma} to double-check the tree-level constraints like tree-level unitarity, unboundedness from below and the presence of other minima as well as the calculation of the $S$ parameter as already mentioned. In order to circumvent the command line input, we have modified {\tt gmcalc} to read in a file with all necessary parameters ($\lambda_i, M_i, \mu_i^2$) as well as the running electroweak VEV as calculated by \SPheno. In addition, we added a function to write the results for the tree-level unitarity check, the check for other minima as well as unbounded-from-below directions into an external file in a SLHA-like format. The different codes are combined in numerical scans using the tool {\tt SSP} \cite{Staub:2011dp}. \\

\subsection{Input parametrisation}
At the tree-level, i.e. with conserved custodial symmetry, and after applying the tadpole conditions, there are seven free Lagrangian parameters 
\begin{eqnarray*}
  & \lambda_1,\ \lambda_2,\ \lambda_3,\ \lambda_4,\ \lambda_5,\ M_1,\ M_2\,.&
 \end{eqnarray*}
In addition, the relative size of the $SU(2)$-doublet and -triplet VEVs, controlled by $s_H$, is a free parameter.
In principle, these parameters could directly be used as input. However, the overall majority of randomly chosen points would then be ruled out by the requirement to have a CP-even scalar mass with $\sim 125$~GeV. Therefore, it is convenient to use $m_h$ directly as input. We have explored two different sets of input parameters:

\begin{enumerate}
 \item {\bf Input I}: here, we use the SM-like Higgs mass $m_h$ as input, together with the mixing angle $\alpha$ between the CP-even neutral $SU(2)$-doublet and -triplet components. In addition, the heavy Higgs mass $m_H$ is used as input to set the overall 
 mass scale of the new scalars. Using those input parameters, $\lambda_1$, $M_1$ and $M_2$ are calculated according to
 \begin{align}
M_1 =   & \frac{3 s_H \sqrt{2-2 s_H^2} \left(t_\alpha^2+1\right) v^2 (2 \lambda_2-\lambda_5)-2 \sqrt{3} m_h^2 t_\alpha+2 \sqrt{3} m_H^2
   t_\alpha}{3 \sqrt{1-s_H^2} \left(t_\alpha^2+1\right) v}\,,\\
M_2 = & \frac{1}{9 s_H^2 \sqrt{1-s_H^2} \left(t_\alpha^2+1\right) v^2} \times \Big[v \big(m_h^2 t_\alpha \left(-3 \sqrt{2-2 s_H^2} s_H t_\alpha+2 \sqrt{3} s_H^2-2 \sqrt{3}\right) \nonumber \,,\\
   & -3 s_H \sqrt{2-2
   s_H^2} \left(t_\alpha^2+1\right) v^2 \left(2 \lambda_2 \left(s_H^2-1\right)+s_H^2 (-(\lambda_3+3 \lambda_4))-\lambda_5
   s_H^2+\lambda_5\right)\big) \nonumber \\
   & +m_H^2 v \left(-2 \sqrt{3} s_H^2 t_\alpha-3 \sqrt{2-2 s_H^2} s_H+2 \sqrt{3} t_\alpha\right)\Big]\,,\\
\lambda_1 =& -\frac{m_h^2+m_H^2 t_\alpha^2}{8 \left(s_H^2-1\right) \left(t_\alpha^2+1\right) v^2}\,,
\end{align}
where $t_\alpha = \tan \alpha$.
The full list of input parameters for this choice is 
  \begin{eqnarray}
  & m_h,\ m_H,\ \alpha,\ \lambda_2,\ \lambda_3,\ \lambda_4,\ \lambda_5,\ s_H\,.&
  \label{eq:input_choice_I}
 \end{eqnarray}
The advantage of this input is that Higgs constraints can easily be kept under control by choosing small or moderate values of $s_H$ and $\alpha$ at the same time. The disadvantage is that the independent handling of $s_H$ and $\alpha$ implies a tuning of the other dependent tree-level parameters. As a result, loop corrections can have a significant impact, as we will see below.
\item {\bf Input II}: here, we use $m_h$ and $m_5$ instead of $\lambda_1$ and $M_1$ as input. Moreover, $M_2$ is set relative to $M_1$ via a dimensionless parameter $r_{12}$. The used relations are 
\begin{align}
M_1 = & \frac{s_H \left(v^2 \left(3 \lambda_5 \left(s_H^2-1\right)-2 \lambda_3 s_H^2\right)+2 m_5^2\right)}{\sqrt{2} v \left((6
   r_{12}-1) s_H^2+1\right)} \,, \\
M_2 = & r_{12} M_1 \,,\\   
\lambda_1 =& \frac{1}{32 \left(s_H^2-1\right) v^2 \left(v \left(\sqrt{2} s_H^3 v
   (\lambda_3+3 \lambda_4)-s_H^2 (M_1+3 M_2)+M_1\right)-\sqrt{2} m_h^2 s_H\right)} \times \nonumber \\
  & \Big[3 s_H \left(s_H^2-1\right) v^2 \left(4 M_1 s_H v (\lambda_5-2 \lambda_2)+2 \sqrt{2} s_H^2 v^2
   (\lambda_5-2 \lambda_2)^2+\sqrt{2} M_1^2\right) \nonumber \\
  & +4 m_h^2 v \left(-\sqrt{2} s_H^3 v (\lambda_3+3 \lambda_4)+s_H^2
   (M_1+3 M_2)-M_1\right)+4 \sqrt{2} m_h^4 s_H\Big]\,.
\end{align}
Thus, the full list of free parameters for this input is
 \begin{eqnarray}
  & m_h, \ m_5,\ \lambda_2,\ \lambda_3,\ \lambda_4,\ \lambda_5,\ r_{12},\ s_H\,.&
  \label{eq:input_choice_II}
 \end{eqnarray}
The advantage of this input is that one has direct control over the SM-like Higgs mass as well as the BSM scale $\simeq m_5$. On the other side, the mixing between the SM-like Higgs and the other states 
is not an input, i.e. it can in principle become very large. This will then cause conflicts with Higgs coupling measurements. 
 \end{enumerate}
Other proposed input sets, which we don't explore further in the following, are: \\
$\{m_h,\lambda_2,\lambda_3,\lambda_4,\lambda_5,M_1,M_2\}$,
$\{m_h,m_H,m_3,m_5,\alpha,M_1,M_2\}$, and $\{m_h,m_5,\lambda_2,\lambda_3,\lambda_4,\,M_1,M_2\}$. \\

We are going to start now to investigate the loop constraints on the GM model using the two input sets defined above. We start with a discussion of the perturbativity constraints before we turn 
to the vacuum stability constraints. First, we concentrate on specific parameter regions to study the different effects. In a second step, we consider the global picture by performing random scans over 
large parameter ranges.

\subsection{Perturbativity constraints}
\subsubsection{Dependence on $s_H$}
\label{sec:results_per_sH}
\begin{figure}[tb]
\centering
\includegraphics[width=0.5\linewidth]{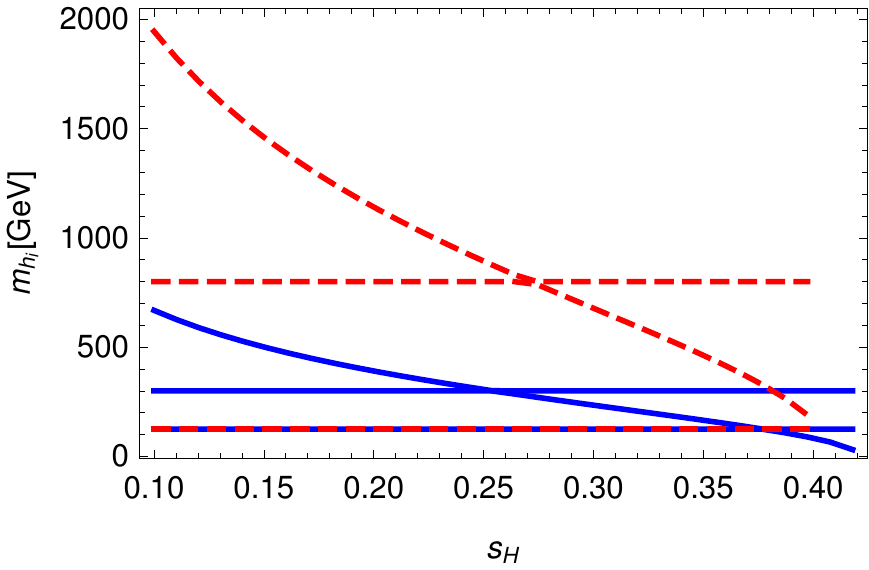}
\caption{The (tree-level) masses of the three CP-even states as function of $s_H$ for two different values of $m_H$: 300~GeV (blue) and 800~GeV (dashed red). The other input parameters are:
$\lambda_2 =0.1$, $\lambda_3 = 0.5$, $\lambda_4=-0.02$, $\lambda_5=0.1$, $\alpha=20\degree$.
}
\label{fig:mhi_sH}
\end{figure}
We start with discussing the role of $s_H$ since it was shown in Ref.~\cite{Braathen:2017izn} that large $s_H$ usually implies large radiative corrections. For this reason we consider the parameter point 
\begin{eqnarray*}
\lambda_2 =0.1\,,\ \lambda_3 = 0.5\,,\ \lambda_4=-0.02\,,\ \lambda_5=0.1\,,\ \alpha=20\degree\,, \quad \text{with} \quad m_H = 300~\text{GeV \, or \,} 800~\text{GeV}\,.
\end{eqnarray*}
The tree-level masses of the three CP-even scalars are shown in Fig.~\ref{fig:mhi_sH}. As can be seen, a large separation between $m_5$ and $m_H$ can be present for both small and large $s_H$. In general, the very heavy states for small values of $s_H$ do also increase the size of the loop effects as we will see. 
\begin{figure}[tb]
\includegraphics[width=0.41\linewidth]{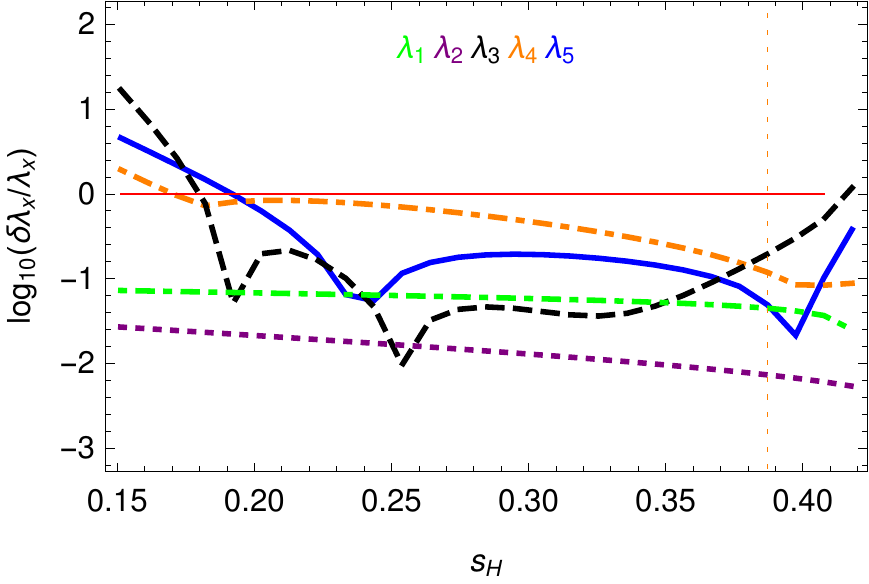} \hfill 
\includegraphics[width=0.41\linewidth]{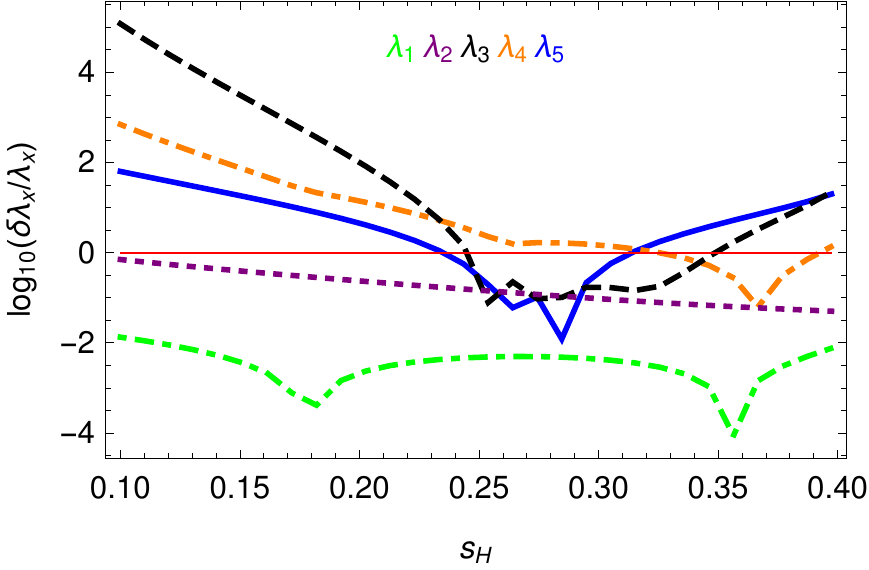} \\
\includegraphics[width=0.41\linewidth]{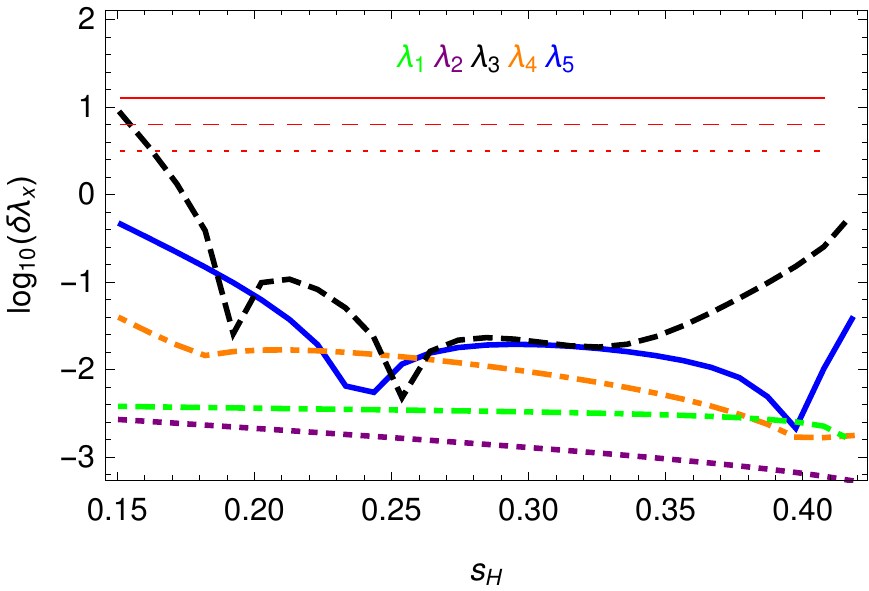} \hfill
\includegraphics[width=0.41\linewidth]{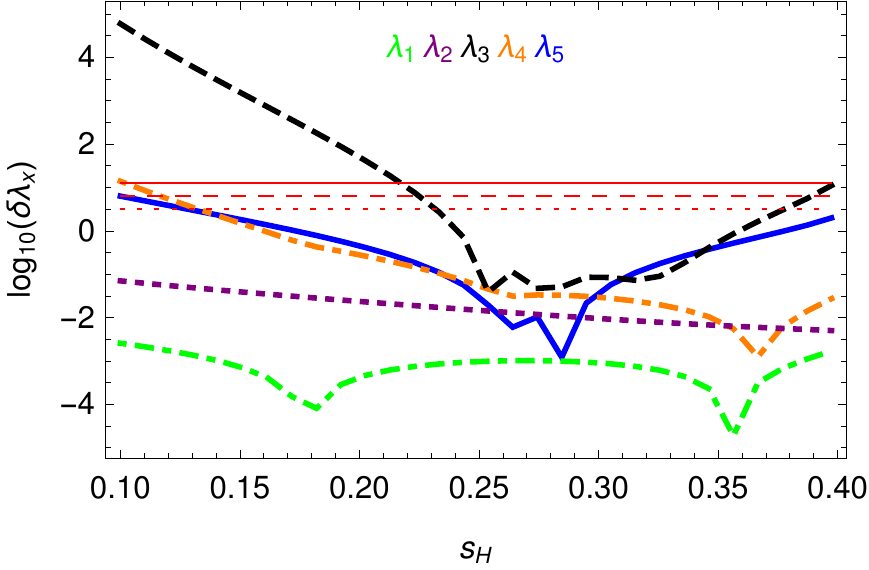} \\
\includegraphics[width=0.41\linewidth]{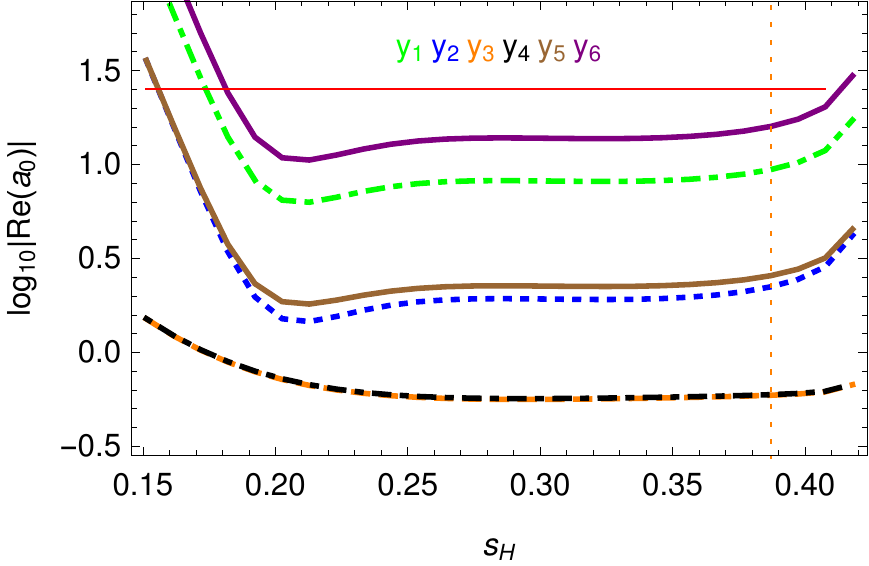} \hfill 
\includegraphics[width=0.41\linewidth]{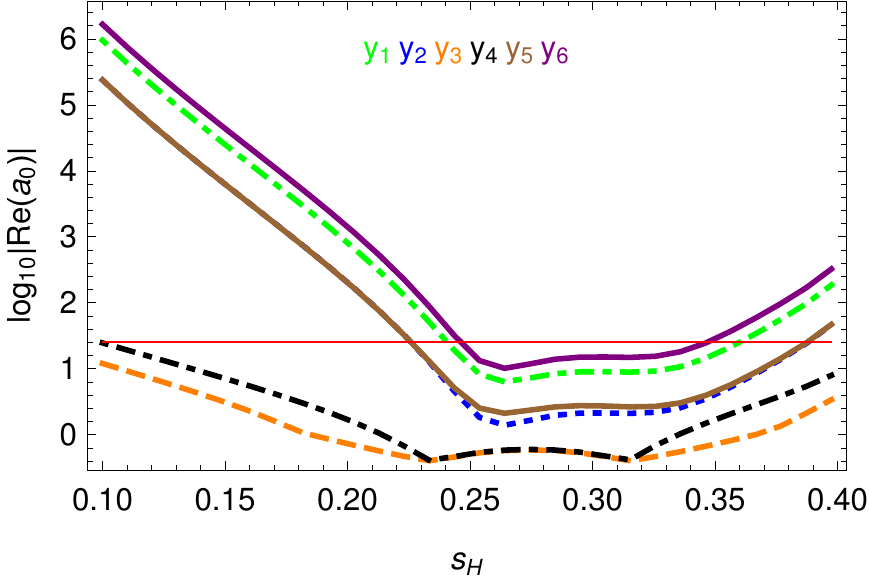} \\
\includegraphics[width=0.41\linewidth]{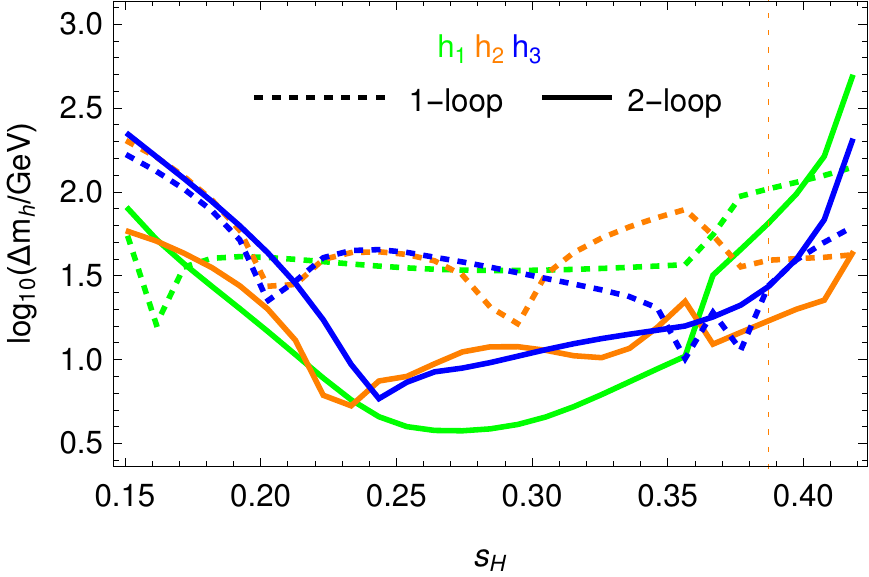}  \hfill
\includegraphics[width=0.41\linewidth]{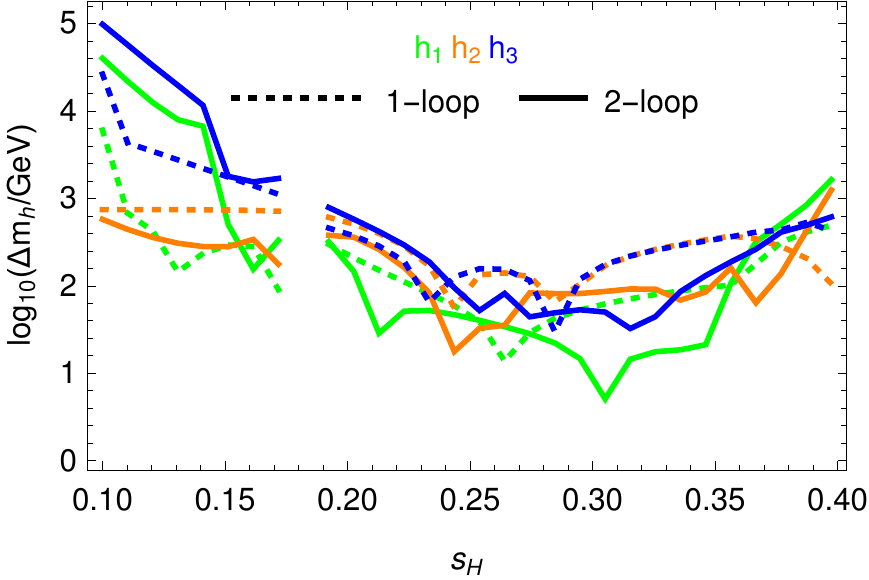} 
\caption{The relative (first row) and absolute (second row) size of the counter-terms to the quartic couplings $\lambda_i$ as a function of $s_H$. The third row gives the absolute value of different eigenvalues of the scattering matrix when using the renormalised quartic couplings as input. The fourth row  shows the size of the one- and two-loop corrections to the 
scalar masses in the \MS scheme. The red line in the second row indicates values of $\pi$, $2\pi$ and $4\pi$ and in the third row of $8\pi$. On the left we set $m_H=300$~GeV, on the right $m_H=800$~GeV. The other parameters are analogous to Fig.~\ref{fig:mhi_sH}.}
\label{fig:per_sH}
\end{figure}
\begin{figure}[tb]
\includegraphics[width=0.45\linewidth]{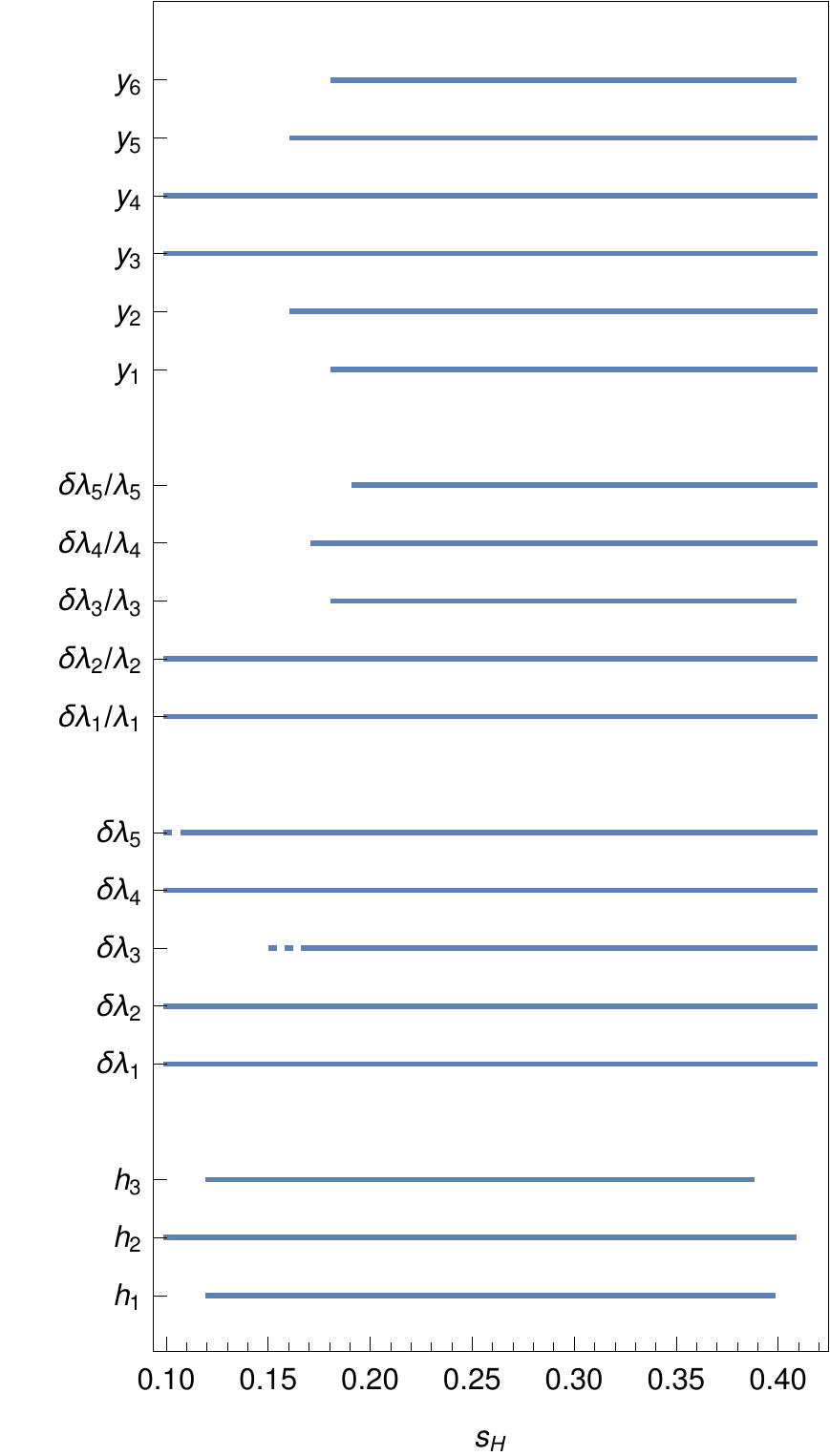} \hfill 
\includegraphics[width=0.45\linewidth]{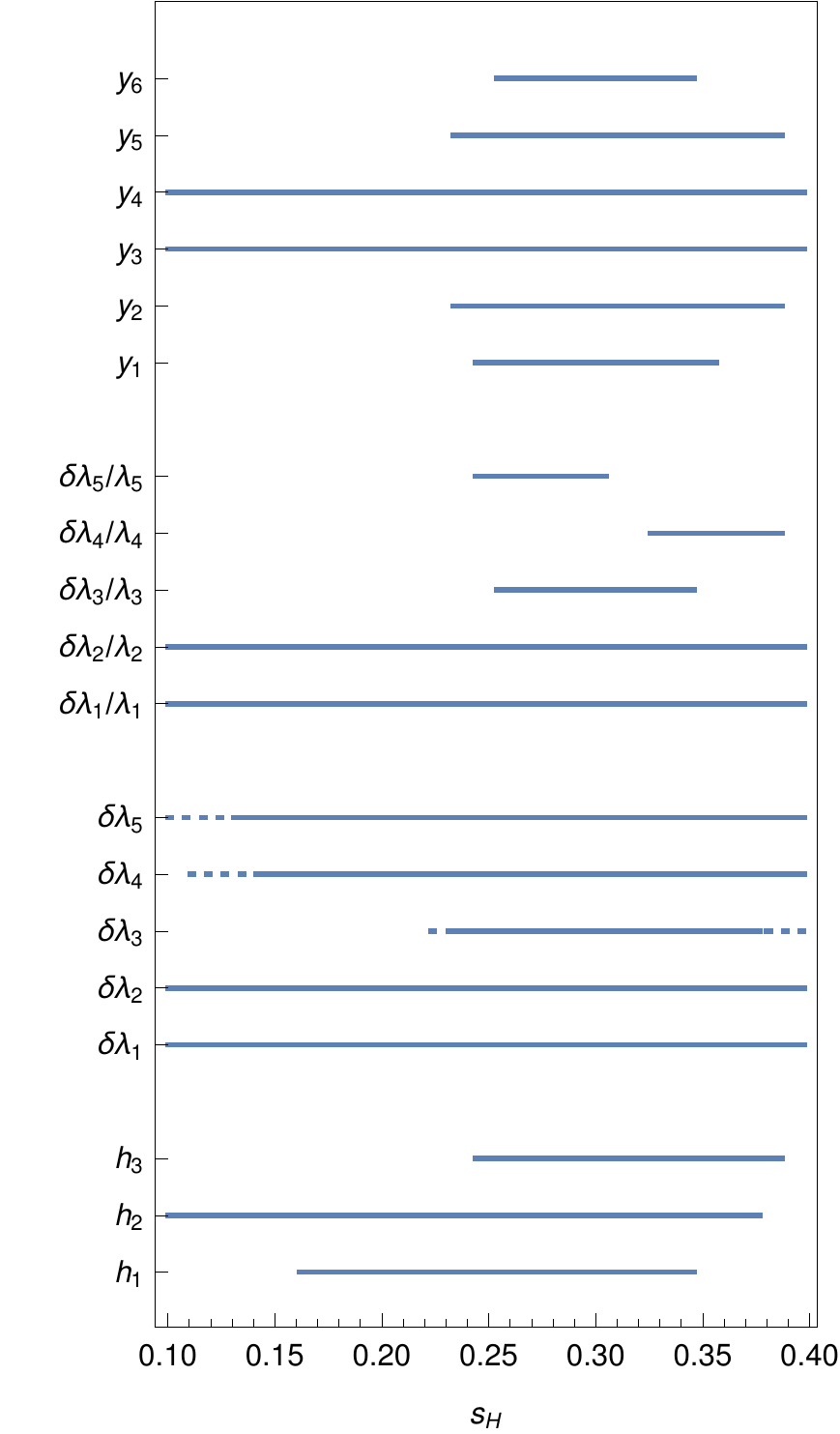} 
\caption{
Comparison of the perturbativity limits of the parameter point of Fig.~\ref{fig:mhi_sH}. The left-hand side corresponds to $m_H=300$~GeV, the right-hand side to $m_H=800$~GeV.  We show here the allowed parameter ranges which fulfil different sets of constraints. (i) $h_i$: 
 $|(m_{h_i}^{2})_{\rm Tree} - (m_{h_i}^{2})_{\rm 1L}| > |(m_{h_i}^{2})_{\rm 2L} - (m_{h_i}^{2})_{\rm 1L}|$; (ii) $\delta \lambda_x$: $|\delta \lambda_x| < \pi \ (2 \pi)$\, [solid line (dashed line)]; (iii) $|\delta \lambda_x / \lambda_x| < 1$., (iv) $y_i$ are the absolute values of different eigenvalues of the scattering matrix when using $\lambda_N+\delta\lambda_{Nx}$ as input.
 }
\label{fig:limits_sH}
\end{figure}

Consequently, we find that perturbativity constraints are not only important for large $s_H$, but that they can also be significant for small $s_H$. This is shown in Fig.~\ref{fig:per_sH} where we show the size of the different loop effects. In the first row of Fig.~\ref{fig:per_sH}, we show the ratio of the different CTs normalised to the tree-level coupling. Here, we have defined 
\begin{align}
\delta \lambda_{n} / \lambda_{n} \equiv \text{Max} \{|\delta\lambda_{na}/\lambda_n|,|\delta\lambda_{nb}/\lambda_n| \}\,, \qquad n =2,5 \,,\\  
\delta \lambda_{n} / \lambda_{n} \equiv \text{Max} \{|\delta\lambda_{na}/\lambda_n|,|\delta\lambda_{nb}/\lambda_n|,|\delta\lambda_{nc}/\lambda_n| \}\,, \qquad n =3,4 \,.
\end{align}
We see that for $m_H=300$~\text{GeV}, the couplings $\lambda_3$, $\lambda_4$ and $\lambda_5$ fail the constraint $\delta \lambda/\lambda<1$ for values of $s_H$ up to 0.2. 
And again for $s_H > 0.4$, $\lambda_3$ is in conflict with this constraint. For $m_H = 800$~\text{GeV}, there is always a contribution which violates this bound  over the entire range of $s_H$. This choice to define perturbativity seems to be quite strong. 
It might also give a `false-positive' result since large $\delta \lambda/\lambda$ can easily occur if the tree-level quartic is very small. This means that there is some tuning of parameters at tree level which gets spoilt by the loop corrections. In this case, it can happen that the perturbative series behaves well and that higher order terms remain small corrections to the one-loop terms. Therefore, a more robust limit is to check the absolute size of the CTs: if those are very large, e.g. $> 4\pi$, then higher order corrections are expected to become more and more important. Therefore, in the second row of Fig.~\ref{fig:per_sH} the absolute size of the counter-terms is shown. Here, we defined 
\begin{align}
\delta \lambda_{n}\equiv \text{Max} \{|\delta\lambda_{na}|,|\delta\lambda_{nb}| \}\,, \qquad n =2,5 \,, \\  
\delta \lambda_{n} \equiv \text{Max} \{|\delta\lambda_{na}|,|\delta\lambda_{nb}|,|\delta\lambda_{nc}| \}\,, \qquad n =3,4 \,.
\end{align}
The qualitative behaviour of the different lines looks very similar to the case of $\delta\lambda/\lambda$. Note that for $m_H=800$~GeV and small $s_H$, the CTs to some quartic couplings can become as large as $\mathcal O(10^4~{\rm GeV})$. This demonstrates how bad the perturbation theory can behave and that an OS calculation, although formally possible, is not well defined in this parameter region. In this example, because of the steep increase of $\delta\lambda_3$ towards small values of $s_H$, the bounds on $s_H$ are rather independent of the choice of the maximal value for the quartic CT, i.e. $|\delta\lambda|<2\pi$ and $|\delta\lambda|<4\pi$ result in approximately the same bounds. \\
Since the maximal value of $|\delta\lambda|$ which we still consider as viable is arbitrary, we also test another condition to get an upper limit of $|\delta\lambda|$ which is the behaviour of the scalar $2\to 2$ scattering. For that, we calculate the eigenvalues of the scattering matrix with the renormalised couplings instead of the tree-level values. This not only gives hints for the perturbative behaviour of the parameter point but also  indicates the robustness of the tree-level unitarity constraints under radiative corrections. We show in the third row of Fig.~\ref{fig:per_sH} the absolute values of the eigenvalues $y_1\, \dots \, y_6$ defined as
{\allowdisplaybreaks
\begin{align}
y_1 =& \text{Max}\Big\{8|2 {\tilde{\lambda}}_{3b}  + {\tilde{\lambda}}_{4b}|,|2 {\tilde{\lambda}}_{3c}  + {\tilde{\lambda}}_{4c}|\,,  \nonumber \\ & 
\hspace{2cm} 2\left(|2 {\tilde{\lambda}}_{3a}  + 2 ({\tilde{\lambda}}_{4a} + {\tilde{\lambda}}_{4b}) + 3 {\tilde{\lambda}}_{3b}|  + \sqrt{(2 \tilde{\lambda}_{3a}-3 \tilde{\lambda}_{3b}+2 \tilde{\lambda}_{4a}-2 \tilde{\lambda}_{4b})^2+8 \tilde{\lambda}_{3c}^2} \right) \Big\} \,, \\
y_2 =& \text{Max}\left\{4|2 {\tilde{\lambda}}_{4b}  + {\tilde{\lambda}}_{3b}|,4| 2 {\tilde{\lambda}}_{4c}  + {\tilde{\lambda}}_{3c}|\right\} \,,\\
y_3 =&  \text{Max}\left\{|4 {\tilde{\lambda}}_{2a}  - {\tilde{\lambda}}_{5a}|, \frac12\left(|4 {\tilde{\lambda}}_{2a} +4 {\tilde{\lambda}}_{2b}   + {\tilde{\lambda}}_{5a}| +  \sqrt{(4 \tilde{\lambda}_{2a}-4 \tilde{\lambda}_{2b}+\tilde{\lambda}_{5a})^2+8 \tilde{\lambda}_{5b}^2}\right)\right\} \,,\\
y_4 =& \text{Max}\left\{2|2 {\tilde{\lambda}}_{2a}  + {\tilde{\lambda}}_{5a}|, |2 {\tilde{\lambda}}_{2a}  + 2 {\tilde{\lambda}}_{2b}   - {\tilde{\lambda}}_{5a} |  + \sqrt{(-2 \tilde{\lambda}_{2a}+2 \tilde{\lambda}_{2b}+\tilde{\lambda}_{5a})^2+8 \tilde{\lambda}_{5b}^2}\right\} \,,\\
y_5 =& 2 \text{Max}\Big\{|2 {\tilde{\lambda}}_{1}  + 2 {\tilde{\lambda}}_{4b}  - {\tilde{\lambda}}_{3b}| + \sqrt{(2 \tilde{\lambda}_1+\tilde{\lambda}_{3b}-2 \tilde{\lambda}_{4b})^2+\tilde{\lambda}_{5a}^2}\,, \nonumber \\
& \hspace{2cm}|2 {\tilde{\lambda}}_{1}  + 2 {\tilde{\lambda}}_{4c}  - {\tilde{\lambda}}_{3c}|  + \sqrt{(2 \tilde{\lambda}_1+\tilde{\lambda}_{3c}-2 \tilde{\lambda}_{4c})^2+\tilde{\lambda}_{5b}^2}\Big\}\,,\\
y_6 =& \text{Max}(x_{1,2,3}) \,,
\end{align}}
where $x_{1,2,3}$ are the solutions of the polynomial of eq.~(\ref{eq:UniPoly}). All $y_i$ should be smaller than $8\pi$. We find that this in general results in stronger constraints than those from the condition $|\delta\lambda|<2\pi$. For $m_H=300$~GeV, these `loop corrected' unitarity constraints are comparable to those from $\delta\lambda/\lambda < 1$, while for $m_H = 800$~GeV, they are a bit weaker and wouldn't rule out the entire parameter range as the condition $|\delta\lambda/\lambda < 1|$ would do. \\

We  now turn to the constraints using the \MS calculation. The size of the one- and two-loop corrections to the CP-even masses is shown in the fourth row in Fig.~\ref{fig:per_sH}. We see that these corrections could cause shifts of hundreds of GeV in the masses, i.e. for small and large $s_H$, they can be as large as the tree-level values. Moreover, we find that the two-loop corrections can be larger than the one-loop corrections. This reflects again a breakdown of perturbation theory. \\

\begin{figure}[tb]
\includegraphics[width=0.42\linewidth]{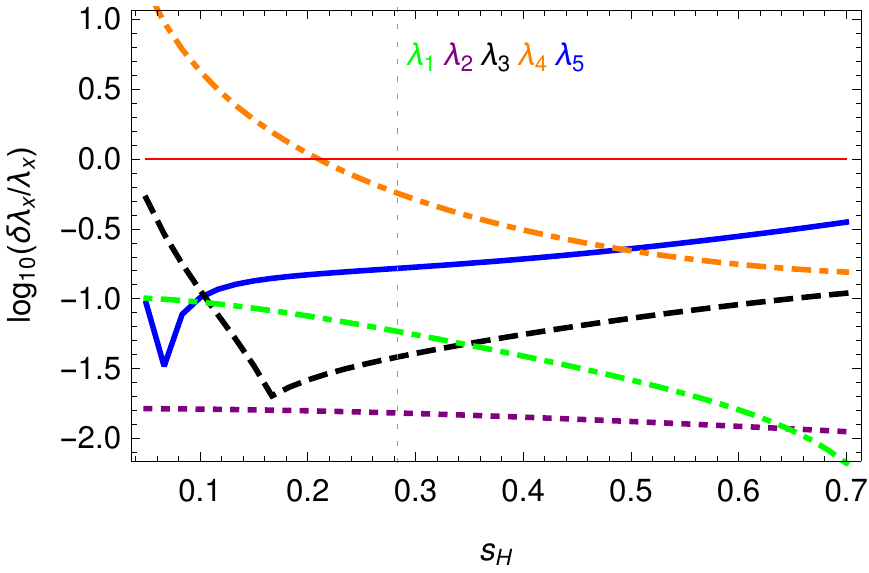} \hfill 
\includegraphics[width=0.42\linewidth]{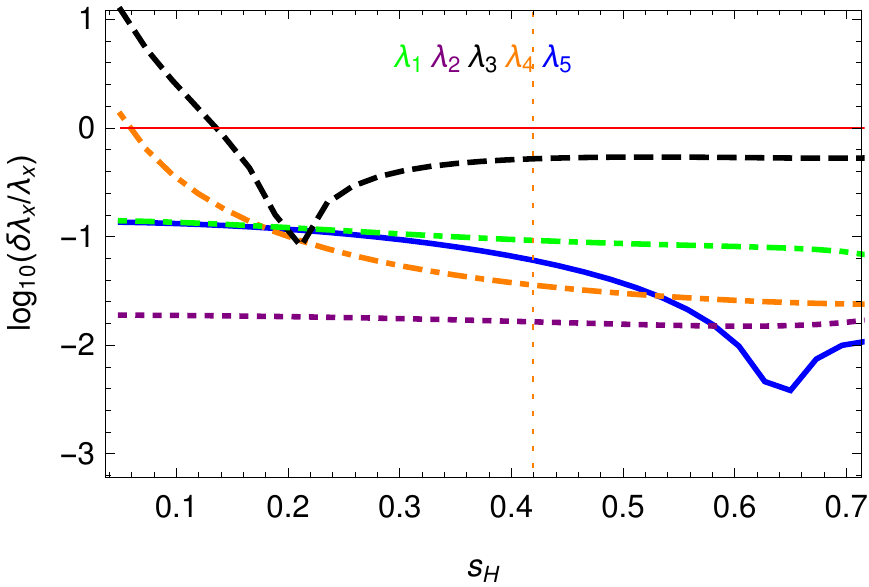}
\includegraphics[width=0.42\linewidth]{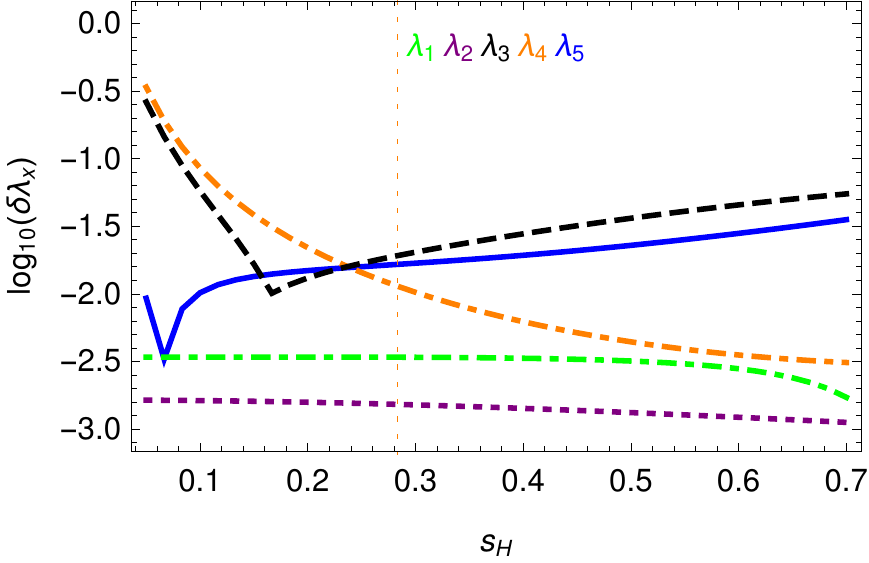} \hfill
\includegraphics[width=0.42\linewidth]{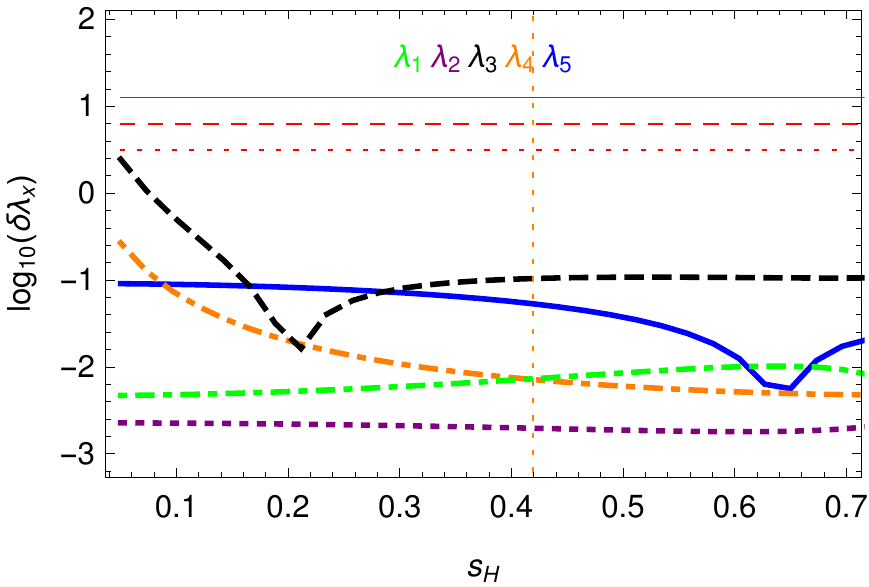} 
\includegraphics[width=0.42\linewidth]{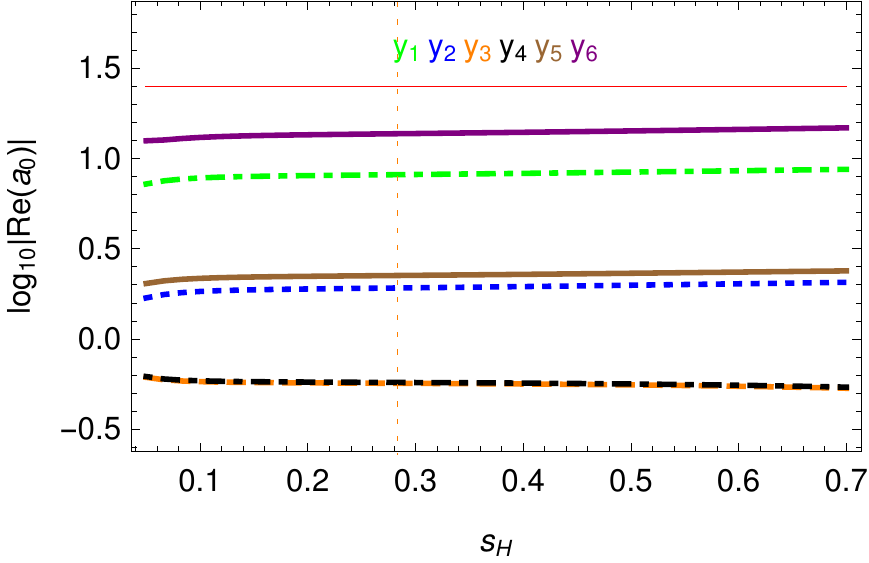} \hfill
\includegraphics[width=0.42\linewidth]{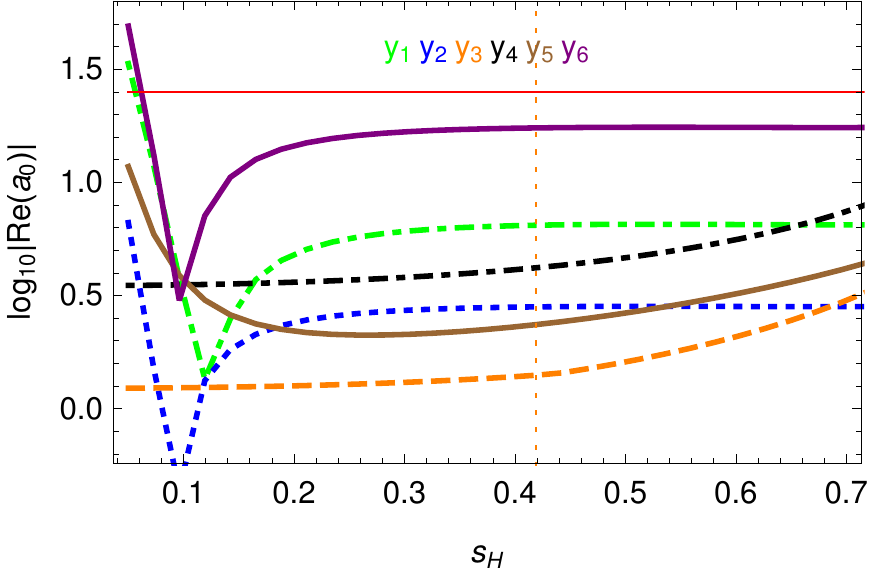} 
\includegraphics[width=0.42\linewidth]{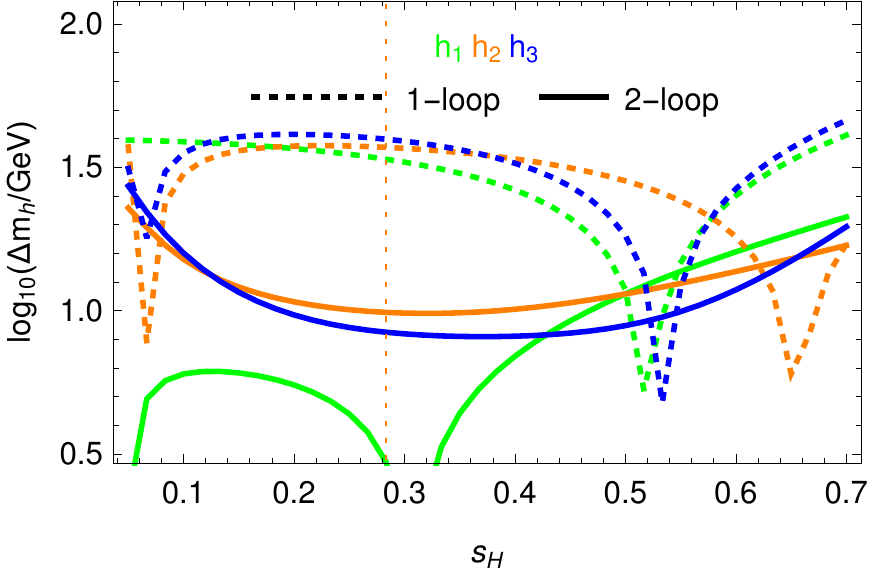}  \hfill
\includegraphics[width=0.42\linewidth]{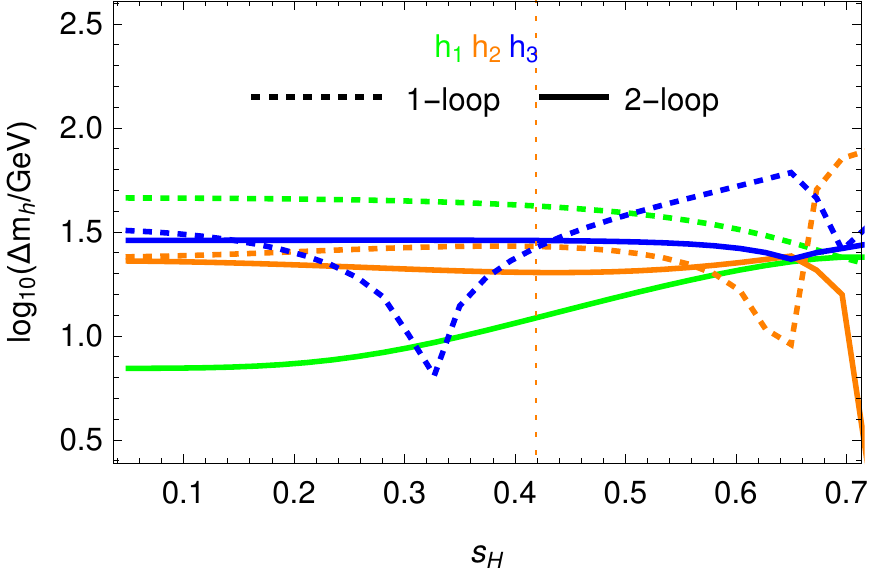}
\caption{On the left: the same as Fig.~\ref{fig:per_sH} for $m_H=300$~GeV with the additional condition $\alpha=75 s_H \frac{\degree}{\text{rad}}$. 
On the right-hand side, we use input choice II as defined in eq.~(\ref{eq:input_choice_II}), i.e. $(m_h,m_5)$ as input instead of $(m_h,m_H,\alpha)$ with $m_5=300~{\rm GeV},\,\lambda_i=0.2$ and $r_{12}=0.15$. 
The vertical orange dashed line shows the {\tt HiggsBounds} limit.}
\label{fig:per_alpha_sH}
\end{figure}

In Fig.~\ref{fig:limits_sH}, we summarise the limits on $s_H$ using the different perturbativity limits. We see that for $m_H=300$~GeV the overall limits from $\delta \lambda_x/\lambda_x$ and $|\delta\lambda_x|$ are quite similar. We further observe that the limits from the one- and two-loop \MS corrections are slightly weaker for small $s_H$ but stronger for large $s_H$. All in all, we find that for $m_H=300$~GeV, a sizeable range of $s_H$ is still allowed by all constraints. This is different to $m_H=800$~GeV where $|\delta\lambda_x/\lambda_x|$ would rule out the entire range, while for the other three sets of constraints still a window in $s_H$ exists where these constraints are fulfilled. It is interesting to note that the constraint from $h_3$ and $\lambda_3$ give quite similar results. Thus, there is an obvious correlation between the size of the one-loop CTs in the OS scheme and the hierarchy between the one- and two-loop corrections in the \MS scheme. \\

Before we move to the impact of the other parameters on the perturbative behaviour of the model, we comment on the dependence on the different input choices. It has already been shown in Ref.~\cite{Braathen:2017izn} that the loop corrections are usually small for $s_H$ if $m_h$ and $m_5$ are used as input, but not $\alpha$. We also find this for our input choice II, cf. eq.~(\ref{eq:input_choice_II}). As shown in Fig.~\ref{fig:per_alpha_sH} on the right column, one can go up to $s_H=0.7$ for $m_5=300$~GeV without running into obvious problems with perturbation theory. However, there is a strong correlation between $s_H$ and $\alpha$ and for large $s_H$ the mixing between the SM-like Higgs and the triplets becomes so large that this causes conflicts with Higgs observables as indicated by the vertical dashed line. 
We find a similar behaviour for input choice I if we impose a correlation between $\alpha$ and $s_H$ `by hand': if we demand $\alpha=75 s_H \frac{\degree}{\text{rad}}$, then we can also go up to very large values for $s_H$ together with $m_H=300$~GeV without running into trouble with perturbativity. However, again the region of $s_H > 0.3$ is ruled out by the Higgs constraints. Hence, the overall picture between both input modes is comparable. We also learn from this comparison that large loop corrections occur if the chosen mixing angle $\alpha$ is far away from a natural value which is correlated with $s_H$. As a  consequence, we use for the further examples with Input I values of $s_H$ between 0.2 and 0.3 and take $\alpha$ between 10$\degree$ and 20$\degree$.

\subsubsection{Dependence on heavy scalar masses}
\begin{figure}[t!]
\centering
\begin{subfigure}[The size of the different loop effects. See Fig.~\ref{fig:per_sH} for more details.]{
\parbox{0.45\linewidth}{
\includegraphics[width=0.95\linewidth]{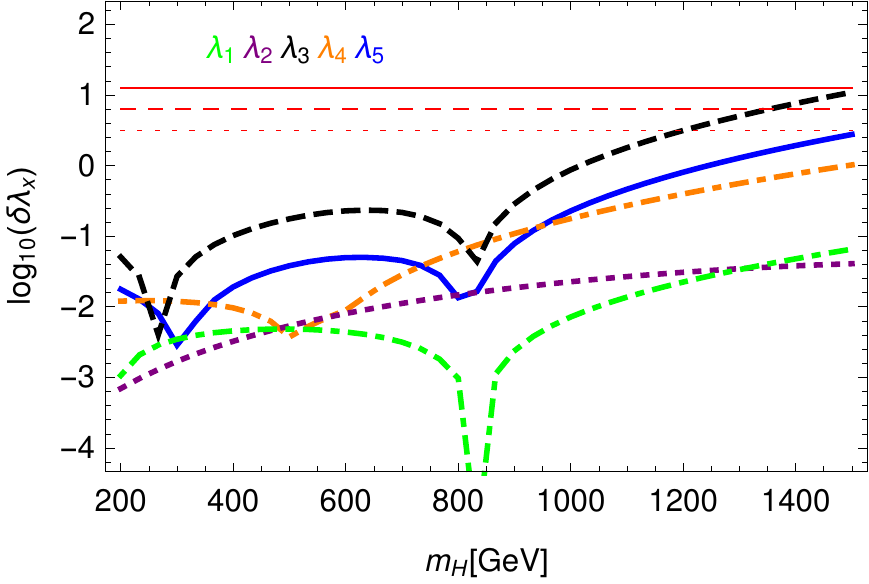} \\
\includegraphics[width=0.95\linewidth]{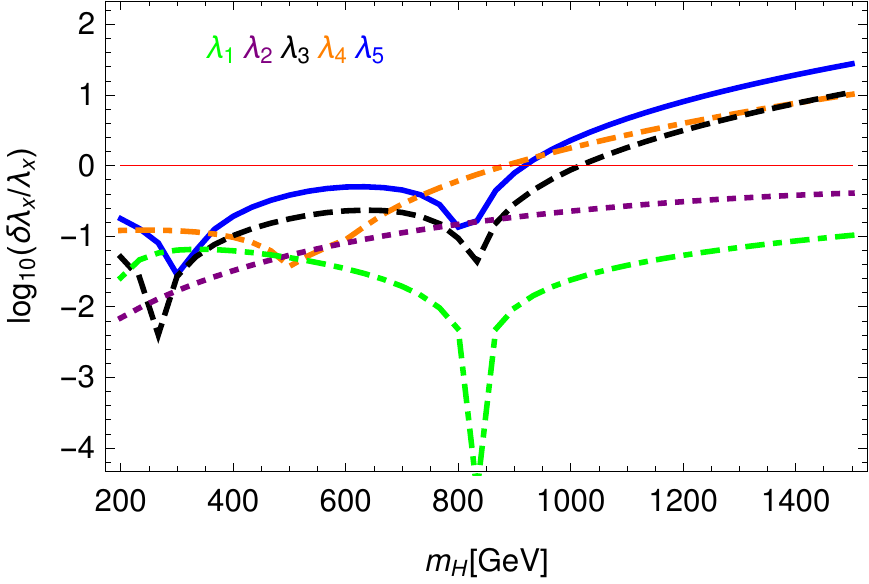} \\
\includegraphics[width=0.95\linewidth]{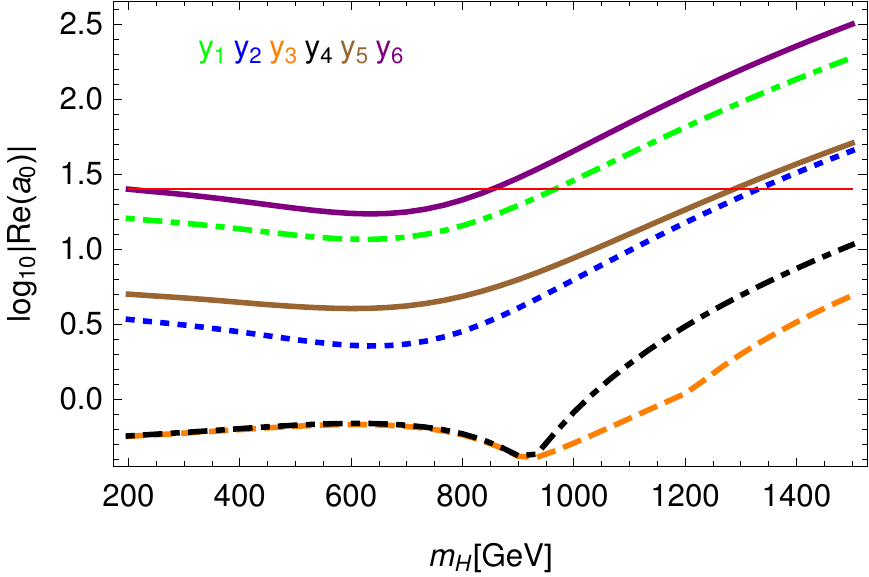} \\
\includegraphics[width=0.95\linewidth]{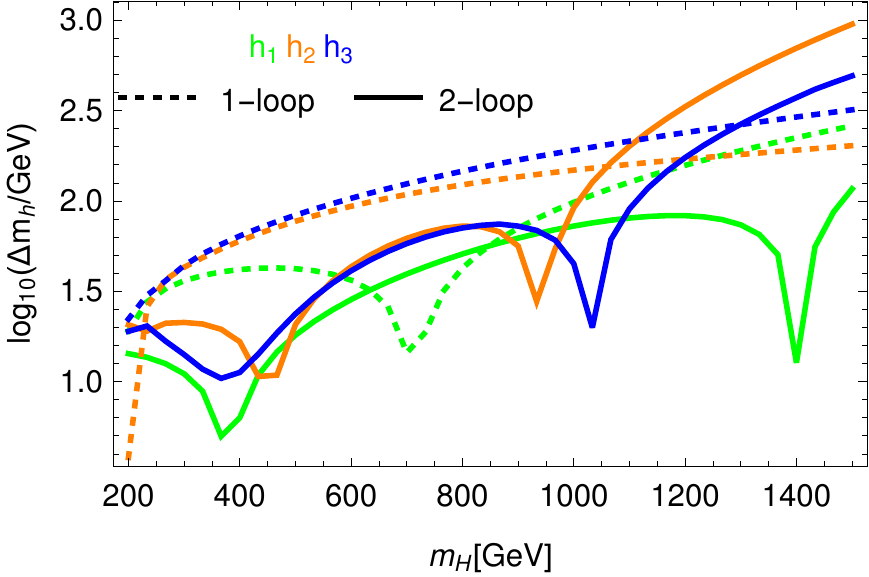}  }}
\end{subfigure}
\hfill
\begin{subfigure}[Comparison of the different perturbativity limits. See Fig.~\ref{fig:limits_sH} for more details.
]{
\parbox{0.45\linewidth}{
    \includegraphics[width=1.00\linewidth]{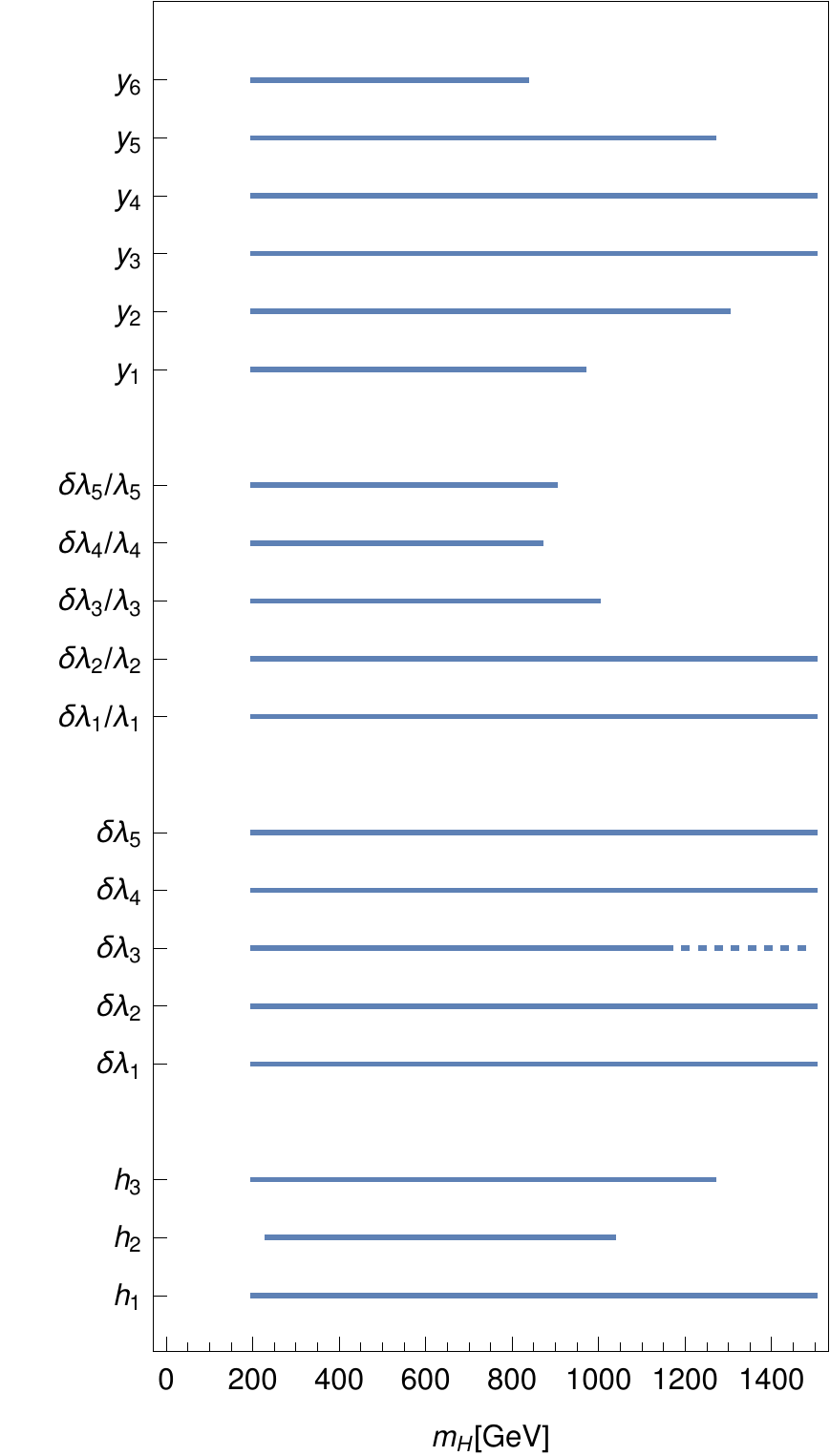}\\[7mm]
  }
 }
\end{subfigure}
 \caption{Perturbativity limits as a function of $m_H$. The other parameters are chosen as
$\lambda_2 =0.1$, $\lambda_3 = 1$, $\lambda_4=-0.1$, $\lambda_5=0.1$, $\alpha=20\degree$, $s_H=0.25$.}
 \label{fig:per_mH}
\end{figure}
We have already seen in the last subsection during the discussion about the dependence of the loop corrections on $s_H$ that the loop corrections usually become more important the heavier the 
new scalars are. The reason is that large scalar masses imply large values for the trilinear couplings $M_1$ and $M_2$ which enter the scalar loop corrections. One finds for instance that the one-loop corrections to the neutral CP-even Higgs with mass $m_5$ scales as
\begin{equation}
\Delta m^2_5 \sim m^2_5 \left(1 + C \frac{m_5^2}{v^2}\right) \,.
\end{equation}
Here, we have neglected all quartic couplings and expressed $M_1$, $M_2$ by $m_5$. The coefficient $C$ is a complicated function of $s_H$ and of the ratio $M_1/M_2$. The important point is that it is usually much larger than $\frac{1}{16\pi^2}$ and can become $\mathcal O(1)$ for large $s_H$ and/or large ratios. \\
We show  the impact in Fig.~\ref{fig:per_mH}{\color{blue}{a}}  where we give the size of the loop corrections using a fully numerical calculation as a function of $m_H$. The other parameters are set to
\begin{eqnarray*}
\lambda_2 =0.1,\ \lambda_3 = 1,\ \lambda_4=-0.1,\ \lambda_5=0.1,\ \alpha=20\degree,\ s_H=0.25\,.
\end{eqnarray*}
In all four different formulations of perturbativity constraints shown in Fig.~\ref{fig:per_mH}{\color{blue}{a}}, we find that large $m_H$ is generically more constrained than lower masses, i.e. the values of the CTs as well as the size of the loop-corrected \MS masses increase with increasing $m_H$, up to a few specific values where cancellations among the different loop contributions are present. 
In the lower panel  of Fig.~\ref{fig:per_mH}{\color{blue}{a}} we present the size of the one- and two-loop corrections to the neutral CP-even \MS masses.
We observe that, for values of $m_H$ above 1.1~TeV (1.3~TeV), the two-loop corrections to $h_2$ ($h_3$) are larger than the one-loop corrections. Thus, in this example, the strongest perturbativity constraints in the \MS scheme are actually not due to the loop corrections to the SM-like state but due to the new scalars. For the SM-like state we see a short range between 600--800 GeV where the two-loop corrections are larger than the one-loop corrections. However, this is obviously because the one-loop corrections are suppressed by an accidental cancellation. Therefore, we extend the \MS perturbativity limit by a threshold for the minimal size of the one-loop corrections, cf. sec.~\ref{sec:constraints_per}: only if the one-loop corrections to the squared masses are larger than $(20 ~\text{GeV})^2$ and if the two-loop corrections are larger than the one-loop corrections, we consider this as breakdown of perturbation theory. \\
For the one-loop CTs in the OS scheme, shown in the upper two panels of Fig.~\ref{fig:per_mH}{\color{blue}{a}}, we find again that the strongest limits on perturbativity stem from the constraint $|\delta \lambda_x /\lambda_x|<1$. Looking at the unitarity constraints, shown in the third panel, we end up with a similar upper bound on $m_H \leq 800~$GeV for this particular parameter point, more than $\sim 200$~GeV lower than the limit from the \MS mass corrections. 
The constraint $|\delta \lambda_x|<\pi$, in turn,  leads to a limit on $m_H$ of 1.2 TeV, which is in-between the one obtained from the loop corrections to $m_{h_2}$  and $m_{h_3}$. Using a weaker upper limit of $2\pi$ or $4\pi$ wouldn't lead to any constraint in the tested parameter range.
Qualitatively, however, we observe a clear correspondence of the perturbativity constraints formulated in the \MS scheme and the ones from the OS CTs.
In Fig.~\ref{fig:per_mH}{\color{blue}{b}}, we show the ranges of allowed $m_H$ for this scenario, analogously to Fig.~\ref{fig:limits_sH}. \\

\subsubsection{Dependence on large quartic couplings}
\begin{figure}[t!]
\centering
\begin{subfigure}[The size of the different loop effects. See Fig.~\ref{fig:per_sH} for more details.]{
\parbox{0.45\linewidth}{
\includegraphics[width=0.95\linewidth]{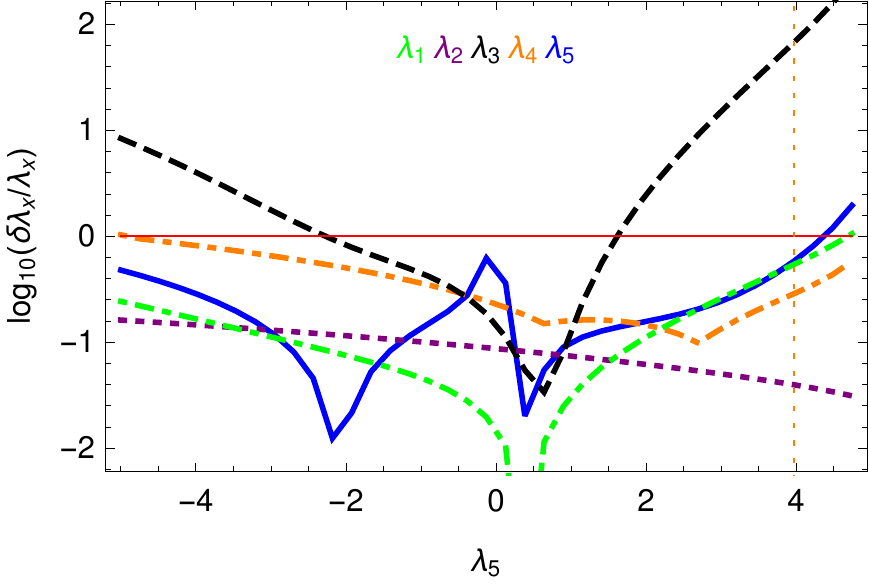} \\
\includegraphics[width=0.95\linewidth]{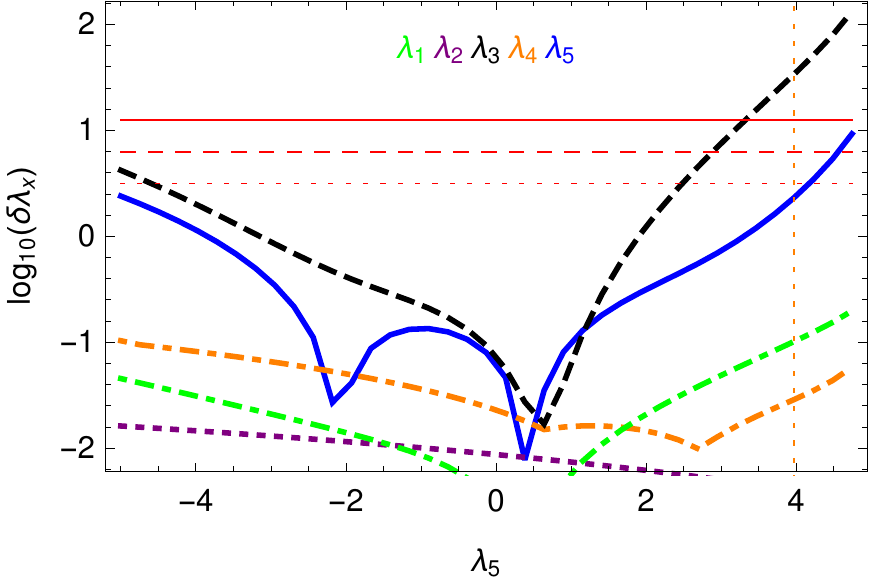} \\
\includegraphics[width=0.95\linewidth]{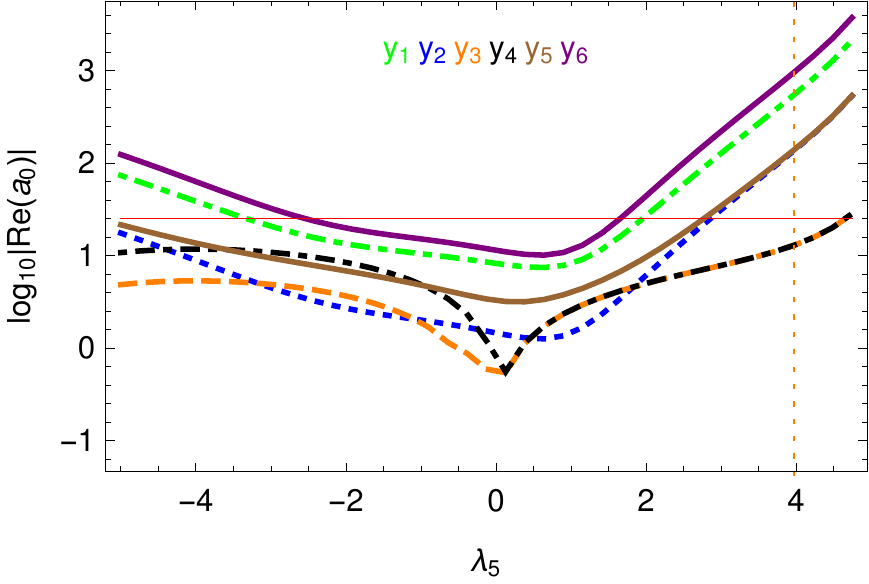} \\
\includegraphics[width=0.95\linewidth]{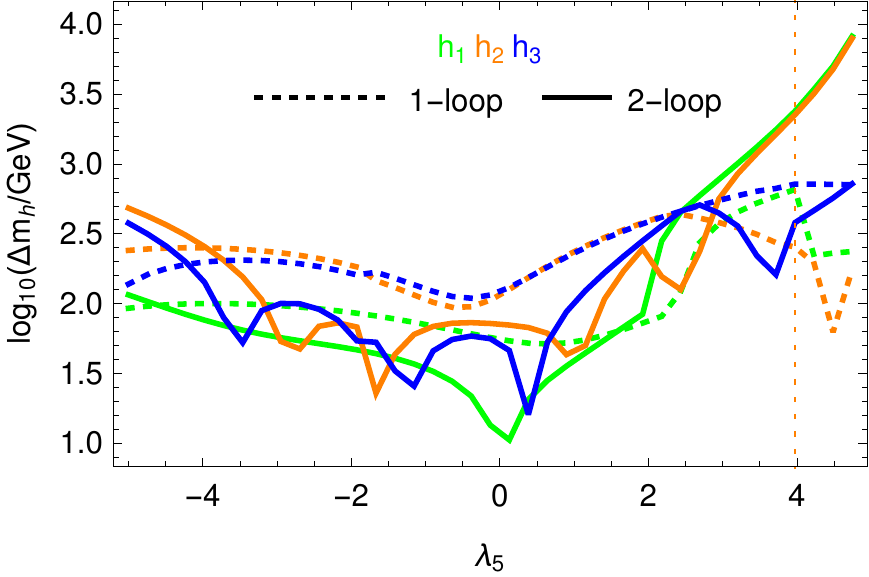}  }}
\end{subfigure}
\hfill
\begin{subfigure}[Comparison of the different perturbativity limits. See Fig.~\ref{fig:limits_sH} for more details.
]{
\parbox{0.45\linewidth}{
    \includegraphics[width=1.00\linewidth]{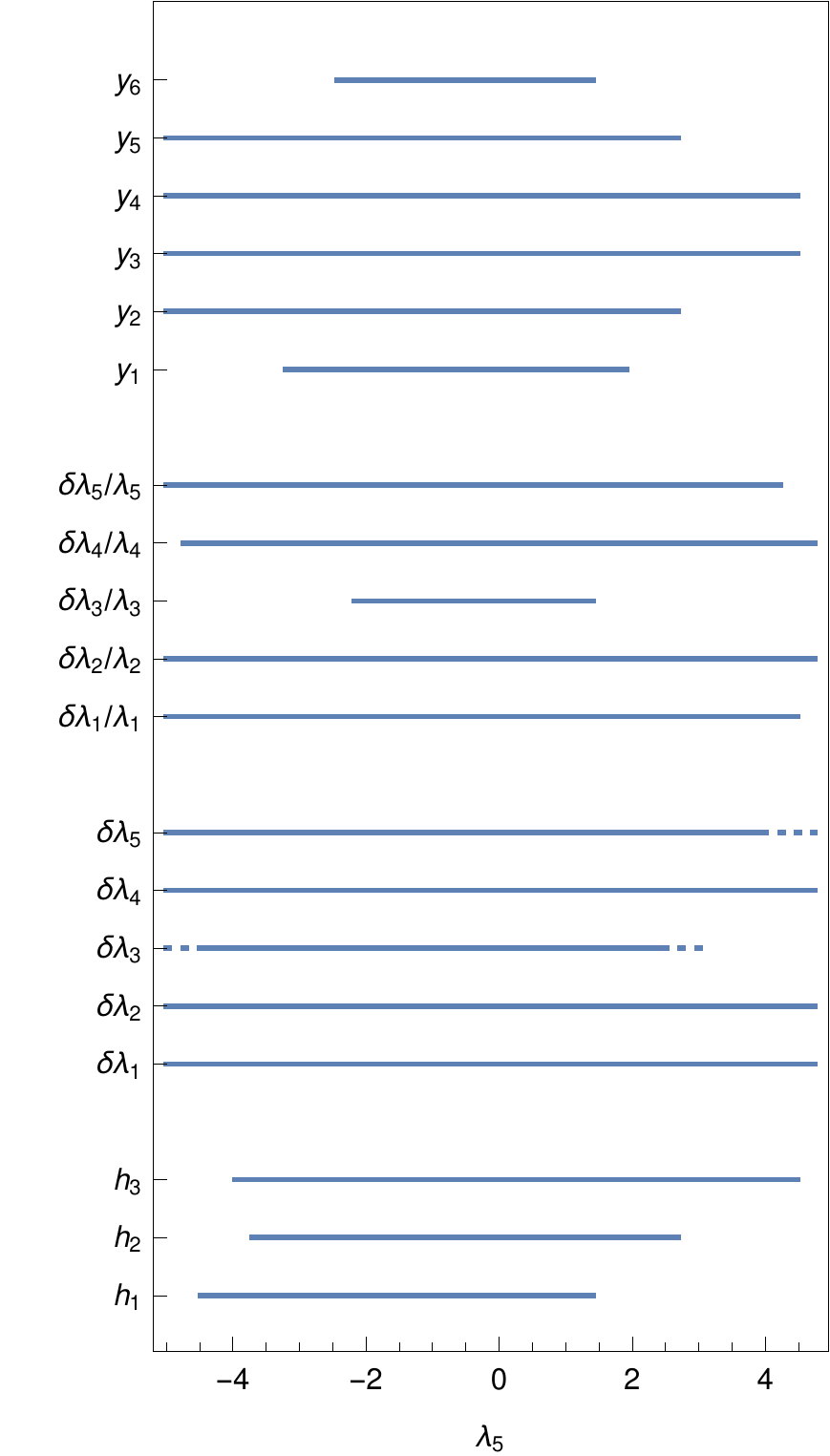}\\[7mm]
  }
 }
\end{subfigure}
 \caption{Perturbativity limits as function of $\lambda_5$. The other parameters are chosen as: $\lambda_2 =0.1$, $\lambda_3 = 0.5$, $\lambda_4=-0.1$,  $\alpha=20\degree$, $s_H=0.3$, $m_H=750$~GeV.}
 \label{fig:per_l5}
\end{figure}
So far, we have not considered the role of the quartic couplings on the perturbativity limits. 
Unlike in the THDM, the quartic couplings in the GM model are usually not taken very large, i.e. $O(10)$. This is due to the tree-level unitarity limits 
in this model which already severely constrain combinations of couplings to be much smaller than $4\pi$, see eq.~({\ref{eq:uni}). For instance, if we assume  only $\lambda_3$ and $\lambda_4$ to be non-negligible at tree-level, then large $\lambda_{3}$ is only allowed if it cancels against a large $\lambda_{4}$, confining both parameters to a narrow strip around $\lambda_3 = -\frac{11 \lambda_4}{7}\pm \frac{2 \pi}{7}$ which is cut off at roughly $\lambda_4 \simeq 2$. 

On $\lambda_5$, however, there are comparably weak tree-level unitarity constraints, i.e. $|\lambda_5| \gg 1$ is easily possible without violating any 
unitarity limit. On the other hand, $\lambda_5$ enters the one- and two-loop corrections, i.e. large effects are expected there. This is depicted in Fig.~\ref{fig:per_l5}{\color{blue}{a}} where the loop effects are plotted as a function of $\lambda_5$. The other parameter values are set to 
\begin{equation}
\lambda_2 =0.1\,,\ \lambda_3 = 0.5\,,\ \lambda_4=-0.1\,,\  \alpha=20\degree\,,\ s_H=0.3\,,\ m_H=750~\text{GeV}\,. 
\end{equation}
For positive values of $\lambda_5$, there is a fast increase in the size of the loop corrections. In particular the two-loop corrections to the SM-like Higgs grow very quickly and are as large as the tree-level mass for $\lambda_5\simeq 2$. For the second-lightest Higgs, the two-loop corrections are larger than the one-loop corrections for $\lambda_5 \simeq 2.25$. This roughly corresponds  to the value at which $\delta \lambda_3 \simeq \pi$, depicting again the correlation between the two-loop corrections in the \MS scheme and OS CTs. For positive values of $\lambda_5$, similar constraints are also found when using the condition $|\delta \lambda_x/\lambda_x|<1$ or the `loop improved' unitarity constraints. For negative values of $\lambda_5$, in turn, these two  sets of conditions are much more restrictive: they would forbid values of $\lambda_5$ below -2, while the other two sets of conditions are fulfilled until $\lambda_5 \simeq -4$. 
\begin{figure}[t!]
\centering
\begin{subfigure}[The size of the different loop effects. See Fig.~\ref{fig:per_sH} for more details. Here,the vertical black lines shows the tree-level unitarity constraints.]{
\parbox{0.45\linewidth}{
\includegraphics[width=0.95\linewidth]{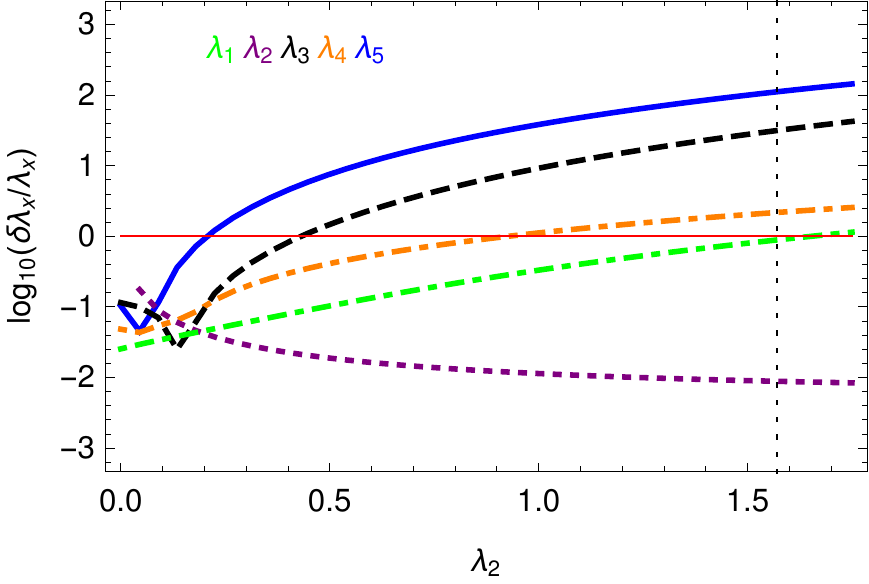} \\
\includegraphics[width=0.95\linewidth]{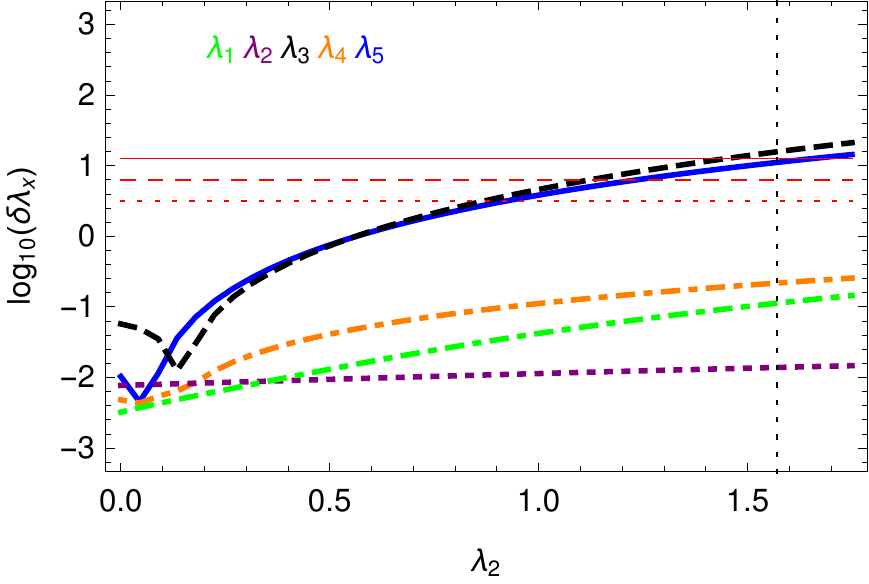} \\
\includegraphics[width=0.95\linewidth]{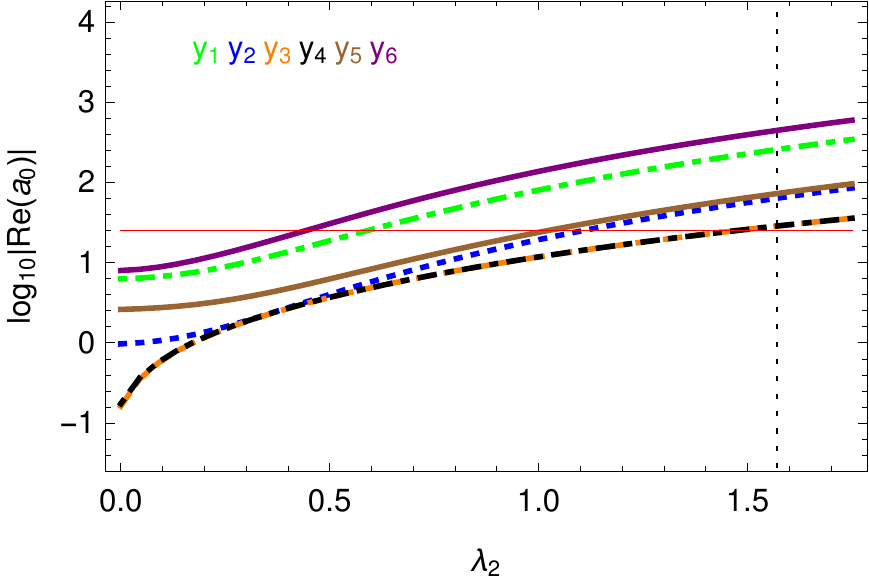} \\
\includegraphics[width=0.95\linewidth]{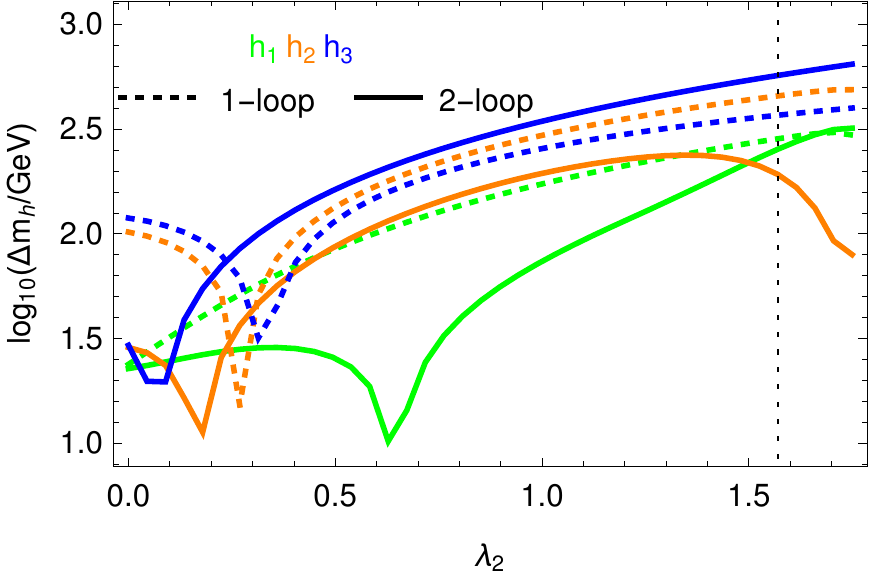}  }}
\end{subfigure}
\hfill
\begin{subfigure}[Comparison of the different perturbativity limits. See Fig.~\ref{fig:limits_sH} for more details.]{
\parbox{0.45\linewidth}{
    \includegraphics[width=1.\linewidth]{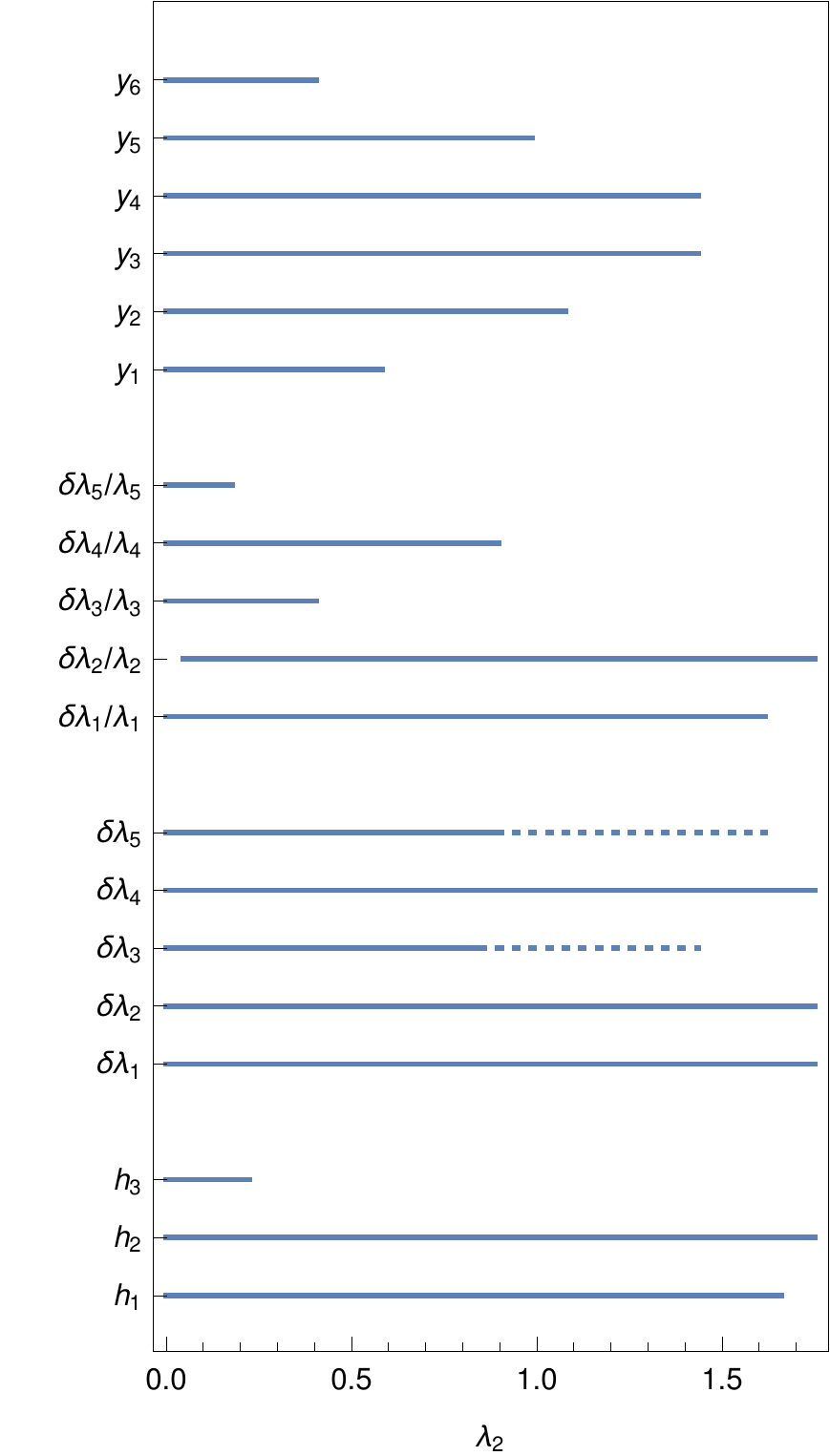} \\[7mm]
  }
 }
\end{subfigure}
 \caption{Perturbativity limits as function of $\lambda_2$.  The other parameters are chosen as: $\lambda_3 = 0.5$, $\lambda_4=-0.1$, $\lambda_5=0.1$,  $\alpha=20\degree$, $s_H=0.25$, $m_H=600$~GeV.}
 \label{fig:per_l2}
\end{figure}

We now turn to the other quartic couplings. Even though those are constrained to smaller values than $\lambda_5$ due to the tree-level unitarity bounds, 
the one-loop corrections to those quartics can turn out to be so large that we end up with stronger perturbativity constraints on $\lambda_{1\cdots 4}$ than on $\lambda_5$. An indication is already seen in Fig.~\ref{fig:per_l5} where it is actually the counter-terms to $\lambda_3$ which become problematic much earlier than those to $\lambda_5$.
In Fig.~\ref{fig:per_l2}{\color{blue}{a}}, we show the loop effects as a function of $\lambda_2$. The other input values are set to
\begin{eqnarray*}
&\lambda_3 = 0.5\,,\ \lambda_4=-0.1\,,\ \lambda_5=0.1\,,\  \alpha=20\degree\,,\ s_H=0.25\,,\ m_H=600~\text{GeV}\,. & 
\end{eqnarray*}
We find almost continuously increasing CTs and loop corrected masses for increasing $\lambda_2$. The two sets of constraints which have been the most restrictive ones on the other examples ($|\delta \lambda_x / \lambda_x| < 1$ and the unitarity constraints) are already violated for $\lambda_2\simeq 0.2$. In this case, this is quite similar to the upper limit which is obtained from the two-loop corrections to $h_3$. On the other hand, when using the absolute size of the CTs or the two-loop corrections of the first two scalar masses as constraints, the limits are much weaker and range between $\lambda_2 \simeq  1 \dots 1.5$, comparable with the tree-level unitarity constraints. The constraints from the different perturbativity requirements are summarised in Fig.~\ref{fig:per_l2}{\color{blue}{b}}.

\subsubsection{Impact on parameter regions}
\begin{figure}[tb]
\includegraphics[width=0.5\linewidth]{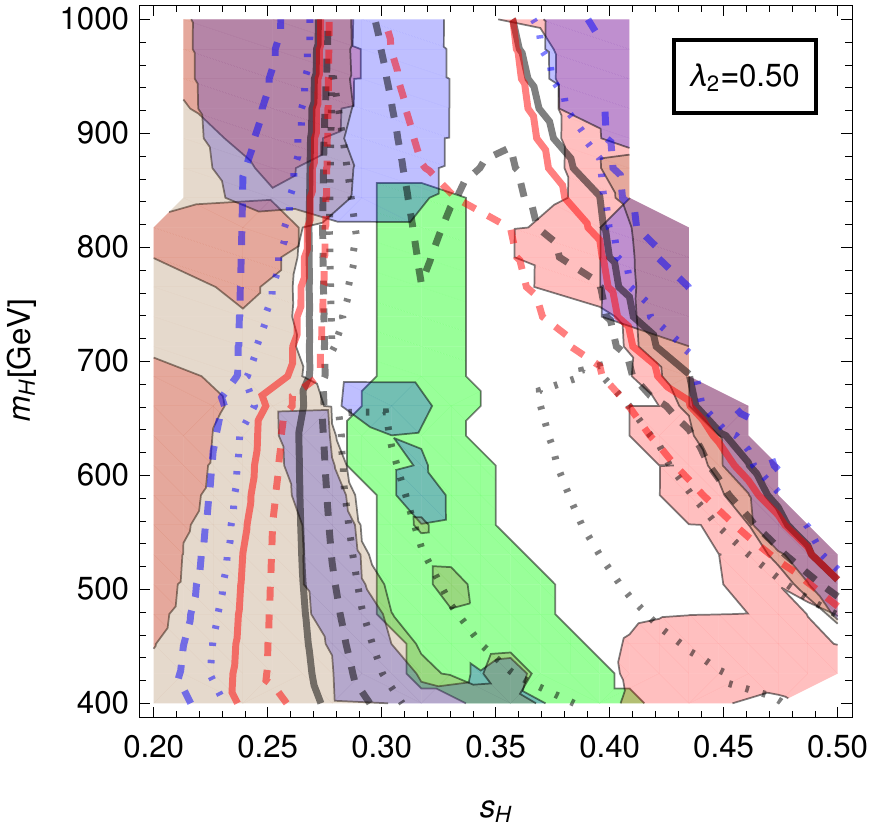} \hfill 
\includegraphics[width=0.5\linewidth]{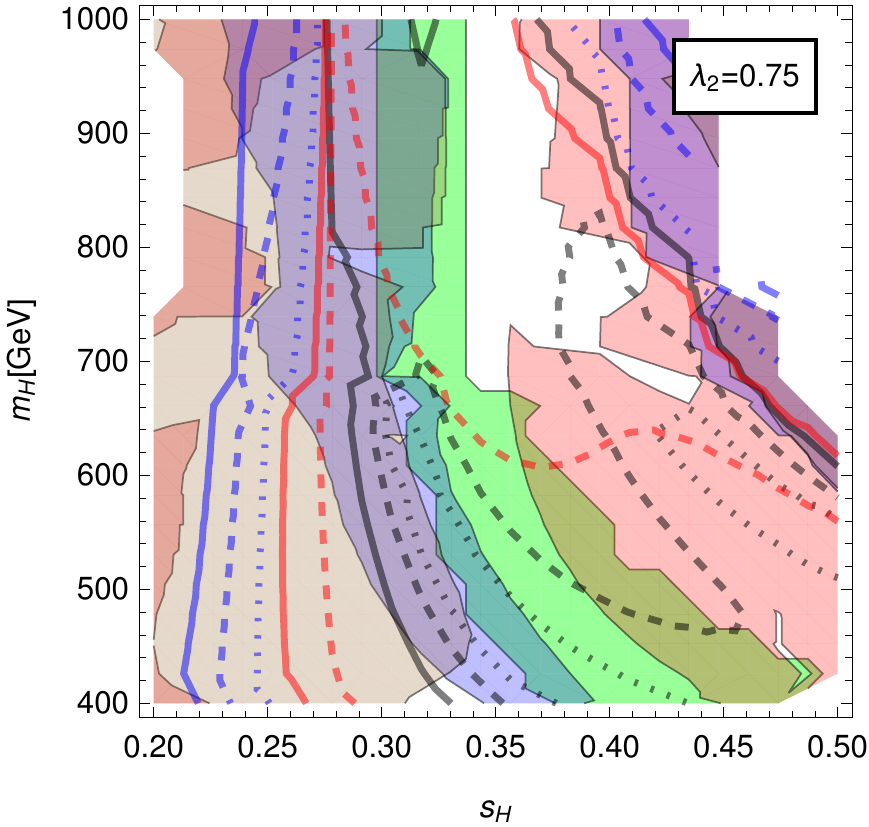} \\
\includegraphics[width=0.5\linewidth]{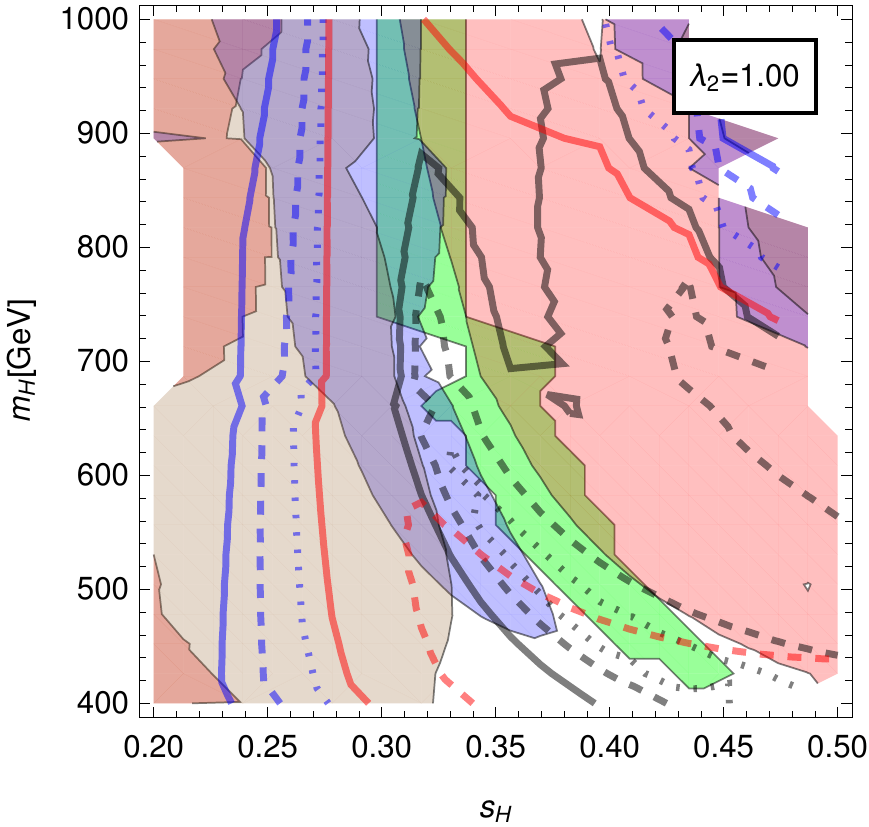}  \hfill
\includegraphics[width=0.5\linewidth]{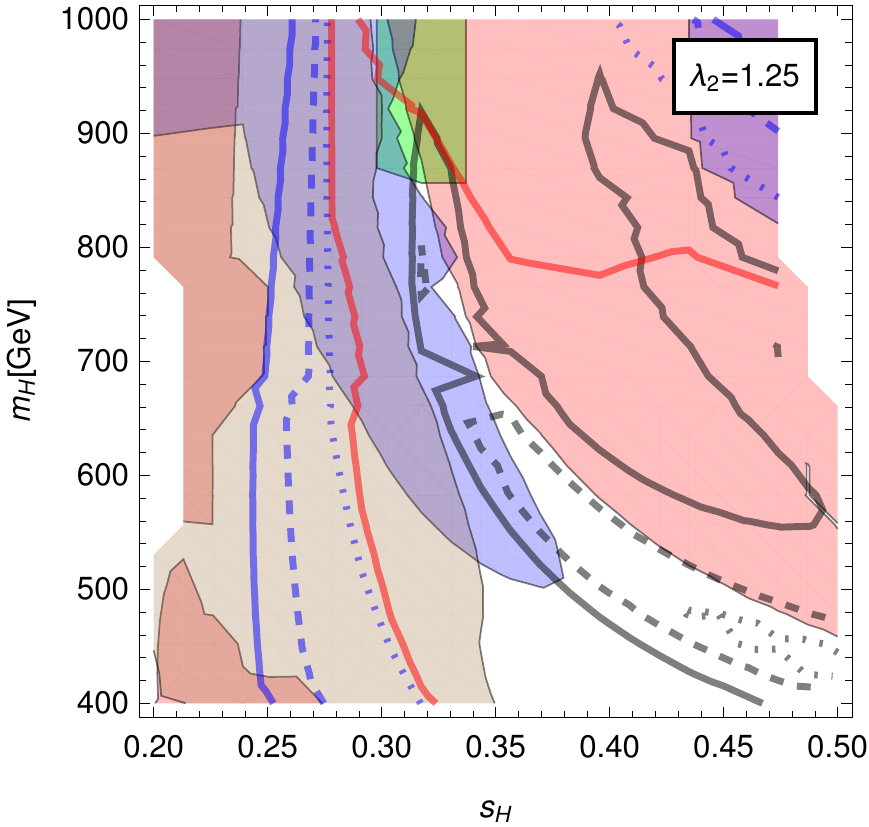} 
\caption{Impact of the perturbativity constraints in the $(m_H,s_H)$ plane for different values of $\lambda_2$ on regions 
which are allowed at tree-level (green areas). The other parameters are set to $\lambda_3 = 0.5$, $\lambda_4=-0.1$, $\lambda_5=0.1$,  $\alpha=20\degree$.
The shaded area indicates the Higgs mass constraints ($|(m_{h_i})^{2})_{\rm Tree} - (m_{h_i}^{2})_{ \rm 1L}| > |(m_{h_i}^{2})_{\rm 2L} - (m_{h_i}^{2})_{ \rm 1L}|$) [red: $h_1$;
blue: $h_2$; brown: $h_3$]. The blue lines are the contours of constant values for $\text{Max}(|\delta \lambda_x|) =c\pi$ [dotted: $c=1$; dashed: $c=2$; full $c=4$].
The black lines indicate $\text{Max}(|\delta \lambda_x/\lambda_x|)>v$ [dotted: $v=1$; dashed: $v=2$; full: $v=4$]. The red lines show when the tree-level 
unitarity limits calculated with $\tilde{\lambda}$'s are violated when setting an upper limit on the scattering eigenvalues of $4\pi$ (dotted), $8\pi$ (dashed) or $16\pi$ (full).}
\label{fig:loop_vs_tree}
\end{figure}

We have seen in the last subsections that the loop corrections in the GM model can become huge, indicating a breakdown of perturbation theory. Of course, it is difficult to define an absolute condition when this breakdown takes places. We have investigated four sets of conditions which ended up in different constraints on the parameters. Which conditions are applied depend on how conservative one wants to be. 
However, the important observation is that at some point all conditions point towards severe conflicts with the perturbative expansion: if the CT to a quartic coupling is O(100) and if the two-loop corrections are larger than the one-loop corrections by an order of magnitude, it is clear that one has entered the strongly coupled regime of the model. \\

We demonstrate at one example the impact of the different perturbativity constraints on the parameter space which seems to be valid at tree-level. We show in Fig.~\ref{fig:loop_vs_tree} an overlay of the allowed parameter space at tree-level and the different loop constraints in the $(m_H,s_H)$ plane for four different values of $\lambda_2$. The other parameters have been set to
\begin{eqnarray*}
\lambda_3 = 0.5\,,\ \lambda_4=-0.1\,,\ \lambda_5=0.1\,,\  \alpha=20\degree\,.
\end{eqnarray*}
The green shaded areas in Fig.~\ref{fig:loop_vs_tree} show the parameter space which is allowed at the tree level.
The areas shaded in red (blue) [brown] indicate when the \MS two-loop correction to $m_{h_1}$ ($m_{h_2}$) [$m_{h_3}$] becomes larger than the corresponding one-loop correction. 
The other contour lines show the OS perturbativity bounds as well as tree-level unitarity bounds calculated with the renormalised couplings $\tilde \lambda$:
the blue lines show the contours of constant Max$(|\delta \lambda_x|)$ of $\pi$ (dotted), $2 \pi$ (dashed) and $4 \pi$ (full line). The black lines show the contours of constant ratios Max$(|\delta \lambda_x/\lambda_x|)$ of 1 (dotted), 2 (dashed) and 4 (full line). Finally, the unitarity bounds are represented by the red contours, showing constant scattering eigenvalues of $4\pi$ (dotted), $8 \pi$ (dashed) and $16 \pi$ (full line).

The first observation is that in neither of the four subfigures, any of the parameter space features generalised scattering amplitudes with $y_i < 4 \pi$. As in the previous examples, we will however always use the less restrictive bound of $8\pi$, as defined in eq.~(\ref{eq:generalized_unitarity_bound}). 
In the left upper panel of Fig.~\ref{fig:loop_vs_tree}, we present the case $\lambda_2=0.5$. We observe that in this case, the loop corrections behave comparatively well -- most of the parameter space which is allowed at tree level is still viable if the higher-order constraints are taken into account. The only exception is the ratio of CTs over the tree-level coupling
which would exclude most of the valid parameter space 
if we were to apply the most restrictive bound of Max$(|\delta \lambda_x/\lambda_x|)<1$. We further observe that the other, less conservative OS bounds agree well with the \MS conditions.

For increasing values of $\lambda_2$, the perturbativity constraints invade more and more the valid tree-level regions. Finally, for $\lambda_2=1.25$, nearly the entire parameter space which is allowed at tree-level is ruled out once the perturbativity constraints are applied.

In all four examples in Fig.~\ref{fig:loop_vs_tree} we see that the strongest constraints always come from the ratios Max$(|\delta \lambda_x/\lambda_x|)<v$, especially if $v=1$ is considered as the maximally-allowed ratio. However, also  using $v=2$ or 4 as bounds, these limits are always stronger than the ones from the absolute values of $|\delta \lambda_x|<c \pi$, even if we apply $c=1$ or $2$ as condition.  The exclusion regions when demanding that the tree-level unitarity constraints calculated with renormalised couplings should be fulfilled are comparable with those from the relative size of the CTs when imposing $<8\pi$ for the maximal eigenvalue of the scattering matrix. If eigenvalues up to $16\pi$ are accepted, the condition becomes more comparable to the ones on the absolute value of the CTs with $c=1$. \\
Similarly, also the hierarchy between the one- and two-loop corrections to the scalar masses leads to quite severe constraints. It is in particular interesting to see that for different parameter points the corrections to different masses are more important. Because of this complementarity, the superposition of the constraints from all three masses cover a significant part of the parameter space. The constraints from the absolute size of the quartic couplings result, for this example, in the weakest limits. 

Quite generically, however, we observe again clear correlations between the size of the CTs and the hierarchy between the one- and two-loop corrections in the \MS scheme.

\begin{figure}[tb]
\centering
\includegraphics[width=0.66\linewidth]{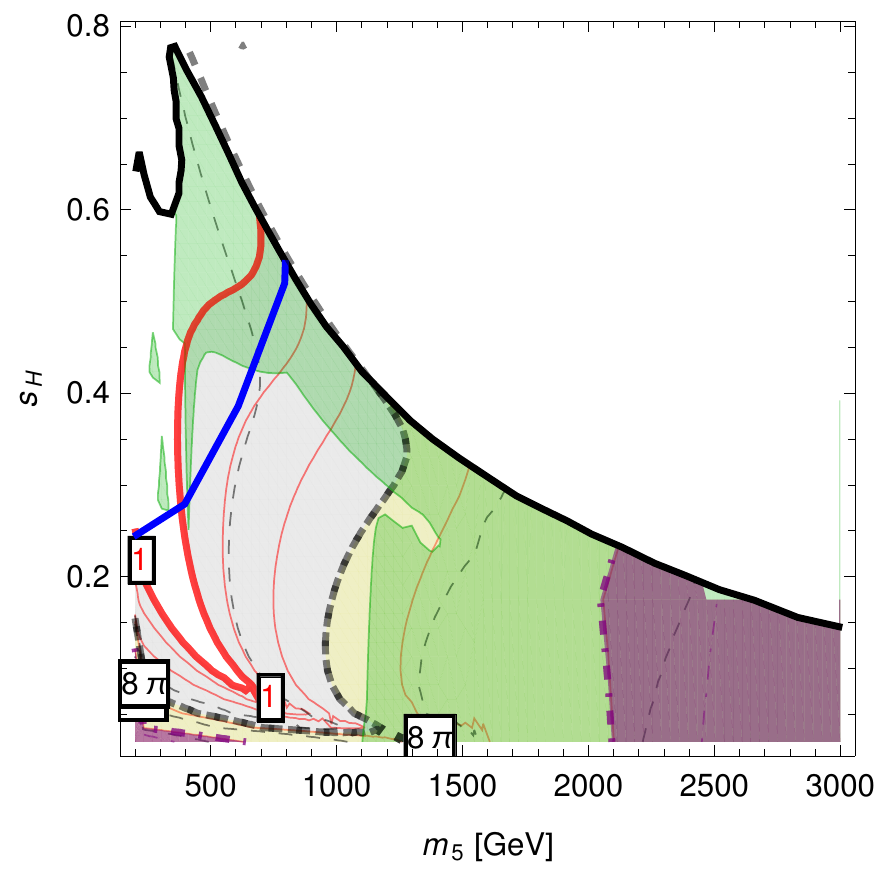}
\caption{
Constraints on the parameter space in the $m_5$-$s_H$ plane using the other parameters according to eq.~(\ref{eq:mh_plane_parameters}). Points above the black (blue) thick line are excluded due to theoretical tree-level bounds (direct LHC searches \cite{Khachatryan:2014sta}) according to Ref.~\cite{Logan:2017jpr}.
The grey shaded area with the red contour lines shows the OS perturbativity constraints based on the relative size of the counter-terms. The red contours correspond to Max$(|\delta \lambda_x/\lambda_x|)=1,\,2,\,4,\,16$ and 64. The yellow shaded area corresponds to the unitarity constraints using renormalised parameters. The black dashed contours correspond to the associated scattering eigenvalues of $4\,\pi$, $8\,\pi$, $16 \, \pi$ and $64\,\pi$. 
As the green contour we show the \MS constraints from the size of the two-loop corrections vs. the one-loop corrections. Regions tainted green do not pass the constraint, i.e. Max$(|(m_{\phi_i}^{2})_{\rm 2L} - (m_{\phi_i}^{2})_{ \rm 1L}|/|(m_{\phi_i}^{2})_{\rm Tree} - (m_{\phi_i}^{2})_{\rm 1L}|)>1$, $i=1,2,3$.
Finally, the purple shaded area corresponds to the perturbativity constraints on the absolute size of the CTs. The purple dot-dashed contours correspond to Max$(|\delta \lambda_x|)=2\,\pi$ and $4\,\pi$. 
}
\label{fig:h5plane}
\end{figure}

Finally, we want to show some constraints on a particular benchmark, the so-called `H5plane', which has been  promoted for the Georgi-Machacek model recently, see Refs.~\cite{deFlorian:2016spz,Logan:2017jpr}. This plane is characterised by only two free parameters $m_5$ and $s_H$. The other parameters are fixed according to
\begin{align}
m_h = 125~{\rm GeV}\,,\quad \lambda_3 = -0.1\,,\quad \lambda_4 = 0.2\,, \quad
\lambda_2 = 0.4 \frac{m_5}{1~{\rm TeV}}\,, \quad M_1 = \frac{\sqrt{2} \, s_H}{v} (m_5^2 + v^2)\,,\quad M_2 = \frac{M_1}{6}\,.
\label{eq:mh_plane_parameters}
\end{align}
Obviously, the input parametrisation for this plane is different from the standard input choices I and II defined above. We have modified our code accordingly.

It has already been shown in Ref.~\cite{Braathen:2017izn} that there are, what appears to be, serious problems with perturbativity for large values of $m_5$. This was shown using the \MS scheme, with the result that much of the parameter space is ruled out once the constraint is taken into account that the 2-loop mass correction to the SM-like Higgs must not be larger than the 1-loop correction. In Fig.~\ref{fig:h5plane}, we now show the constraints arising from all other perturbativity conditions. As can be seen from there, the perturbativity constraints cut deeply into the parameter space of the H5plane. In particular, demanding in the OS scheme that $|\delta \lambda_x/\lambda_x|<1$ (thick red contour line) only leaves valid parameter space below $m_5 \lesssim 500~$GeV. Only considering the perturbative unitarity cuts using renormalised couplings and demanding $8\,\pi$ to be the upper limit (thick black dashed contour line), instead leaves points up to about 1.2~TeV. The loosest constraints come from the absolute size of the CTs. Only the parameter space below $m_5 \simeq 2.1~{\rm TeV}~(2.5~{\rm TeV})$ is ruled out by these considerations if the cut $|\delta \lambda_x|<2\,\pi~(4\,\pi)$ is applied.
In addition to the \MS constraints already discussed in Ref.~\cite{Braathen:2017izn} for the lightest neutral scalar, we include the same condition for the two heavier neutral scalars. In fact, it turns out that the 2-loop corrections to the second mass eigenstate (corresponding to $m_5$) are the most dangerous. The parameter space ruled out because of Max$(|(m_{\phi_i}^{2})_{\rm 2L} - (m_{\phi_i}^{2})_{ \rm 1L}|/|(m_{\phi_i}^{2})_{\rm Tree} - (m_{\phi_i}^{2})_{\rm 1L}|)>1$ is shown in green.

Also shown in the figure are the constraints extracted from Ref.~\cite{Logan:2017jpr}: only points below the black solid line are allowed from theoretical tree-level constraints. Note that for large $s_H$, this line is in agreement with the loop-level unitarity bound -- considerable deviations however appear below $s_H \lesssim 0.4$. The parameter space above the blue line is excluded from the LHC direct searches for doubly-charged scalars \cite{Khachatryan:2014sta}. In total, depending on how conservative a perturbativity cut is applied, the left-over parameter space of the H5plane is either small or even just a tiny strip.

\subsection{Vacuum stability}
\label{sec:result_vacuum}
So far, we have discussed one-loop perturbativity constraints on the parameter space of the GM model, a new kind of constraint which, to the best of our knowledge, has not been discussed in literature in the context of the GM model -- or any other BSM model -- before. 
However, also the impact of loop-corrections on well-known constraints which already exist at tree level is expected to be significant. Here we turn to the discussion of the vacuum stability constraints and show how the impact of the loop corrections can alter the conclusions drawn on that basis. A discussion of these effects for the THDM was done in Ref.~\cite{Staub:2017ktc}, and we can find here quite similar features for the GM model. 

\subsubsection{Stabilising UFB directions}
\begin{figure}[tb]
\centering
\includegraphics[width=0.66\linewidth]{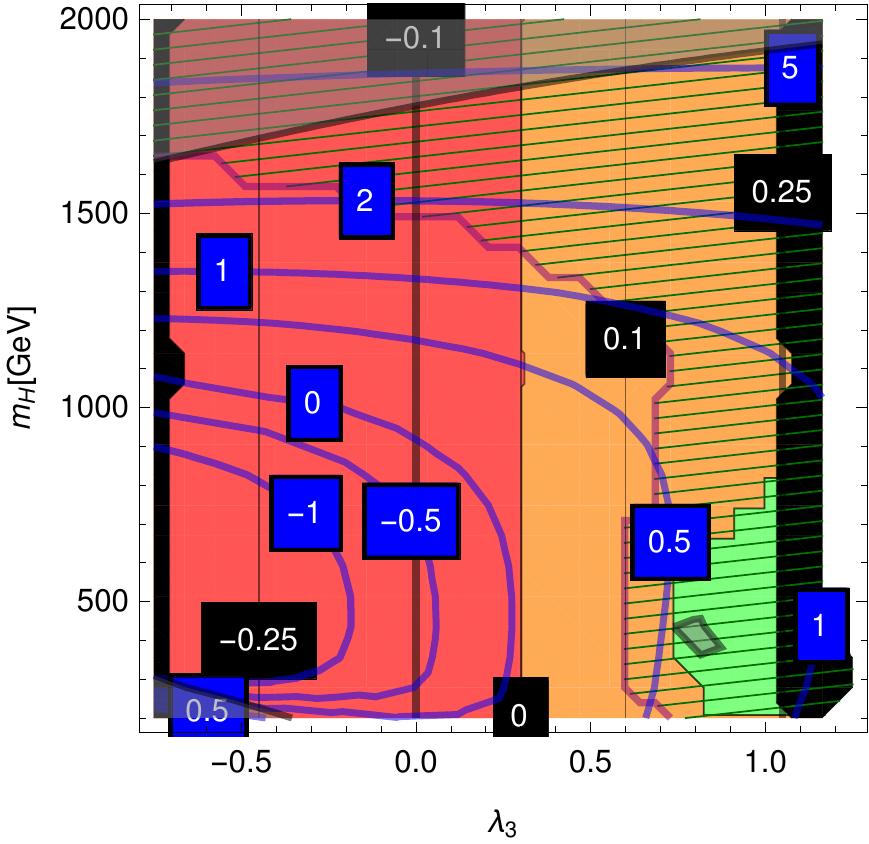}  
\caption{Comparison of the vacuum stability at tree-level and one-loop level. The shaded areas represent the stability of the potential at tree-level: unbounded from below direction exist (red), other minima deeper than the ew vacuum exist (orange), the ew vacuum is stable (green).  The green hatching indicates a stable ew vacuum at one-loop whereas no hatching corresponds to a metastable vacuum at one-loop. Black lines show the value of constant $\lambda_3/3+\lambda_4$ at tree level,  and the blue ones of $\text{min}\{\lambda_{3a} + 2 \lambda_{3b} + \lambda_{4a} + 4 \lambda_{4b} +   4 \lambda_{4c},\  \lambda_{3a}+\lambda_{4a},\ \lambda_{3b}+2\lambda_{4b},\ \lambda_{3b}+\lambda_{4b},\ \lambda_{3c}+\sqrt{2}\sqrt{(\lambda_{3a} +\lambda_{4a})(\lambda_{3b}+2\lambda_{4b})} + 2 \lambda_{4c}\}$ at one-loop. The black area is forbidden by the tree-level unitarity conditions. 
The grey shaded area indicates the perturbativity constraints. The other parameter values are 
$\lambda_2 = 0.1$, $\lambda_4=-0.1$, $\lambda_5=0.1$,  $\alpha=15\degree$, $s_H=0.23$.}
\label{fig:ufb_l3}
\end{figure}
We start with unbounded-from-below directions which exist for the tree-level potential for many different field combinations. We have already discussed in sec.~\ref{sec:constraints_ufb} that these directions are very often expected to disappear once loop effects are included. While we have focused in sec.~\ref{sec:constraints_ufb} on the RGE-improved potential containing only quartic couplings -- which is a valid approach in the limit of very large scales -- we use here the one-loop effective potential. There are mainly two reasons for that: (i) the running of the quartic couplings is usually very fast, i.e. the scale at which the couplings or combinations of them change their sign is not for away from the input scale. Thus this scale is often not much higher than the scale of the dimensionful parameters in the potential. 
(ii) Even if the UFB conditions are satisfied at higher scales, this doesn't mean that those points are necessarily stable: it can and will happen that the potential in the UFB directions is deformed to a local minimum which is deeper than the electroweak one. In order to check this, one needs to find all minima of the effective potential and compare their depths. 
We do this explicitly at one example in Fig.~\ref{fig:ufb_l3} 
where we compare the vacuum stability at tree level and at the one-loop level as a function of tree-level input value $\lambda_3$ as well as $m_H$. We used as further input 
\begin{eqnarray*}
\lambda_2 = 0.1\,,\ \lambda_4=-0.1\,,\ \lambda_5=0.1\,,\  \alpha=15\degree\,,\ s_H=0.23\,.
\end{eqnarray*}
At tree level, the most constraining condition $\lambda_3/3+\lambda_4<0$ for the presence of a UFB direction, cf. eq.~(\ref{eq:UFB_secon_to_last_line}), therefore becomes
\begin{equation}
\lambda_3 < 0.3\,.
\end{equation}
This rules out a large fraction of parameter space in the shown plane. Moreover, also for $\lambda_3 > 0.3$, other and deeper minima than the ew one are present at tree-level. As a consequence, the tree-level potential is only stable in a small region with $\lambda_3 \simeq 1$ and $m_H < 900$~GeV. At the one-loop level, there are finite corrections to the quartic couplings. Therefore, the most constraining conditions to not have a UFB direction in the combined tree-level and CT potential become
\begin{align}
\label{eq:loop_ufb}
\text{min}&\{\lambda_{3a} + 2 \lambda_{3b} + \lambda_{4a} + 4 \lambda_{4b} +   4 \lambda_{4c},\quad \lambda_{3a}+\lambda_{4a}, \quad \lambda_{3b}+2\lambda_{4b},\quad \lambda_{3b}+\lambda_{4b}, \nonumber \\ &~\, \lambda_{3c}+\sqrt{2}\sqrt{(\lambda_{3a} +\lambda_{4a})(\lambda_{3b}+2\lambda_{4b})} + 2 \lambda_{4c}\} > 0\,,
\end{align}
with $\lambda_{Nx} = \lambda_N + \delta\lambda_{Nx}$. The contours for constant values of eq.~(\ref{eq:loop_ufb}) are also shown in Fig.~\ref{fig:ufb_l3}: negative values appear only for a rather small region with mainly $\lambda_3 < 0$ and $m_H<1$~TeV. Thus, the UFB directions in the other parts of the plane disappeared already just because of the CTs independently of the other one-loop corrections.\footnote{One can check that the additional loop corrections are also positive.} As expected, not the entire region where the loop potential doesn't have a UFB direction also provides a stable ew vacuum. There is still a non-negligible region where the ew minimum is not the global minimum of the scalar potential. Nevertheless, the region with the ew minimum as the global minimum is significantly larger than at tree-level. \\

\begin{figure}[tb]
\centering
\includegraphics[width=0.66\linewidth]{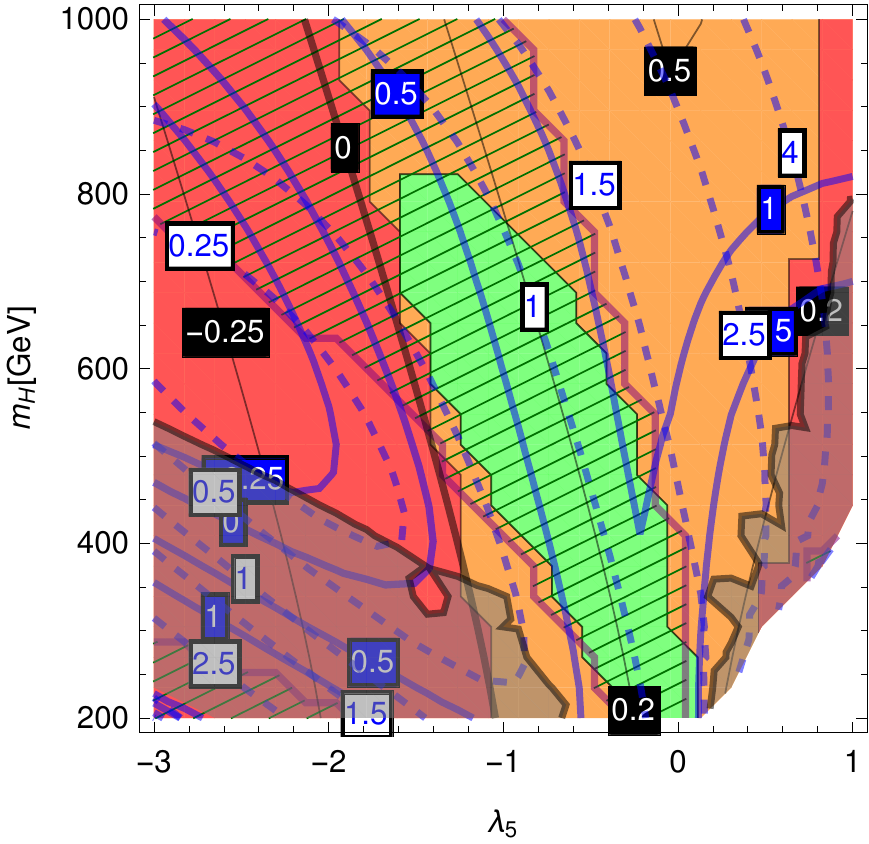}  
\caption{Comparison of the vacuum stability at tree-level and one-loop level. The shaded areas represent the stability of the potential at tree-level: unbounded from below direction exist (red), other minima deeper than the ew vacuum exist (orange), the ew vacuum is stable (green). Hatched regions feature a stable ew minimum at one-loop. Black lines show the value of constant $\lambda_2 - \frac14 |\lambda_5| + \sqrt{2 \lambda_1 (2\lambda_4+\lambda_3)}$ using the tree-level input values, and the full blue ones of 
$\text{Min}\{4 \lambda_{2a} - |\lambda_{5a}| + 4 \sqrt{2 \lambda_1 (2 \lambda_{4b} + \lambda_{3b})}, 
2 \lambda_{2a} + \lambda_{2b} + 
 2 \sqrt{\lambda_{1} (\lambda_{3a} + 2 \lambda_{3b} + \lambda_{4a} + 4 (\lambda_{4b} + \lambda_{4c}))} - \frac12 \lambda_{5a} - \lambda_{5b}  \}$. 
For the dashed blue lines, the term of eq.~(\ref{eq:CW_l5}) was added in addition to the tree-level + CT potential and the UFB condition of $4 \lambda_{2a} +\lambda_{5a} + 4 \sqrt{2 \lambda_1 (2 \lambda_{4b} + \lambda_{3b})}>0$ has been re-derived. 
The grey shaded area indicates the perturbativity constraints. The other parameter values are 
$\lambda_2 = 0.1$, $\lambda_3=0.5$, $\lambda_4=-0.1$,   $\alpha=20\degree$, $s_H=0.33$.}
\label{fig:ufb_l5}
\end{figure}
A similar mechanism also works for other UFB directions. In Fig.~\ref{fig:ufb_l5} we show the $\lambda_5-m_H$ plane and consider the condition 
\begin{equation}
\lambda_2 - \frac14 |\lambda_5| + \sqrt{2 \lambda_1 (2\lambda_4+\lambda_3)} > 0 
\label{eq:UFB_condition_Abs_lam5}
\end{equation}
which is violated in the depicted $(\lambda_5,m_H)$ plane at tree-level roughly for $\lambda_5 < -1$ (up to some corrections stemming from the changes in $\lambda_1$ in this plane).
We show for comparison again the contours of the equivalent conditions for UFB directions in the tree-level plus CT potential, i.e.
\begin{align}
\label{eq:ufb_l5}
\text{min}\{4 \lambda_{2a} - |\lambda_{5a}| + 4 \sqrt{2 \lambda_1 (2 \lambda_{4b} + \lambda_{3b})}, & \nonumber \\
& \hspace{-4cm} 2 \lambda_{2a} + \lambda_{2b} + 
 2 \sqrt{\lambda_{1} (\lambda_{3a} + 2 \lambda_{3b} + \lambda_{4a} + 4 (\lambda_{4b} + \lambda_{4c}))} - \frac12 \lambda_{5a} - \lambda_{5b}  \} < 0 
\end{align}
The difference between these lines is not as 
pronounced as in Fig.~\ref{eq:loop_ufb} -- however, the regions with a stable minimum at the one-loop level are still significantly larger than at tree-level. This means that also in this example  
the loop corrections from the Coleman-Weinberg potential are very important to stabilise the UFB directions. In the entire plane the more stringent condition of eq.~(\ref{eq:ufb_l5}) is
 $4 \lambda_{2a} - |\lambda_{5a}| + 4 \sqrt{2 \lambda_1 (2 \lambda_{4b} + \lambda_{3b})}$, which corresponds to the direction $\langle H^0\rangle=\langle \chi^-\rangle= \langle \phi^-\rangle=\langle \chi^{--}\rangle=\langle \phi^0\rangle=0, \langle H^+\rangle=x \langle \chi^0\rangle$. In this direction, 
the terms proportional to $\langle \chi^0 \rangle^4$ in the CW potential which don't come together with a logarithm are
\begin{align}
\label{eq:CW_l5}
V_{\rm CW} \sim & \frac{1}{32 \pi^2} \Big(64 (\lambda_2^2 + 8 \lambda_3^2 + 20 \lambda_3 \lambda_4 + 17 \lambda_4^2) + 
  4 \lambda_5^2 +   4 (8 \lambda_2 (6 \lambda_1 + 2 \lambda_2 + 7 \lambda_3 + 11 \lambda_4) + \nonumber \\
  &    2 (2 \lambda_1 + 4 \lambda_2 - \lambda_3 + 2 \lambda_4) \lambda_5 +      3 \lambda_5^2) x^2 + 
  3 (64 \lambda_1^2 + 12 \lambda_2^2 + \lambda_5^2) x^4\Big)\,.
\end{align}
In the next step, for achieving an insight into the question where the UFB directions go at one-loop, we add these terms to the tree-level + counter-term potential according to eq.~(\ref{eq:one_loop_potential}) and re-derive the UFB condition of eq.~(\ref{eq:UFB_condition_Abs_lam5}) for the modified potential. The corresponding dashed blue contour lines are shown in Fig.~\ref{fig:ufb_l5}. We see that the corresponding values are positive over the entire parameter range -- which means that the tree-level UFB direction becomes bounded from below at the one-loop level.

\subsubsection{Stabilising meta-stable regions}

\begin{figure}[tb]
\includegraphics[width=0.5\linewidth]{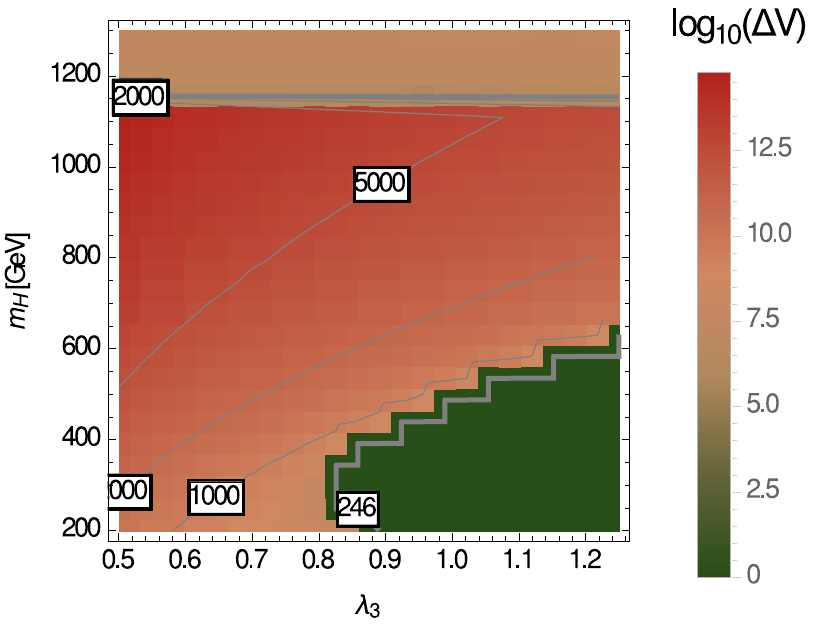} \hfill 
\includegraphics[width=0.5\linewidth]{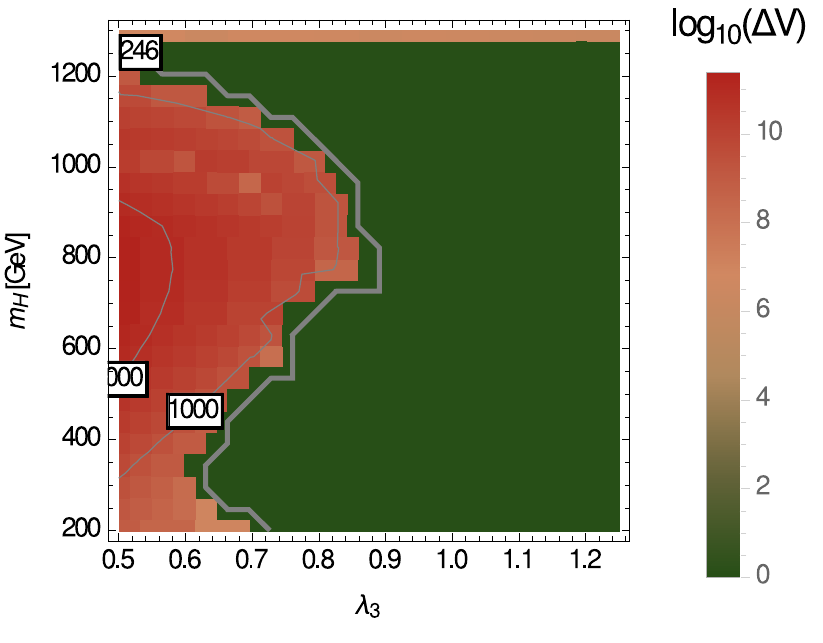} 
\caption{The scalar potential at the tree-level (left) and one-loop level (right): the colour indicates the difference between the panic minimum and the electroweak minimum. The contour lines are constant values for $\sqrt{v_\phi^2+4 (v_\eta^2 +v_\chi^2)}$. The other parameters are set to $\lambda_2 = 0.1$,  $\lambda_4=-0.1$, $\lambda_5=0.1$,  $\alpha=20\degree$, $s_H=0.33$.}
\label{fig:vacuum_l3}
\end{figure}

If no UFB direction exists, the ew vacuum could still be unstable due to the existence of other minima deeper than the one with correct EWSB.
This case was to some extent already mentioned in the last subsection. Here we are going to investigate it in more depth. 
In Fig.~\ref{fig:vacuum_l3}, we show the 
difference $\Delta V$ between the ew minimum and the panic vacuum, $\Delta V = V_{\rm EWSB}-V_{\rm panic}$. The panic vacuum is defined as the minimum which is deeper than and closest in field space to the ew vacuum, i.e. the one to which the ew vacuum configuration would tunnel to eventually. $\Delta V$ is a positive-semidefinite quantity:
if the desired (i.e. ew) vacuum configuration corresponds to the global minimum of the potential, then such a panic vacuum does not exist (or is exactly as deep as the ew minimum) and we define $\Delta V =0$. Positive values of $\Delta V$ are, in turn, reached if there is a non-ew panic vacuum.
The  contour lines drawn in the figure correspond to lines of constant $\sqrt{v_\phi^2+4 (v_\eta^2 +v_\chi^2)}$. On the left-hand plane of Fig.~\ref{fig:vacuum_l3}, we present the result at the tree-level, and on the right-hand side with the full one-loop corrections included. The other parameters were set to
\begin{eqnarray*}
\lambda_2 = 0.1\,,\  \lambda_4=-0.1\,,\ \lambda_5=0.1\,,\  \alpha=20\degree\,,\ s_H=0.33\,.
\end{eqnarray*}
We see that at tree-level, only in one corner of the depicted parameter region, the ew minimum is also the global minimum of the potential. The depth\footnote{The `depth' is the difference of the potential with a given VEV configuration compared to the potential with all VEVs vanishing.} of this minimum is about $-10^7 \text{GeV}^4$. For $\lambda_3<0.8$ and/or $m_H>600$~GeV, minima occur which are deeper than the ew one by several orders of magnitude. Note that the sum of all VEVs   $\sqrt{v_\phi^2+4(v_\eta^2 +v_\chi^2)}$  for these other minima is in the TeV range, i.e. clearly larger than the ew scale. Nevertheless, those minima are not `too far' away in field space, meaning that the tunnelling  from the ew to the panic vacuum is very fast. We are going to quantify this statement below. \\
Turning to the vacuum structure of the one-loop potential, shown on the right in Fig.~\ref{fig:vacuum_l3}, we observe that the ew vacuum corresponds to the global minimum of the potential for a much larger region of the depicted parameter space. Up to $m_H \simeq 1200$~GeV, values of $\lambda_3$ exist for which the ew potential is stable. Obviously, the one-loop corrections can be large and positive: they manage to overcompensate the potential difference at tree-level of 10 orders of magnitude. This large value is not surprising because one could estimate the corrections to be  $O(\frac{m_H^4}{16\pi^2})$. 
Only for small values of $\lambda_3$, the other minima remain deeper than the ew one. \\
\begin{figure}[tb]
\centering
\includegraphics[width=0.66\linewidth]{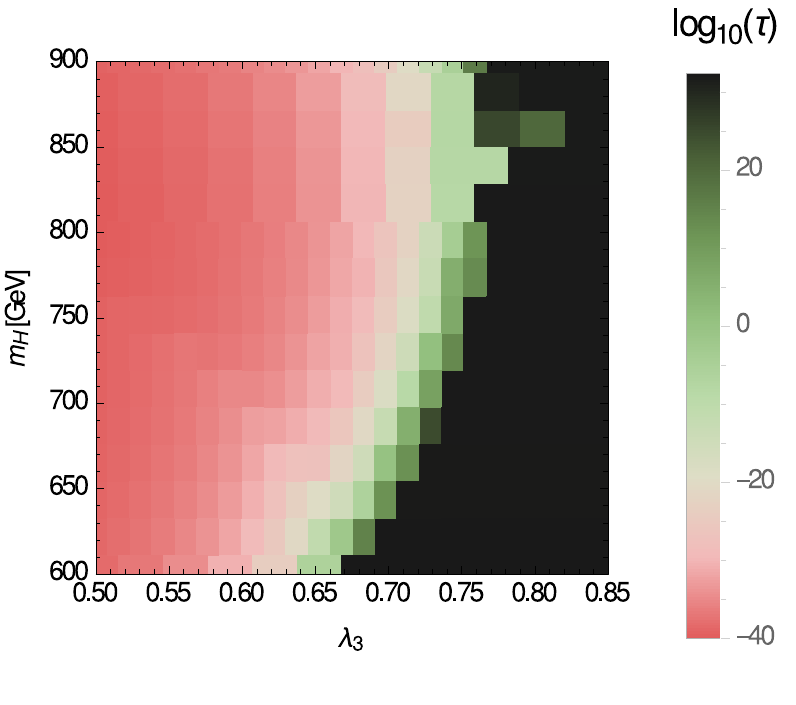}  
\caption{Zoom into the parameter plane of the right-hand side of Fig.~\ref{fig:vacuum_l3} and presenting the calculated life-time of the electroweak vacuum normalised to the life-time of the universe at zero temperature.}
\label{fig:lifetime_l3}
\end{figure}
We  conclude this discussion with a brief analysis of the tunnelling rate. We have calculated the tunnelling from the desired to the panic vacuum using the code  {\tt CosmoTransitions} \cite{Wainwright:2011kj} in combination with \Vevacious. The result is shown in Fig.~\ref{fig:lifetime_l3}, where we have zoomed into the region $600~\text{GeV}<m_H<900~\text{GeV}$ of the right-hand side of Fig.~\ref{fig:vacuum_l3}. We find that there is only a narrow band where the life-time of the ew minimum is comparable to the age of universe when calculating the tunnelling rate at zero temperature. These points could in principle be considered as `long-lived', i.e. viable. However, once the thermal corrections are included as explained in Ref.~\cite{Camargo-Molina:2014pwa}, they also usually become short-lived. Therefore, from now on, we are going to consider all points which feature a deeper panic vacuum as not viable. This is similar to the THDM where it was also found that the tunnelling to deeper minima is always very fast on cosmological time scales \cite{Barroso:2013awa,Staub:2017ktc}.

\subsubsection{De-stabilising stable regions}
\label{sec:results_vac_destab}
\begin{figure}[tb]
\centering
\includegraphics[width=0.66\linewidth]{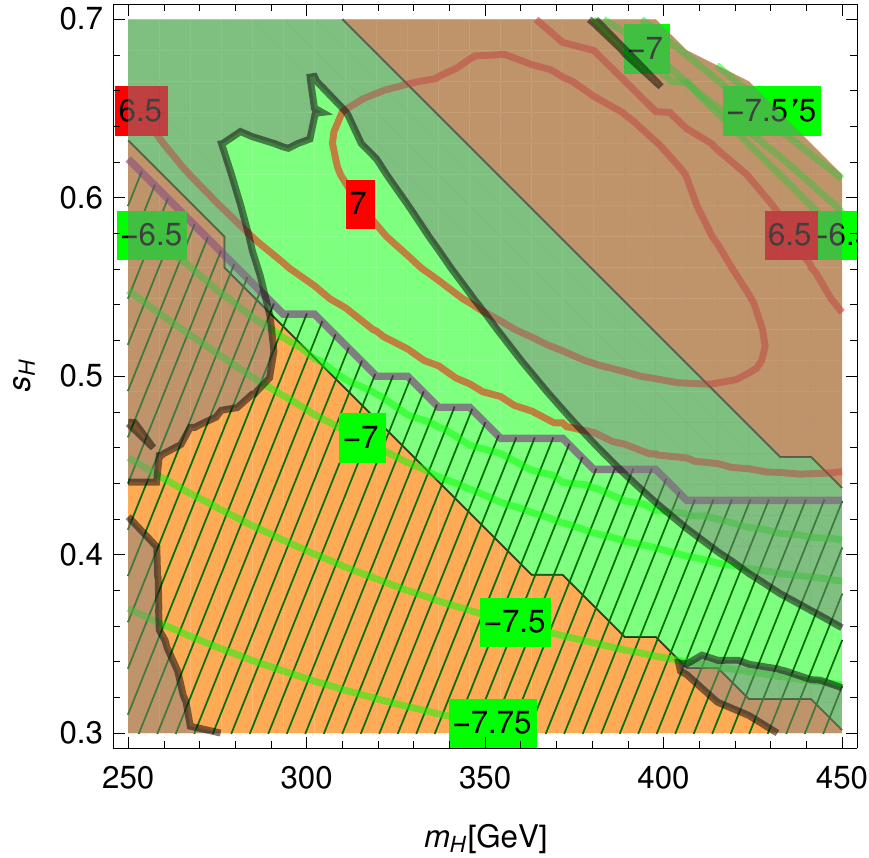}  
\caption{Comparison of the vacuum stability at tree level and the one-loop level. The shaded areas represent the stability of the potential at tree level: unstable (orange) or stable (green). 
The green hatched area is stable at the one-loop level. The green, purple and red contour lines show the depth of the loop-corrected ew minimum: the values are $\text{sign}(V)\log_{10}|V/\text{GeV}^4|$. 
The grey shaded area indicates the perturbativity constraints where we have used the \MS condition. The other parameter values are 
$\lambda_2 = 1$, $\lambda_3=0.9$, $\lambda_4=-0.2$, $\lambda_5=0.2$,  $\alpha=15\degree$.}
\label{fig:destabilize}
\end{figure}

So far, we have discussed the situation that the stability of the potential is {\it increased} once loop corrections are included. 
We found this situation to appear far more often than the opposite effect.
However, it can also happen in some cases that the stability of the ew potential is {\it decreased} via the loop corrections. The reason is that the loop corrections are often positive. Thus, if they are larger than the depth of  the ew minimum at tree-level, they can push it to positive values. Consequently, the minimum with all VEVs set to zero is deeper at the loop-level. In other words, it can happen that the ew symmetry is restored by the loop corrections. We show an example of this situation in Fig.~\ref{fig:destabilize}. Here, the stability at tree level and the one-loop level is shown in the $(m_H,s_H)$ plane for 
\begin{eqnarray*}
 \lambda_2 = 1\,,\ \lambda_3=0.9\,,\ \lambda_4=-0.2\,,\ \lambda_5=0.2\,,  \alpha=15\degree\,.
\end{eqnarray*}
The tree-level vacuum is stable in a strip around the line $(s_H=0.7,m_H=250~\text{GeV})$ to $(s_H=0.35,m_H=450~\text{GeV})$. However, once the loop corrections are included, the  depth of the ew potential in a large fraction of this area becomes positive. 
This is indicated by the green, purple and red contour lines which show the logarithmic values for negative, zero, and positive potential depth of the ew minimum.
For $m_H=250$~GeV, all points which are stable at the tree level are actually unstable at the loop level. For $m_H=350$~GeV, still more than half of the $s_H$ range which is allowed at tree level is forbidden after the inclusion of loop corrections. Only for $m_H\geq 450$~GeV, all the points which are stable at tree level are also stable at the one-loop level. In general, we find that the restoration of the ew symmetry at the loop-level appears mainly in parameter regions with small $m_H$ and not too small $s_H$.  On the other side, we find that the entire area which is `below' the green band, i.e. where the tree-level ew vacuum is unstable, is stabilised at the loop level. Hence, the overall picture that loop corrections increase the parameter space in which the vacuum is stable, still holds.

\subsection{The global picture}
We have discussed so far the perturbativity constraints as well as the loop-improved vacuum stability constraints at selected examples. As the final step, we want to obtain an impression of the `global picture', i.e. the impact of these constraints in a wide fraction of the full parameter space of the GM model. For this purpose, we performed random scans for both input choices according to eqs.~(\ref{eq:input_choice_I}) and (\ref{eq:input_choice_II}), in the following parameter ranges:
\begin{enumerate}
 \item Input I: 
 \begin{eqnarray*}
  &  \lambda_2,\lambda_3,\lambda_4 \in [-1,1]\,, \qquad \lambda_5 \in [-2.5,2.5]\,,& \\
 & m_H \in [200~\text{GeV}, 2000~\text{GeV}]\,,\qquad \alpha \in [3\degree,50\degree]\,,\qquad s_H \in [0.05,0.95]\,.&
 \end{eqnarray*}
 \item Input II: 
 \begin{eqnarray*}
&  \lambda_2,\lambda_3,\lambda_4 \in [-1,1]\,, \qquad \lambda_5 \in [-2.5,2.5]\,,&\\
&m_5 \in [200~\text{GeV}, 2000~\text{GeV}]\,, \qquad  r_{12} \in [10^{-2},10^2]\,, \qquad s_H \in [0.05,0.95]\,.&
 \end{eqnarray*}
\end{enumerate}
The values of the $\lambda$'s were chosen to ensure that a large fraction of them are in agreement with tree-level unitarity conditions, i.e. to make the scan more efficient. The scans were carried out until for each input choice, 250,000 points were collected which (i) have a tachyon-free particle spectrum, (ii) which pass the tree-level unitarity constraints and (iii) which are in agreement with the $S$ parameter. Afterwards, these points were confronted with different cuts.  First, the experimental constraints on the Higgs couplings were applied using {\tt HiggsBounds}. In the second step, the consequences of the tree-level constraints (UFB, other minima) were compared to the loop-improved vacuum stability constraints and to the perturbativity constraints in addition. In the case of the perturbativity cuts, we made two choices:
\begin{enumerate}
 \item {\it weak condition} (``Weak P.''): the condition that corrections to the scalar masses calculated in the \MS scheme must be smaller at two-loop than at one-loop is applied. In addition, an upper limit for the absolute value of the CTs for the quartic couplings of $2\pi$ has been set. 
 \item {\it strong condition} (``Strong P.''):  the condition that the generalised tree-level unitarity conditions must be fulfilled for the renormalised quartic couplings is applied.
\end{enumerate}
A summary of the number of points surviving the different cuts is given in Table~\ref{tab:summary_random}. 
Passing the {\tt HiggsBounds}  constraints (column `{\tt HB}') is taken as a prerequisite for the subsequent columns. The numbers in the columns ``Tree-Level'' as well as in ``Loop-Level'' show a cutflow in itself, i.e. points which pass ``Tree-Level Stability'' also pass the {\tt HB} as well as Tree-Level UFB  constraints, and points which pass ``Loop-Level Weak/Strong P.'' also pass the {\tt HB} and Loop-Level Stability constraints. Points which pass the loop-level constraints do not necessarily also pass the tree-level constraints and vice versa, in accordance with the previous observations that tree-level constraints are often changed significantly at the loop level. In the last column ``Agreement'', we compare the numbers from the tree- and loop-level constraints: first, we show the number of points which pass both tree and loop constraints. In brackets, we show which percentage of  points which  pass tree-level constraints  also pass the one-loop constraints and vice versa. For instance, 54\% of those points which pass the tree-level cuts are still viable after imposing the loop-level ``Stability'' and ``Weak P.'' conditions. In turn, out of all points which pass the latter, only 35\% would have also been considered viable at tree-level.
A cross-check of the viability of the results is that the fraction of points which have passed the ``Strong P.'' cut and are also in agreement with the tree-level conditions is always larger than the analogue with the ``Weak P.'' cuts. This means that the stronger constraints are  more stringent in filtering out points at which the loop corrections have a large impact. 

\begin{table}[t]
{\small
\begin{tabular}{c|c||cc||ccc||cc}
\hline 
&                             &\multicolumn{2}{|c||}{Tree-Level} &\multicolumn{3}{|c||}{Loop-Level} & \multicolumn{2}{|c}{Agreement}  \\
\hspace{-.2cm}Input & \hspace{-.2cm}{\tt HB} & UFB & \hspace{-.25cm} Stability & Stability & \hspace{-.25cm}Weak P. & \hspace{-.25cm}Strong P. & Weak P. & Strong P.\\
\hline 
 I  & 121034 & 23417 & 2114 & 12749 & 3220 & 1558 & 1150 (54\%, 35\%) & 662 (31\%, 42\%)\\
II  & 159588 & 27316 & 14288 & 34143 & 14222 & 5704 & 7105 (49\%, 49\%) & 3258 (22\%, 57\%)\\
\hline
\end{tabular}
}
\caption{Summary of the results of a random scan using Input I and II. A sample with 250,000 points each has been generated which pass 
the tree-level unitarity constraints. The numbers give the points surviving the different cuts. As weak perturbativity cut we applied the \MS condition
based on the hierarchy of the one- and two-loop corrections to the masses of the CP-even neutral scalars. The strong cut uses the generalised unitarity constraints with renormalised couplings. 
Under `Agreement', we list the number of points which pass both the tree- and loop-level constraints. We show in brackets how many of the points passing the tree-level 
constraints also pass  the loop constraints and vice versa.}
\label{tab:summary_random}
\end{table}

The overall result is that the theoretical constraints at tree level have a misidentification rate of about 50\%. That means that roughly half of the points which seem to be viable when applying tree-level conditions are in conflict with the loop conditions. On the other side, also a large fraction of points which look fine at the loop level would have been discarded if only tree-level conditions had been used. If the strong perturbativity cut is applied, only one third (Input I) respectively one fourth (Input~II) of the points which appear to be valid at tree level also pass the loop constraints. 
Since these are quite large effects, it is important to understand them in some more detail. In the following subsections, we are therefore going discuss the effects and try to pin down where the differences between tree- and loop-level are most pronounced.

\subsubsection{Perturbativity}
\begin{figure}[tb]
\centering
\includegraphics[width=0.4\linewidth]{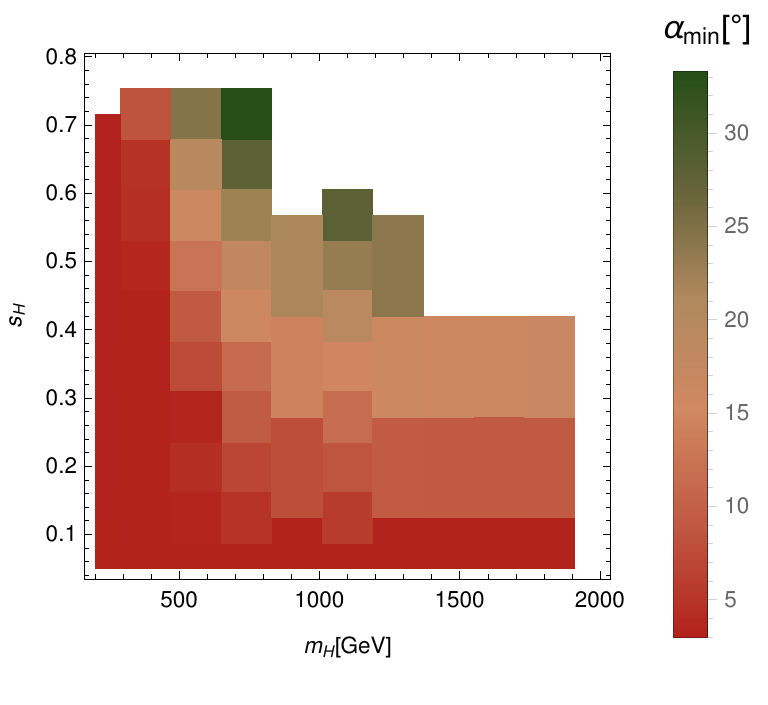} \hspace{1cm}
\includegraphics[width=0.4\linewidth]{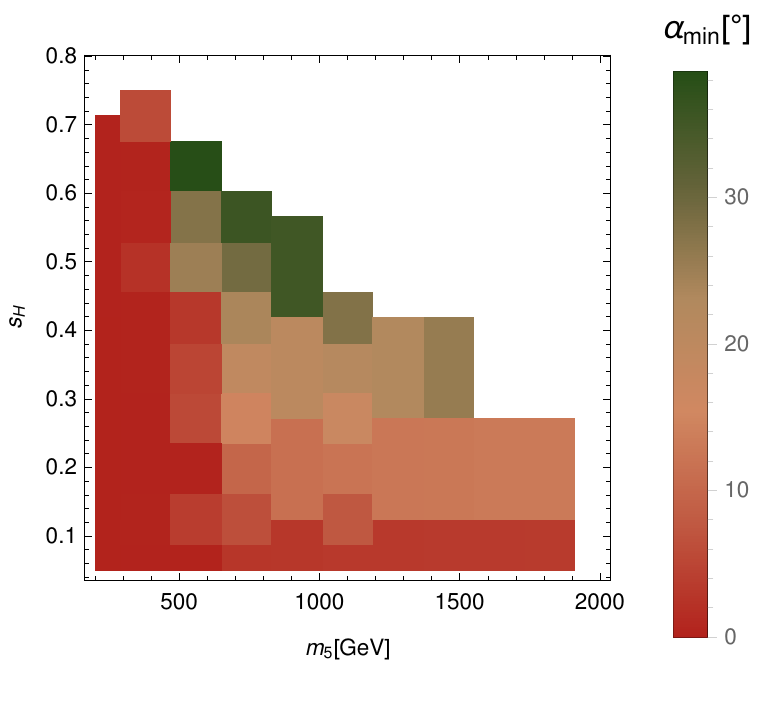} \\
\includegraphics[width=0.4\linewidth]{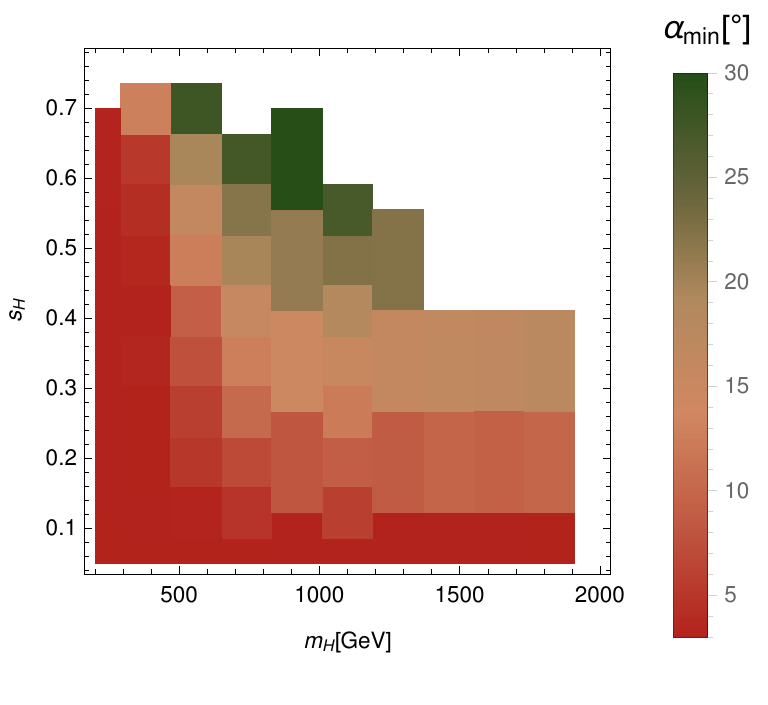} \hspace{1cm}
\includegraphics[width=0.4\linewidth]{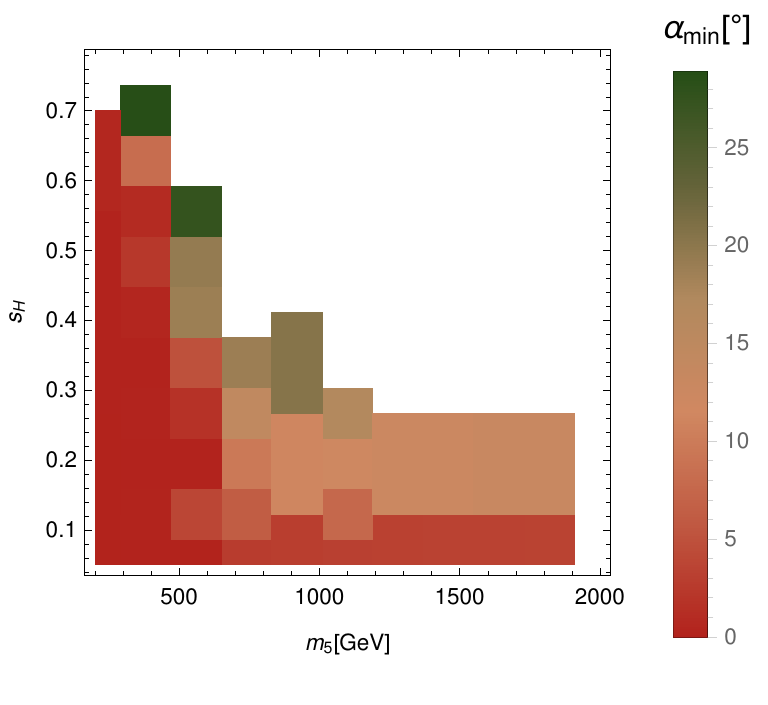} \\
\includegraphics[width=0.4\linewidth]{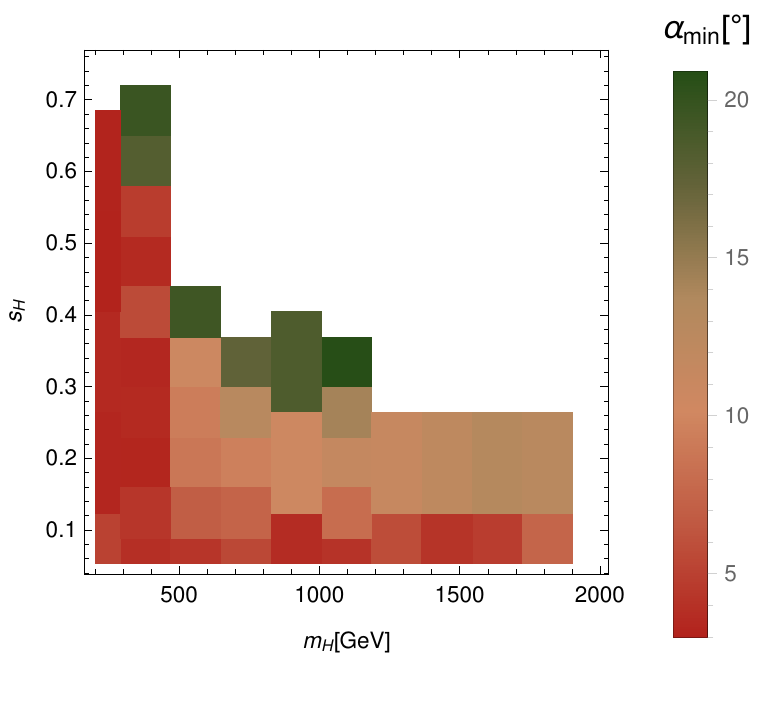} \hspace{1cm}
\includegraphics[width=0.4\linewidth]{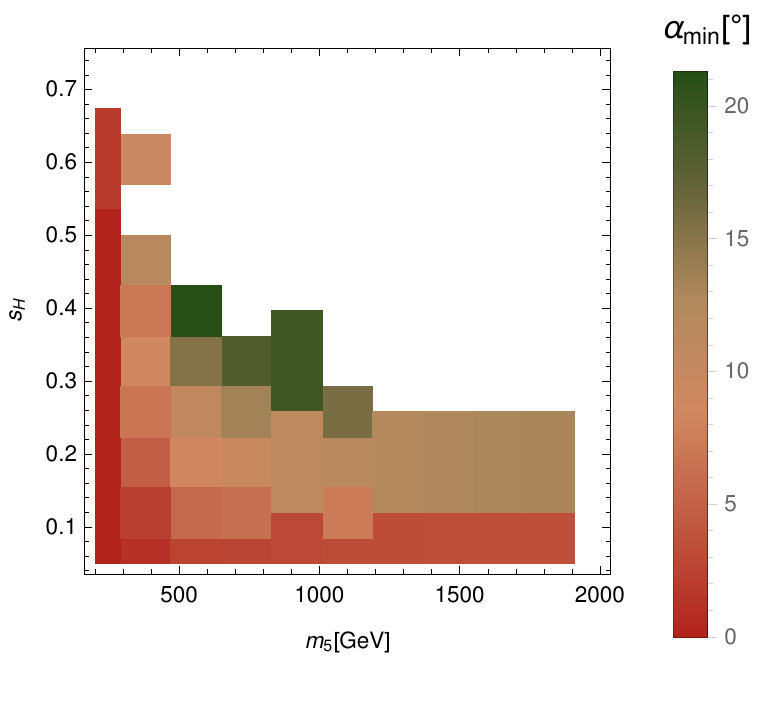} 
\caption{Results of the parameter scan. First row: the minimal value of $\alpha$ found in the $(m_H,s_H)$-plane for Input~I (left) and the $(m_5,s_H)$-plane for Input~II (right). In the second row, we show the same figures after applying the \HB cuts, and in the third row 
after both \HB and weak perturbativity cuts.}
\label{fig:RandomHBalpha}
\end{figure}
The main difference between the two input modes I and II is the treatment of $\alpha$: while it is an input parameter for Input I, it is dynamically calculated for Input II from the other 
parameters by diagonalisation the scalar mass matrix. One finds approximately the following dependence of $\alpha$ on the other parameters
\begin{equation}
\sin2\alpha \sim s_H\sqrt{1-s_H^2} \, \frac{m_3^2}{m_h^2-m_H^2} \,.
\end{equation}
Therefore, usually a correlation between $s_H$ and $\alpha$ is visible and $\alpha$ is naturally large for large $s_H$. On the other side, for Input~I one can, in principle, choose $\alpha$ arbitrarily small independently of $s_H$. The only constraint is that this could lead to tachyons in the scalar sector. To get a picture from the possible values of $\alpha$ for the two input modes, we present in the first row of Fig.~\ref{fig:RandomHBalpha} the minimal value of $\alpha$  which we found in our random scans in the $(m_H,s_H)$ respectively $(m_5,s_H)$ planes. As expected, for Input~II, the minimal values of $\alpha$ tend to be larger for large $s_H$ than for Input~I. Of course, since both inputs are related by just a re-parametrisation, they should be equivalent at the end. However, in order to keep $\alpha$ small for large $s_H$, some tuning in the other parameters is required which is not easily achieved in a random scan. Thus, from this point of view, Input~I looks much more promising to find points which pass the constraints on the Higgs couplings. This is also confirmed in the second row of Fig.~\ref{fig:RandomHBalpha} where we show the minimal value of $\alpha$ in the same planes after the Higgs constraints have been applied. There are parameter regions like  $s_H>0.3, m_H>1200$~\text{GeV} which are hardly accessible via Input~II when choosing the input randomly, but which are populated for Input~I. However, we had already observed in sec.~\ref{sec:results_per_sH} that the perturbativity constrains are in particular strong for cases in which some tension between the values of $s_H$ and $\alpha$ is present. This is also nicely confirmed in our random scan. If we apply in addition to the Higgs coupling constraints the perturbativity constraints, as done in the last row in Fig.~\ref{fig:RandomHBalpha}, we can observe that the two input modes show quite similar results even in the random scans. All regions where only for Input~I, points had survived the \HB cuts, are ruled out by the (weak) perturbativity constraints. Thus, the accessible regions as well as the minimal value of $\alpha$ in these regions look very similar once the \HB constraints are combined with the perturbativity constraints. \\
\begin{figure}[tb]
\includegraphics[width=0.5\linewidth]{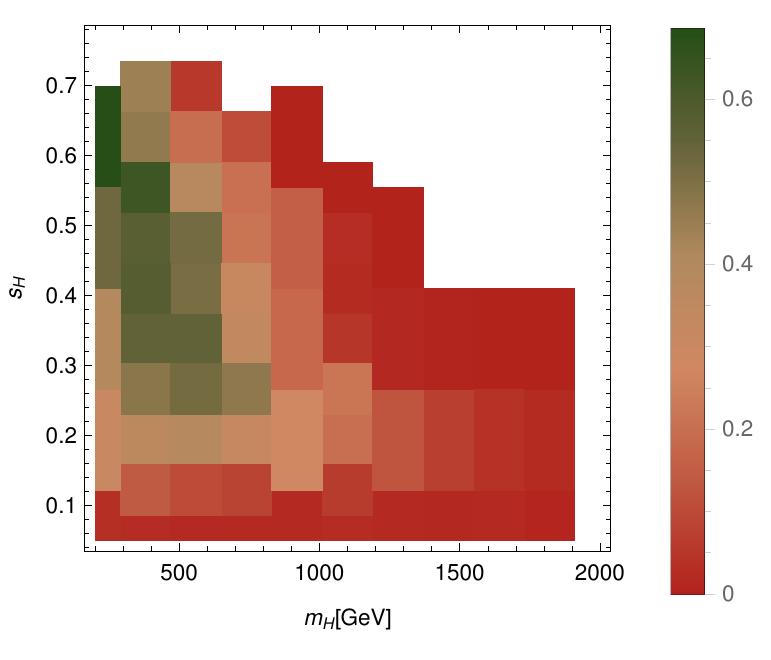} \hfill
\includegraphics[width=0.5\linewidth]{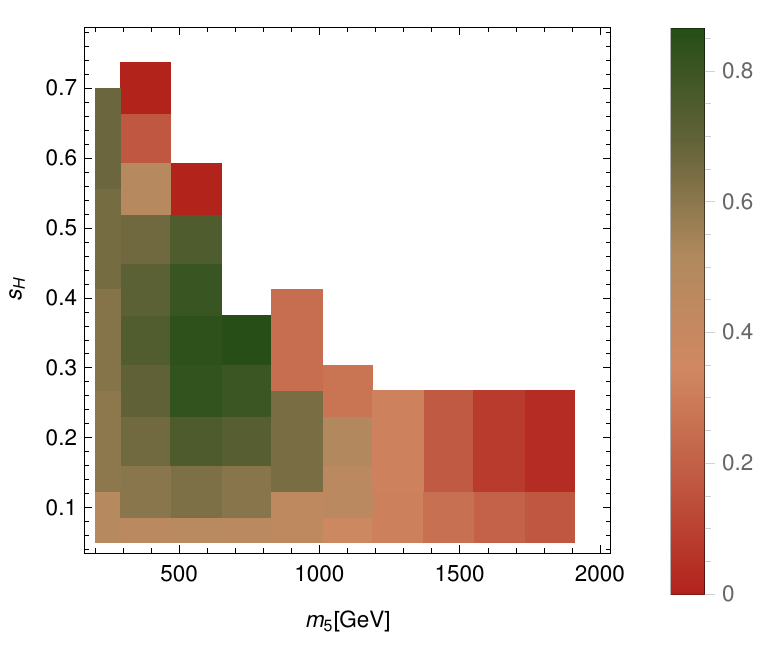} \\
\includegraphics[width=0.5\linewidth]{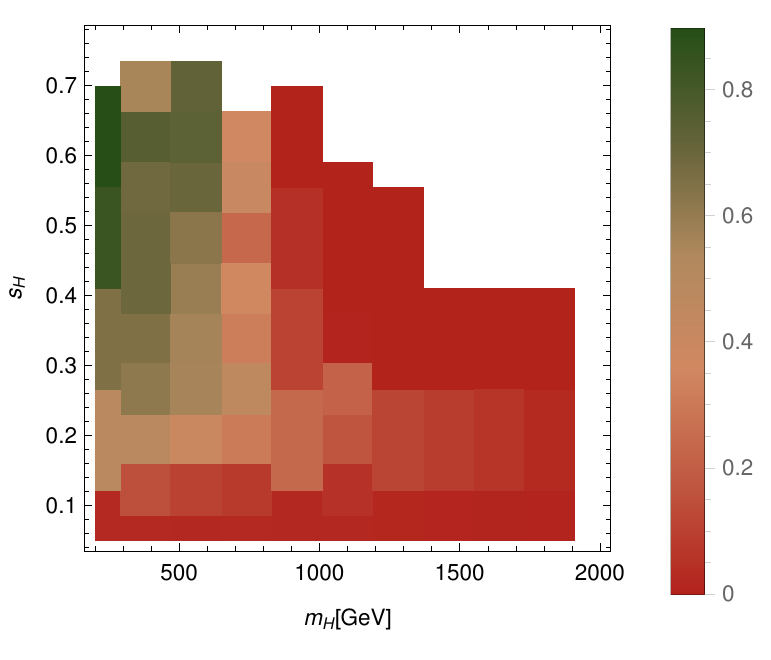} \hfill
\includegraphics[width=0.5\linewidth]{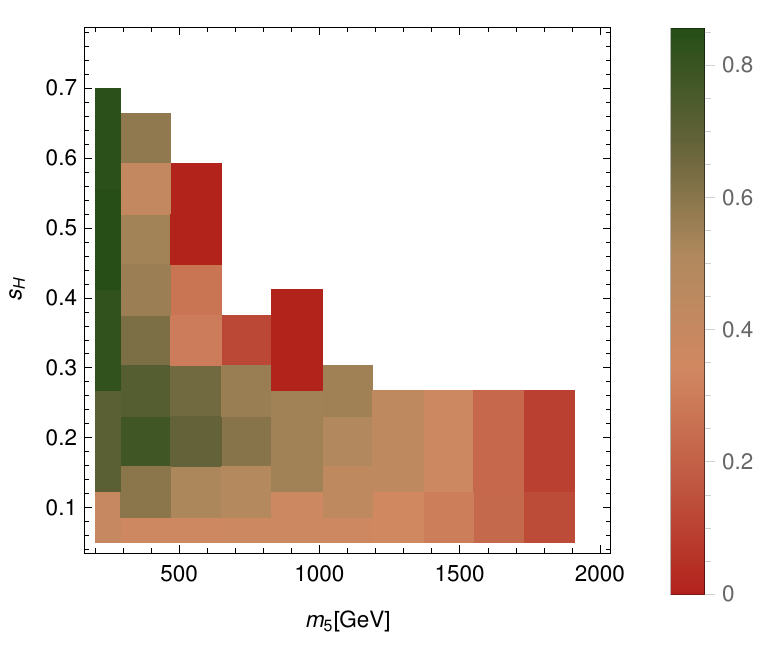}
\caption{Fraction of points passing the weak (first row) or strong (second row) perturbativity cut. Only points which have initially passed the \HB checks are taken into account. The left column is for Input~I, the right one for Input~II.}
\label{fig:RandomPermSH}
\end{figure}
We can confirm this picture also from another point of view. In Fig.~\ref{fig:RandomPermSH}, we show the fraction of points which pass the weak or strong perturbativity constraint in the $(m_{H,5},s_H)$ planes for Input~I respectively Input~II. One finds that these cuts are particularly strong for Input I in the case of (i) large $s_H$ and $m_H$ or (ii) very small $s_H$. In these parameter regions, up to 100\% of the points are ruled out. Is is remarkable that the `strong' cut as we called it since it affects on average more points, can be even less restrictive than the `weak' cut in some regions of parameter space. This is especially the case for Input I and small $m_H$ in combination with large $s_H$. However, one should note that the number of points per bin in these regions is also smaller than in other regions. For Input II, we find in contrast only a soft dependence on $s_H$ as long as it is below 0.5 (strong cut) or 0.3 (weak cut). On the other side, a strong tendency is visible that the perturbativity cuts are more important for increasing $m_5$. 

In general, it is worth noticing that even in those parameter regions where the perturbativity cuts are the least constraining, they still affect at least 15\% of the shuffled points.

\subsubsection{Vacuum stability constraints}
\begin{figure}[tb]
\includegraphics[width=0.5\linewidth]{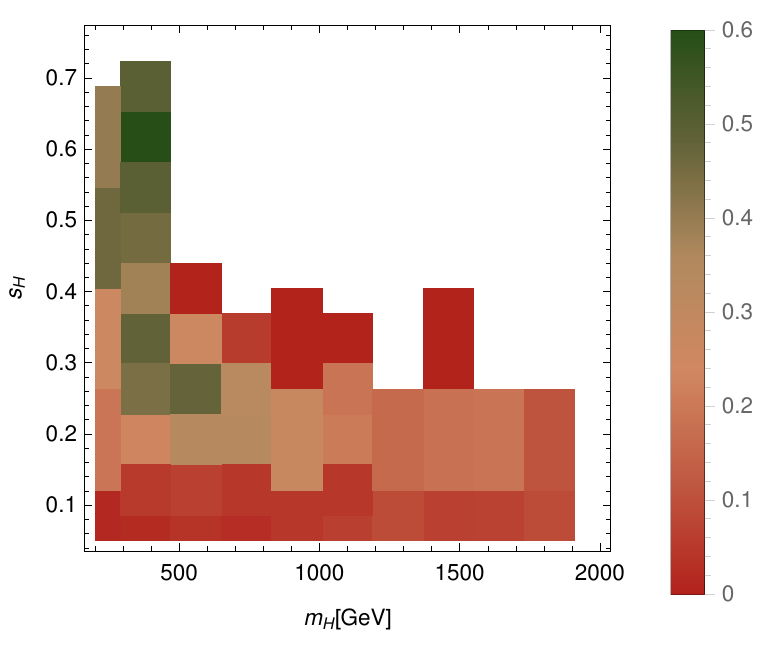} \hfill
\includegraphics[width=0.5\linewidth]{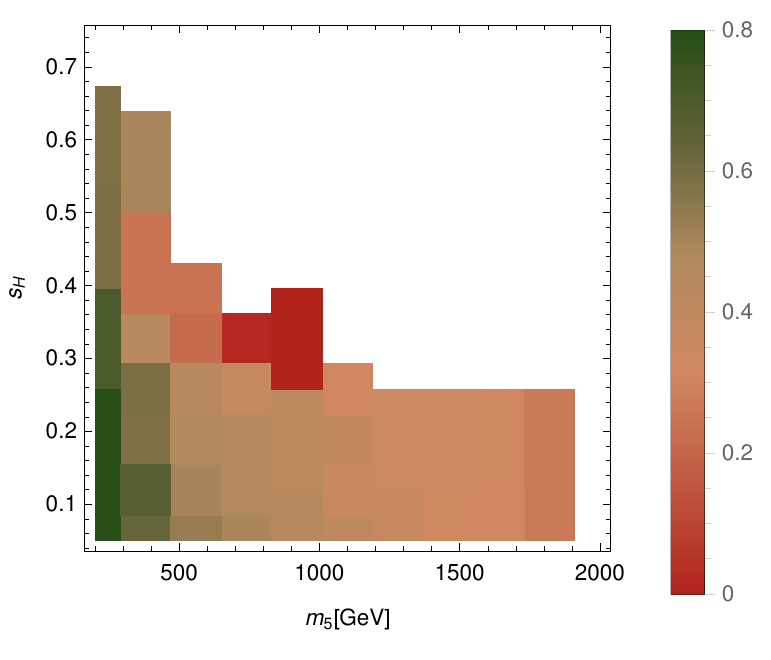} \\
\includegraphics[width=0.5\linewidth]{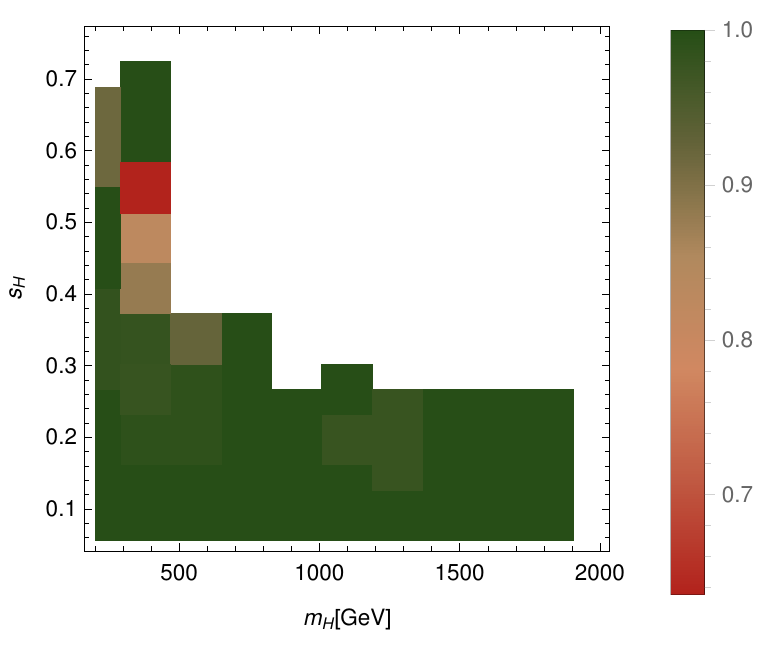} \hfill
\includegraphics[width=0.5\linewidth]{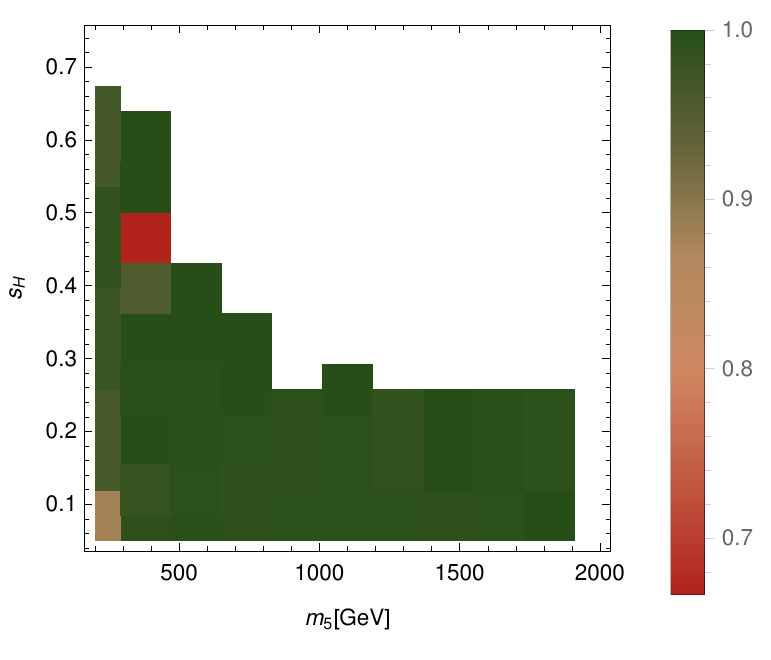}
\caption{First row: fraction of those points which are stable at one-loop and also pass the tree-level constraints. Second row: fraction of those points which are stable at tree-level and  
pass also the loop constraints. The left columns shows the results for Input~I, the right column for Input~II.}
\label{fig:RandomVacuumTreeLoop}
\end{figure}
We now turn to the vacuum stability constraints and discuss the main differences between the tree-level and loop-improved constraints. A comparison between the tree-level and one-loop constraints in the $(m_H,s_H)$ and $(m_5,s_H)$ plane for Input~I and II, respectively, is shown in Fig.~\ref{fig:RandomVacuumTreeLoop}. In this figure, we consider all points which are either stable at the tree- or one-loop-level and show the fraction of points which are also stable at loop- or tree-level, respectively. Thus, these plots present 
\begin{equation}
R = \frac{\#_{|\text{stable at loop}|\text{stable at tree}}}{\#_{|\text{stable at loop}}}  \hspace{1cm} \text{or} \hspace{1cm}
R = \frac{\#_{|\text{stable at tree}|\text{stable at loop}}}{\#_{|\text{stable at tree}}}
\end{equation}
For Input I, we find that at most 60\% of the points per bin which are stable at the one-loop level would also have passed the tree-level constraints. In other words, the misidentification rate that stable points are considered as unstable from tree-level considerations (`false negative') is at least as large as 40\%. For Input~II, the `false negative' misidentification rate can go down to 20\%. However, this is only the case for a small band of very small $m_5$ of 200-300~GeV. Otherwise, this misidentification rate is often 70\% and more. For Input I, parameter regions exist where nearly all points with a stable vacuum at the loop-level would be ruled out by the tree-level constraints. This is in particular the case for very small $s_H$ or for somewhat large $s_H$ ($\sim 0.4$) together with $m_H$ above 500~GeV. For Input~II, there  is also one spot where the misidentification rate is close to 100\% at $s_H \sim 0.35$, $m_5\sim800$-$1000$~GeV.  \\
In contrast, the large majority of points which pass the tree-level constraints is also stable at the loop-level. The main exception is for Input~I in the limit of large $s_H$ and small $m_H$ where the loop corrections restore the ew symmetry as explained in sec.~\ref{sec:results_vac_destab}. In these regions, the misidentification rate that points are assumed to be stable based on tree-level consideration is up to 30\%. In all other parameter regions, it is significantly smaller and at most a few percent. Since the effect of symmetry restoration comes with large $s_H$ together with rather small values of $\alpha$, it is much less pronounced for Input~II. Consequently, the `false positive' misidentification rate is, with the exception of one bin, always below 10\% in the $(m_5,s_H)$ plane. 
\begin{figure}[tb]
\includegraphics[width=0.5\linewidth]{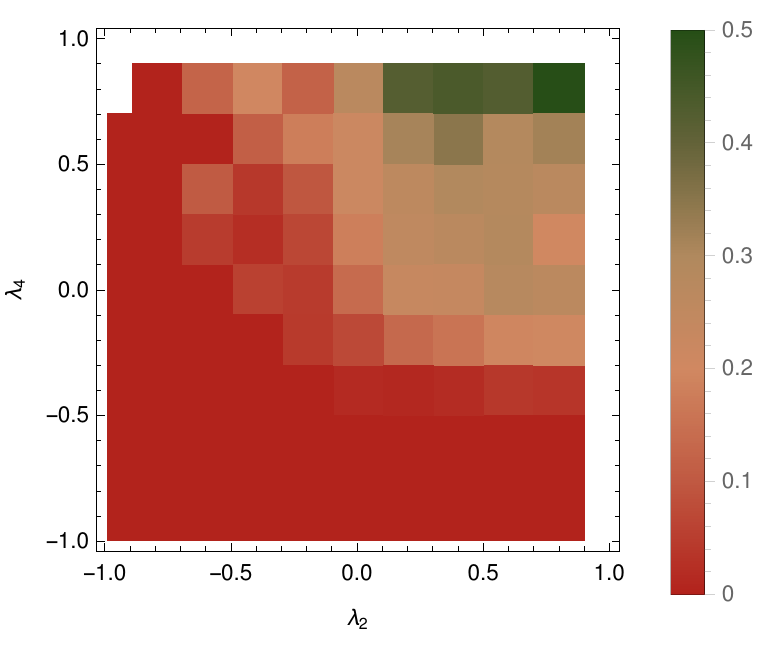} \hfill
\includegraphics[width=0.5\linewidth]{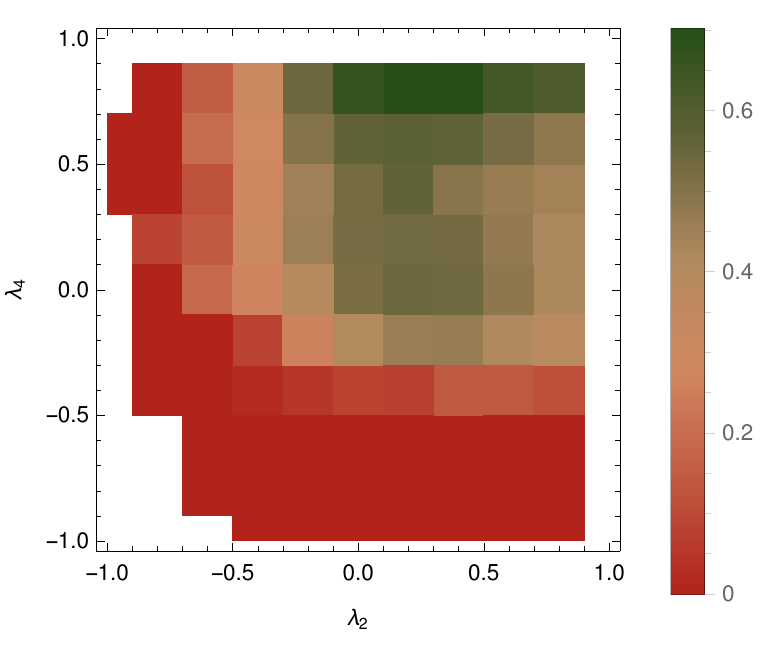} \\
\includegraphics[width=0.5\linewidth]{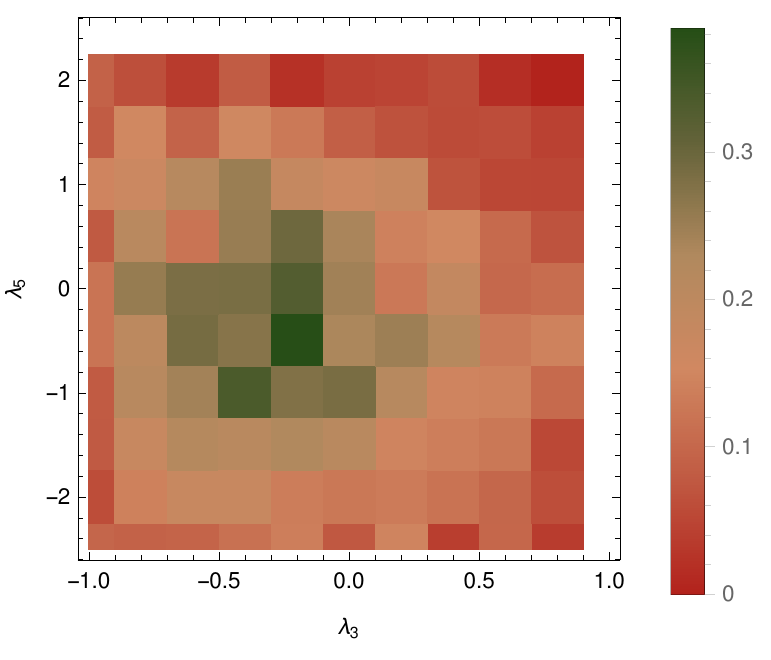} \hfill 
\includegraphics[width=0.5\linewidth]{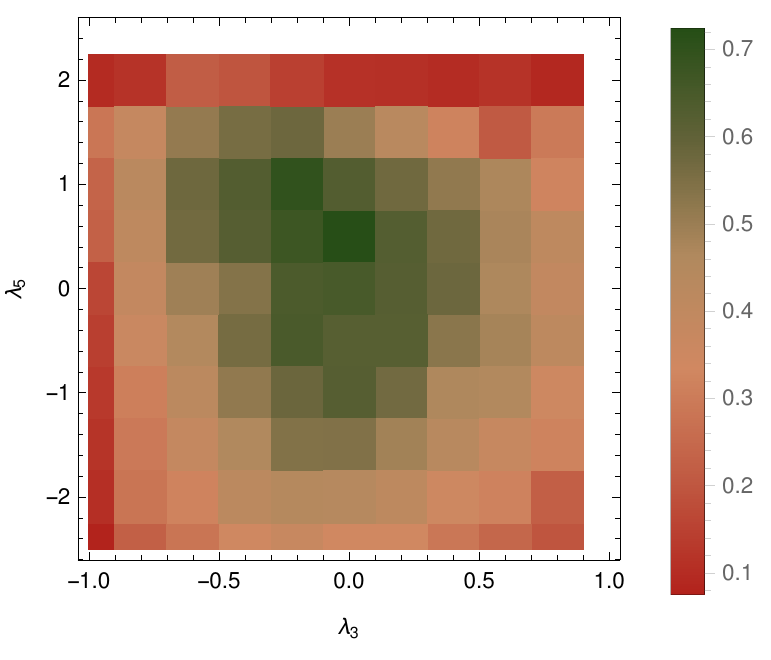}
\caption{Fraction of those points out of the parameter scan which are stable at one-loop and also pass the tree-level constraints. The results are shown in the $(\lambda_2,\lambda_4)$ (upper row) and the $(\lambda_3,\lambda_5)$ plane (lower row). The left column uses the points with the input choice I, the right column uses Input~II.}
\label{fig:RandomVacuumTreeLoopLambda}
\end{figure}
Since for the vacuum stability constraints, and in particular for the check against UFB directions, the quartic couplings are crucial, we show in Fig.~\ref{fig:RandomVacuumTreeLoopLambda} the `false negative' misidentification rate also in the $(\lambda_2,\lambda_4)$ and $(\lambda_3,\lambda_5)$ planes for Input~I and II. 
The overall situation for both input modes is quite comparable. One finds in particular for negative $\lambda_2$ and/or $\lambda_4$ that a large majority is mis-categorised. Only for $\lambda_2 \to 1$, $\lambda_4 \to 1$, more than 50\% of the points which are stable at the loop-level are also considered `stable' from tree-level checks.
The reason is that, although the loop corrections are in principle more important for larger quartic couplings, large $\lambda_{2}$ and $\lambda_4$ make it less likely that a given parameter point fails one of the tree-level checks for UFB directions. \\
In the lower row of Fig.~\ref{fig:RandomVacuumTreeLoopLambda}, which gives the results for the $(\lambda_3,\lambda_5)$ plane, the picture changes. Here we find that most of the points for which 
the tree-level checks also return `stable' if the point is deemed stable at one-loop reside around small absolute values of $\lambda_3$ and $\lambda_5$.
In this case, the misidentification goes down  below 60\% for Input~I and  even below 30\% for Input~II. On the other side, if either $|\lambda_3|$ or $|\lambda_5|$ are large, less than 20\% of the points are categorised correctly. The reason is that for large couplings, the loop corrections are even more important. In contrast to $\lambda_2$ or $\lambda_4$, the quartic coupling $\lambda_5$ enters the UFB checks as $|\lambda_5|$, i.e. UFB directions are as likely for large positive $\lambda_5$ as they are for large negative values. 
In the case of $\lambda_3$, points with large negative tree-level values for that coupling are likely to have a UFB direction at tree-level which becomes bounded from below after including the loop corrections, analogous to $\lambda_2$ and $\lambda_4$; recall also the example of Fig.~\ref{fig:ufb_l3}. At large positive $\lambda_3$, in turn, a point which is considered unstable at tree level because of a deeper non-ew minimum of the scalar potential can become stable at one-loop after the inclusion of the large loop corrections as we have seen at the example of Fig.~\ref{fig:vacuum_l3}.

Finally, we want to remark again what is particularly noticeable in Fig.~\ref{fig:vacuum_l3}: one clearly sees that, if a point passes the loop-level constraints, there is in general a higher chance that it also passes all tree-level tests if Input~II is used rather than Input~I. This is a result of the smaller amount of parameter tuning which is needed in average (i.e. for a randomly chosen parameter point) if $\alpha$ is calculated rather than used as an input.

\subsubsection{Maximal $s_H$}
\begin{figure}[tb]
\includegraphics[width=0.5\linewidth]{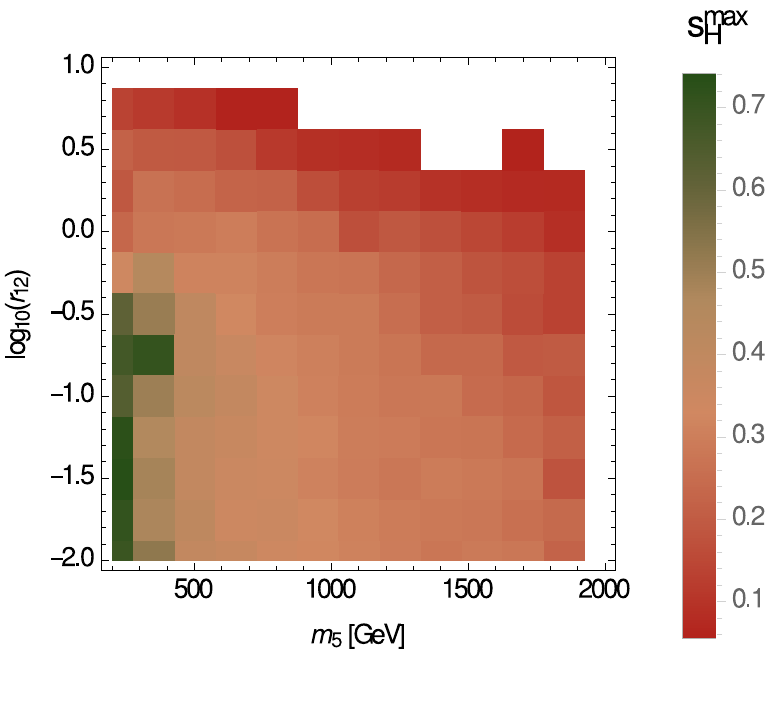} \hfill
\includegraphics[width=0.5\linewidth]{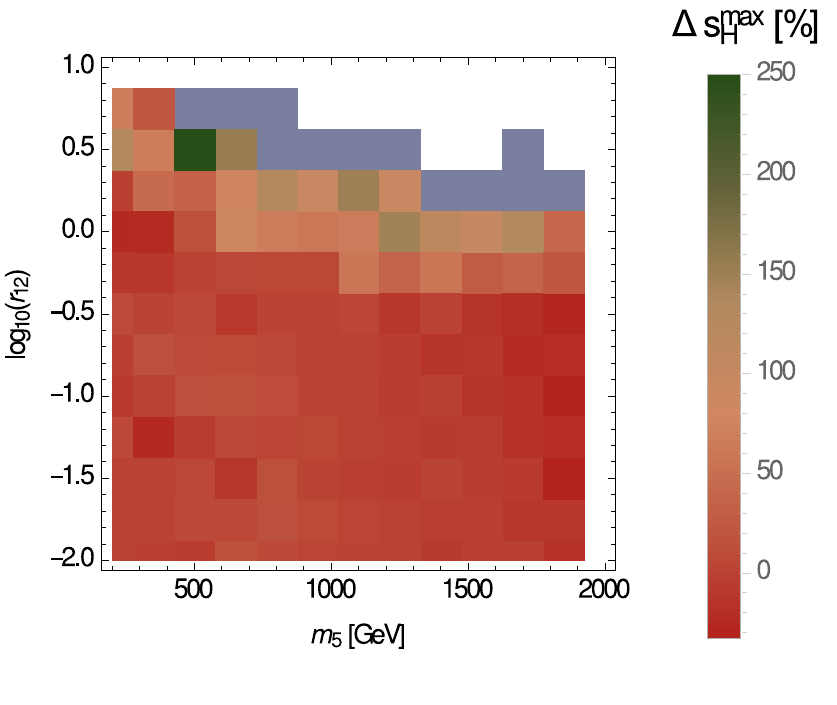} \\
\includegraphics[width=0.5\linewidth]{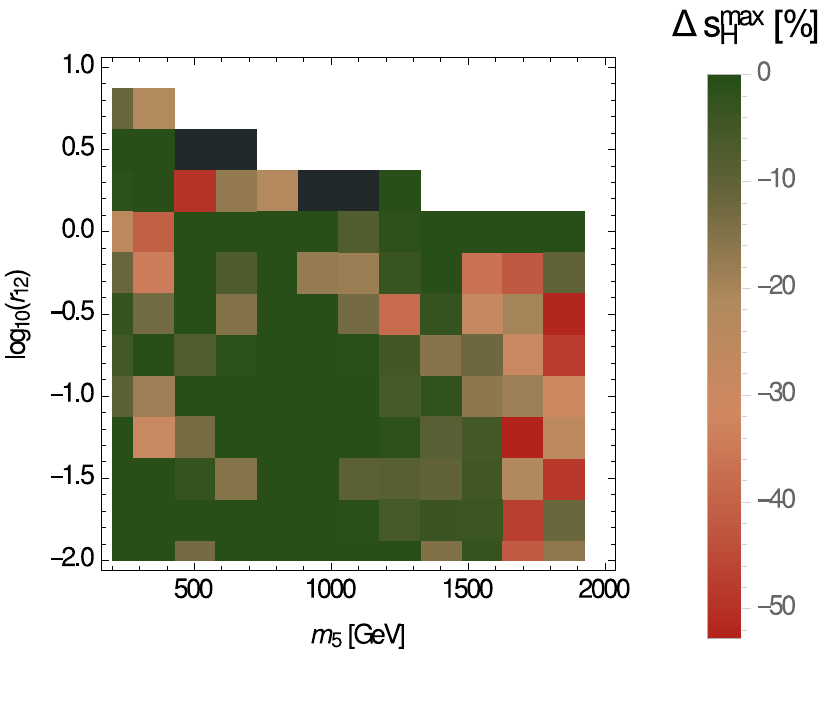} \hfill
\includegraphics[width=0.5\linewidth]{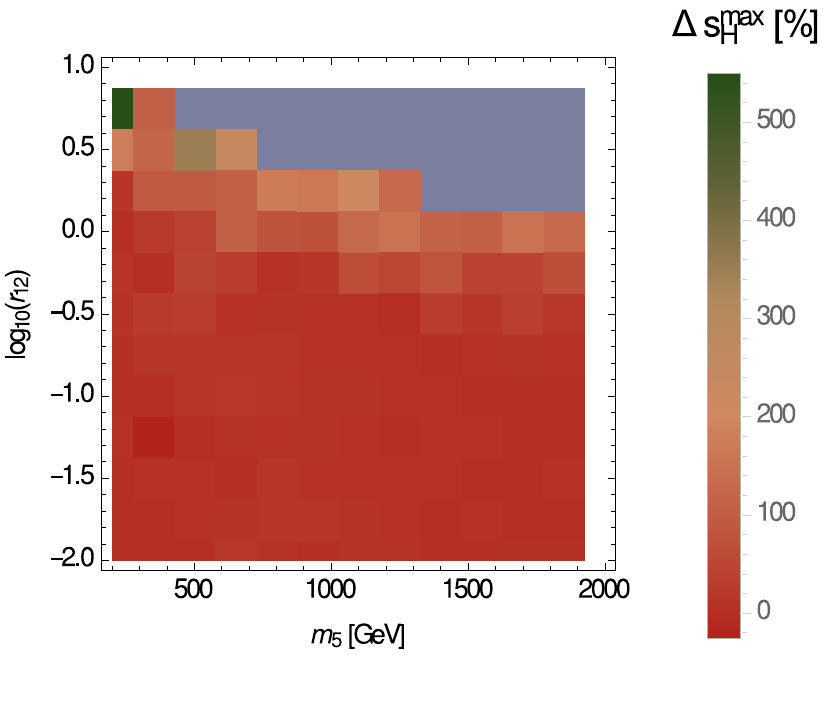}
\caption{Maximally allowed value for $s_H$ using different sets of constraints.
Upper left figure: maximal value of $s_H$ in the $(m_5,r_{12})$ plane for Input II after the {\tt HiggsBounds} cut and all loop constraints. For the perturbativity cuts, we have used the weak conditions. Upper right plane: the relative difference 
to the maximal value when using only tree-level constraints and not checking for perturbativity (depicted is $100 \frac{s_H^{\rm Loop}-s_H^{\rm Tree}}{s_H^{\rm Tree}}$). In the second row, the difference is shown if tree-level vacuum stability constraints are combined with the weak perturbativity cuts (left), or the one-loop vacuum stability is checked but the perturbativity constraints are not applied (right). 
The purple regions are not accessible if the tree-level cuts are applied. In the black bins, no point was found in our random scan 
after applying the one-loop perturbativity constraints on points with a stable tree-level vacuum.
}
\label{fig:RandomMaxSHcomparison}
\end{figure}
We want to close the discussion of our random scans by checking the maximal allowed value of $s_H$ which is possible in the different parameter regions when applying the various tree- and loop-level constraints defined above. Here, we concentrate on Input~II because of the larger number of valid points when using both tree- and loop-level checks. I.e., we consider a comparison between results of the lowest order and higher orders of perturbation theory as more robust for this input choice. In the first row of Fig.~\ref{fig:RandomMaxSHcomparison}, we show the maximal value of $s_H$ which we find after applying the weak perturbativity constraints and  checking the stability of the one-loop effective potential. We see that very large values are only possible for $r_{12}<0$ (i.e. opposite signs for $M_1$ and $M_2$) and small $m_5$ below 500~GeV. For increasing $r_{12}$ or $m_5$, the maximal possible value of $s_H$ quickly goes  down to 0.3 and less. We can now compare these values with $s_H^{\rm max}$ which we would find in the same plane 
if we apply the tree-level vacuum conditions and if we don't check for perturbativity. The results are depicted as well in the first row of Fig.~\ref{fig:RandomMaxSHcomparison}, on the right-hand side. First of all, a band with large $r_{12}$, shown in purple, wouldn't have been accessible at all under these conditions. Close to this region, i.e. still for $r_{12}\gtrsim 1$, the relative changes in $s_H^{\rm max}$ can be 150\% and more. Of course, in total numbers this means for this parameter region rather moderate shifts of 0.01--0.10 in the maximal possible value of $s_H$. In the other parameter regions, the relative differences in $s_H^{\rm max}$ are not too pronounced; they lie between -25\% and +100\%. Nevertheless, these are effects which are not negligible even if one has averaged over thousands of points, i.e. the general behaviour of the model is affected. Of course, for a single point which should be for instance used for collider studies, it is even more important to apply robust checks to test whether the point is allowed or not.\\
In the second row of Fig.~\ref{fig:RandomMaxSHcomparison}, we show the difference in $s_H^{\rm max}$ when using the tree-level checks  and either the weak perturbativity constraints (lower left plane) or the loop-corrected vacuum stability checks (lower right plane). If one would use the tree-level vacuum stability conditions and apply the weak perturbativity conditions in addition, the sample of points considered as `valid' will shrink. Therefore, also  $s_H^{\rm max}$  becomes smaller. The differences can be up to -50\% for large $m_5$. On the other side, if only the loop-improved vacuum stability checks were used without perturbativity checks (right-hand plot), the sample of `valid' points would increase significantly, resulting in larger positive shifts of $s_H^{\rm max}$ up to 500\% for large $r_{12}$. Moreover, a rather large region in the parameter space would seem accessible in this case: the region at large $r_{12}$ and large $m_5$ which is purple in the lower right plane but white in the upper right figure appears to be allowed if the perturbativity constraints are disregarded.

This shows that the two proposals to improve the theoretical checks for parameter points in the GM model  are complementary, and we stress that a check of the stability at the loop level is only meaningful if the perturbative series is trustworthy.

\section{Summary and conclusions}
\label{sec:conclusions}
We have investigated in this work the perturbative behaviour of the Georgi-Machacek model, focussing on the scalar sector.  An on-shell renormalisation of the scalar masses and mixing angles is formally possible once the custodial symmetry is given up at the loop level. However, it has been shown that in large  regions of parameter space, very large loop corrections can appear, pointing towards the breakdown of perturbation theory. Therefore, these regions are most likely strongly coupled although naive limits for the quartic couplings would not indicate this. 
For a definite answer to the question whether the perturbative expansion works or not, it would be necessary to check for the decrease in the scale dependence of different physical processes -- which is so far hardly possible in most BSM models since one would need to calculate the two-loop results for several decay or scattering processes. \\

We have therefore proposed a set of easily-accessible checks which should help to get an impression of how well the perturbative expansion works. These checks either include conditions on the size of the counter-terms to the quartic couplings which should be fulfilled in an on-shell scheme, or use the hierarchy between the one- and two-loop corrections to the scalar masses in the \MS scheme. 
The methodology developed in this work therefore solely relies on 
calculations which are, as of late, possible in an automated way for many BSM models via the  {\tt Mathematica} package \SARAH.
The perturbativity conditions proposed here are thus a lot more sophisticated than the simple tree-level checks and at the same time already easily calculable with the help of modern particle physics computer tools.

We have shown at the example of the Georgi-Machacek model that those comparably easily-accessible conditions are, in many parameter regions, sufficient to be certain that perturbation theory is not applicable: we have found examples in which the counter-terms to quartic couplings are larger than 100 or in which the two-loop mass corrections are several orders of magnitude larger than the one-loop corrections. There are, however,  also regions where the results are not that clear. For instance if the two-loop corrections are comparable to the one-loop corrections or if counter-terms are of order one. Here, it is still a matter of taste or conservatism if one assumes that Born-level or one-loop results for masses or processes are still reliable. 
Although in these cases, we cannot provide a definite yes/no answer to the question of perturbativity, we think that this work has pointed out the potential problems with perturbativity arising in particular in non-supersymmetric models in the presence of sizeable quartic couplings. We further hope that the presented ans\"atze are a step forward towards more reliable checks of this issue in the future. \\

In the specific case of the Georgi-Machacek model, we could identify several parameter regions where the loop corrections tend to be large. In particular, we found that a check for perturbativity is inevitable if (i) the new scalar masses are large, i.e. 1~TeV or more, (ii) if there is a mismatch between the actual values of $s_H$ and the scalar mixing angle $\alpha$ either due to the choice of the input parameters or due to accidental cancellations, or (iii) if quartic couplings become large (although still well below the limits from the tree-level unitarity conditions). \\

The second main outcome of this work is that we have revised the checks for vacuum stability, which have so far only been done at tree-level, by including loop corrections. We have shown that those tree-level checks are usually not reliable in the Georgi-Machacek model.
In particular, in many cases, a vacuum configuration which appears to feature an unstable electroweak vacuum at the tree level turns out to be stable once the one-loop corrections are included. This is the case as (i) a change in hierarchy between different minima of the scalar potential can easily occur at the one-loop level, and (ii) because field directions which appear to be unbounded from below at the tree level turn out to be bounded after the inclusion of the loop corrections. 
The opposite case that the vacuum is destabilised at the loop level is much less likely.
Depending on the parameter regions, we found misidentification rates of the tree-level checks compared to the more reliable loop-level checks of up to 100\%. In the entire parameter region which we have checked, the misidentification rate was always above 15\%. Therefore, we conclude that, for a reliable prediction if a point has a stable or unstable electroweak vacuum, tree-level conditions are not sufficient. On the other side, one needs to be sure that perturbation theory is working if loop corrected checks for the vacuum stability shall be applied. This shows the strong connection between the two topics  discussed in this work.

\section*{Acknowledgements}
FS thanks Johannes Braathen and Mark D. Goodsell for the joint efforts to make two-loop corrections 
in non-supersymmetric available via \SARAH/\SPheno, which has been crucial for this work, 
as well as for discussions about perturbativity 
constraints in non-supersymmetric models. 
MEK is supported by the DFG Research Unit 2239 ``New Physics at the LHC''. 
FS is supported by the ERC Recognition Award ERC-RA-0008 of the Helmholtz Association.

\allowdisplaybreaks

\begin{appendix}

\section{Vertices}
\label{app:vertices}
The Feynman rules for the GM model are partially given in Refs.~\cite{Gunion:1989we,Godfrey:2010qb,Hartling:2014zca}. These 
works focused on the couplings which are important for collider phenomenology like cubic scalar couplings, or the couplings of the 
scalars to vector bosons or fermions. However, other couplings which are important for loop corrections like the quartic scalar 
couplings or the interactions with ghosts have not yet been given explicitly in literature. Therefore, we provide in the following {\it all} couplings
of the scalars in the GM model in the limit of conserved custodial symmetry and also of no CP violation, i.e. all parameters are taken to be 
real. The expressions for the general case are available via \SARAH. \\

We define the following rotation matrices
\begin{align}
Z^H =&  \left(\begin{array}{ccc}
         1 & 0 & 0\\ 
         0 & -\sqrt{\frac23} & \frac{1}{\sqrt{3}} \\
         0 & \frac{1}{\sqrt{3}} & \sqrt{\frac23} \\         
            \end{array} \right)
\times
\left(\begin{array}{ccc} 
      c_\alpha & s_\alpha  & 0\\
     -s_\alpha & c_\alpha & 0 \\
      0 & 0& 1 
       \end{array}\right)  \\
Z^A =& \left(\begin{array}{cc} 
      c_H & s_H \\
     -s_H & c_H  
       \end{array}\right)
\\
Z^P = & \left(\begin{array}{ccc}
         1 & 0 & 0\\ 
         0 & \frac{1}{\sqrt{2}} & \frac{1}{\sqrt{2}} \\
         0 & -\frac{1}{\sqrt{2}} & \frac{1}{\sqrt{2}} \\         
            \end{array} \right)
\times
\left(\begin{array}{ccc} 
      c_H & s_H  & 0\\
     -s_H & c_H & 0 \\
      0 & 0& 1 
       \end{array}\right)                 
\end{align}
which diagonalise the mass matrices given in eqs.~(\ref{eq:mass_matrix_h})--(\ref{eq:mass_matrix_hp}) in the limit of the conserved custodial symmetry.

In the following, the vertices are categorised according to inclusion of scalar (S), fermionic (F) or vector (V) particles. 
The conventions are as follows:
Chiral vertices are parametrised as
 \begin{equation*}
  \Gamma^L_{F_i F_j S_k} P_L + \Gamma^R_{F_i F_j S_k} P_R \,,
 \end{equation*}
where  $P_{L,R}$ are the usual polarisation projectors.
The momentum flow in scalar-scalar-vector (SSV) vertices is
 \begin{equation*}
 \Gamma_{S_i S_j V^\mu_c}  (p^\mu_{S_j} - p^\mu_{S_i})\,, 
 \end{equation*}
where $p_\mu$ are the momenta of the external fields.

\subsection*{SSS}
\begin{align} 
\Gamma_{A^0_{{i}}A^0_{{j}}h_{{k}}}  = & \, 
\frac{i}{2} \Big(Z^A_{{i 2}} \Big(-2 Z^A_{{j 2}} \Big(\Big(4 {\lambda_2}  - {\lambda_5} \Big)v Z_{{k 1}}^{H}  + 4 \sqrt{2} \Big(2 {\lambda_4}  + {\lambda_3}\Big){v_{\chi}} Z_{{k 3}}^{H}  + \Big(-6 M_2  + 8 {\lambda_4} {v_{\chi}} \Big)Z_{{k 2}}^{H} \Big) \nonumber \\ &+ \sqrt{2} Z^A_{{j 1}} \Big(\Big(2 {\lambda_5} {v_{\chi}}  + {M_1}\Big)Z_{{k 1}}^{H}  + 2 {\lambda_5} v_\phi Z_{{k 2}}^{H} \Big)\Big)+Z^A_{{i 1}} \Big(\sqrt{2} Z^A_{{j 2}} \Big(\Big(2 {\lambda_5} {v_{\chi}}  + {M_1}\Big)Z_{{k 1}}^{H}  + 2 {\lambda_5} v_\phi Z_{{k 2}}^{H} \Big) \nonumber \\ &- Z^A_{{j 1}} \Big(16 {\lambda}_{1} v_\phi Z_{{k 1}}^{H}  + \Big(4 {\lambda_5} {v_{\chi}}  + 8 {\lambda_2} {v_{\chi}}  - {M_1} \Big)Z_{{k 2}}^{H}  + \sqrt{2} \Big(8 {\lambda_2} {v_{\chi}}  + {M_1}\Big)Z_{{k 3}}^{H} \Big)\Big)\Big) 
\\ 
\Gamma_{A^0_{{i}}H^+_{{j}}H^+_{{k}}}  = & \, 
\frac{1}{2} \Big(Z^A_{{i 1}} \Big(\Big(-2 {\lambda_5} {v_{\chi}}  + {M_1}\Big)Z_{{j 1}}^{+} \Big(- Z_{{k 3}}^{+}  + Z_{{k 2}}^{+}\Big) - Z_{{j 2}}^{+} \Big(\Big(-2 {\lambda_5} {v_{\chi}}  + {M_1}\Big)Z_{{k 1}}^{+}  + 2 {\lambda_5} v_\phi Z_{{k 3}}^{+} \Big) \nonumber \\ &+ Z_{{j 3}}^{+} \Big(\Big(-2 {\lambda_5} {v_{\chi}}  + {M_1}\Big)Z_{{k 1}}^{+}  + 2 {\lambda_5} v_\phi Z_{{k 2}}^{+} \Big)\Big)+\sqrt{2} Z^A_{{i 2}} \Big({\lambda_5} v_\phi Z_{{j 1}}^{+} \Big(- Z_{{k 3}}^{+}  + Z_{{k 2}}^{+}\Big) \nonumber \\ 
 &- Z_{{j 2}}^{+} \Big(2 \Big(2 {\lambda_3} {v_{\chi}}  + 3 M_2 \Big)Z_{{k 3}}^{+}  + {\lambda_5} v_\phi Z_{{k 1}}^{+} \Big) + Z_{{j 3}}^{+} \Big(4 {\lambda_3} {v_{\chi}} Z_{{k 2}}^{+}  + 6 M_2 Z_{{k 2}}^{+}  + {\lambda_5} v_\phi Z_{{k 1}}^{+} \Big)\Big)\Big) 
\\ 
\Gamma_{h_{{i}}h_{{j}}h_{{k}}}  = & \, 
\frac{i}{2} \Big(Z_{{i 1}}^{H} \Big(Z_{{j 2}}^{H} \Big(2 \sqrt{2} {\lambda_5} v_\phi Z_{{k 3}}^{H}  + \Big(4 {\lambda_5} {v_{\chi}}  -8 {\lambda_2} {v_{\chi}}  + {M_1}\Big)Z_{{k 1}}^{H}  -8 {\lambda_2} v_\phi Z_{{k 2}}^{H} \Big)\nonumber \\ 
 &+Z_{{j 1}}^{H} \Big(-48 {\lambda}_{1} v_\phi Z_{{k 1}}^{H}  + \Big(4 {\lambda_5} {v_{\chi}}  -8 {\lambda_2} {v_{\chi}}  + {M_1}\Big)\Big(\sqrt{2} Z_{{k 3}}^{H}  + Z_{{k 2}}^{H}\Big)\Big)\nonumber \\ 
 &+Z_{{j 3}}^{H} \Big(2 v_\phi \Big(\Big(-4 {\lambda_2}  + {\lambda_5}\Big)Z_{{k 3}}^{H}  + \sqrt{2} {\lambda_5} Z_{{k 2}}^{H} \Big) + \sqrt{2} \Big(4 \Big(-2 {\lambda_2}  + {\lambda_5}\Big){v_{\chi}}  + {M_1}\Big)Z_{{k 1}}^{H} \Big)\Big)\nonumber \\ 
 &+Z_{{i 3}}^{H} \Big(Z_{{j 1}}^{H} \Big(2 v_\phi \Big(\Big(-4 {\lambda_2}  + {\lambda_5}\Big)Z_{{k 3}}^{H}  + \sqrt{2} {\lambda_5} Z_{{k 2}}^{H} \Big) + \sqrt{2} \Big(4 \Big(-2 {\lambda_2}  + {\lambda_5}\Big){v_{\chi}}  + {M_1}\Big)Z_{{k 1}}^{H} \Big)\nonumber \\ 
 &+2 \Big(Z_{{j 3}}^{H} \Big(-12 \sqrt{2} \Big(2 {\lambda_4}  + {\lambda_3}\Big){v_{\chi}} Z_{{k 3}}^{H}  + \Big(-4 {\lambda_2}  + {\lambda_5}\Big)v Z_{{k 1}}^{H}  + \Big(6 M_2  -8 {\lambda_4} {v_{\chi}} \Big)Z_{{k 2}}^{H} \Big)\nonumber \\ 
 &+Z_{{j 2}}^{H} \Big(2 \Big(3 M_2  -4 {\lambda_4} {v_{\chi}} \Big)Z_{{k 3}}^{H}  -8 \sqrt{2} {\lambda_4} {v_{\chi}} Z_{{k 2}}^{H}  + \sqrt{2} {\lambda_5} v_\phi Z_{{k 1}}^{H} \Big)\Big)\Big)\nonumber \\ 
 &+Z_{{i 2}}^{H} \Big(Z_{{j 1}}^{H} \Big(2 \sqrt{2} {\lambda_5} v_\phi Z_{{k 3}}^{H}  + \Big(4 \Big(-2 {\lambda_2}  + {\lambda_5}\Big){v_{\chi}}  + {M_1}\Big)Z_{{k 1}}^{H}  -8 {\lambda_2} v_\phi Z_{{k 2}}^{H} \Big)\nonumber \\ 
 &-2 \Big(4 Z_{{j 2}}^{H} \Big(2 {v_{\chi}} \Big(3 \Big({\lambda_3} + {\lambda_4}\Big)Z_{{k 2}}^{H}  + \sqrt{2} {\lambda_4} Z_{{k 3}}^{H} \Big) + {\lambda_2} v_\phi Z_{{k 1}}^{H} \Big) \nonumber \\ 
 &+ Z_{{j 3}}^{H} \Big(2 \Big(-3 M_2  + 4 {\lambda_4} {v_{\chi}} \Big)Z_{{k 3}}^{H}  + 8 \sqrt{2} {\lambda_4} {v_{\chi}} Z_{{k 2}}^{H}  - \sqrt{2} {\lambda_5} v_\phi Z_{{k 1}}^{H} \Big)\Big)\Big)\Big) 
\\ 
\Gamma_{h_{{i}}H^+_{{j}}H^+_{{k}}}  = & \, 
\frac{i}{2} \Big(\sqrt{2} Z_{{i 3}}^{H} \Big(Z_{{j 3}}^{+} \Big(\Big(4 {\lambda_3} {v_{\chi}}  + 6 M_2 \Big)Z_{{k 2}}^{+}  -8 \Big(2 {\lambda_4}  + {\lambda_3}\Big){v_{\chi}} Z_{{k 3}}^{+}  + {\lambda_5} v_\phi Z_{{k 1}}^{+} \Big)\nonumber \\ 
 &+Z_{{j 2}}^{+} \Big(2 \Big(2 {\lambda_3} {v_{\chi}}  + 3 M_2 \Big)Z_{{k 3}}^{+}  -8 \Big(2 {\lambda_4}  + {\lambda_3}\Big){v_{\chi}} Z_{{k 2}}^{+}  + {\lambda_5} v_\phi Z_{{k 1}}^{+} \Big)\nonumber \\ 
 &+Z_{{j 1}}^{+} \Big(-2 \Big(4 {\lambda_2}  + {\lambda_5}\Big){v_{\chi}} Z_{{k 1}}^{+}  + {\lambda_5} v_\phi \Big(Z_{{k 2}}^{+} + Z_{{k 3}}^{+}\Big)\Big)\Big)\nonumber \\ 
 &+Z_{{i 1}}^{H} \Big(Z_{{j 2}}^{+} \Big(\Big(2 {\lambda_5} {v_{\chi}}  + {M_1}\Big)Z_{{k 1}}^{+}  + 2 {\lambda_5} v_\phi Z_{{k 3}}^{+}  -8 {\lambda_2} v_\phi Z_{{k 2}}^{+} \Big)\nonumber \\ 
 &+Z_{{j 1}}^{+} \Big(-16 {\lambda}_{1} v_\phi Z_{{k 1}}^{+}  + \Big(2 {\lambda_5} {v_{\chi}}  + {M_1}\Big)\Big(Z_{{k 2}}^{+} + Z_{{k 3}}^{+}\Big)\Big)\nonumber \\ 
 &+Z_{{j 3}}^{+} \Big(\Big(2 {\lambda_5} {v_{\chi}}  + {M_1}\Big)Z_{{k 1}}^{+}  + 2 v_\phi \Big(-4 {\lambda_2} Z_{{k 3}}^{+}  + {\lambda_5} Z_{{k 2}}^{+} \Big)\Big)\Big)\nonumber \\ 
 &- Z_{{i 2}}^{H} \Big(\Big(8 {\lambda_2} {v_{\chi}}  + {M_1}\Big)Z_{{j 1}}^{+} Z_{{k 1}}^{+}  + 8 {v_{\chi}} \Big(Z_{{j 2}}^{+} \Big(2 \Big({\lambda_3} + {\lambda_4}\Big)Z_{{k 2}}^{+}  - {\lambda_3} Z_{{k 3}}^{+} \Big) \nonumber \\ 
 &+ Z_{{j 3}}^{+} \Big(2 \Big({\lambda_3} + {\lambda_4}\Big)Z_{{k 3}}^{+}  - {\lambda_3} Z_{{k 2}}^{+} \Big)\Big)\Big)\Big) 
\\ 
\Gamma_{h_{{i}}{H}^{--}{H}^{--}}  = & \, 
-i \Big(\Big(4 {\lambda_2}  + {\lambda_5}\Big)v Z_{{i 1}}^{H}  + 4 \sqrt{2} \Big(2 {\lambda_4}  + 3 {\lambda_3} \Big){v_{\chi}} Z_{{i 3}}^{H}  + \Big(6 M_2  + 8 {\lambda_4} {v_{\chi}} \Big)Z_{{i 2}}^{H} \Big) 
\\ 
\Gamma_{H^+_{{i}}H^+_{{j}}{H}^{--}}  = & \, 
i \Big(Z_{{i 3}}^{+} \Big(\Big(-4 {\lambda_3} {v_{\chi}}  + 6 M_2 \Big)Z_{{j 2}}^{+}  + 8 {\lambda_3} {v_{\chi}} Z_{{j 3}}^{+}  + {\lambda_5} v_\phi Z_{{j 1}}^{+} \Big)+Z_{{i 2}}^{+} \Big(\Big(-4 {\lambda_3} {v_{\chi}}  + 6 M_2 \Big)Z_{{j 3}}^{+}  \nonumber \\ 
 &+ 8 {\lambda_3} {v_{\chi}} Z_{{j 2}}^{+}  + {\lambda_5} v_\phi Z_{{j 1}}^{+} \Big)+Z_{{i 1}}^{+} \Big(\Big(-2 {\lambda_5} {v_{\chi}}  + {M_1}\Big)Z_{{j 1}}^{+}  + {\lambda_5} v_\phi \Big(Z_{{j 2}}^{+} + Z_{{j 3}}^{+}\Big)\Big)\Big) 
\end{align} 

\subsection*{SSSS}
\begin{align} 
\Gamma_{A^0_{{i}}A^0_{{j}}A^0_{{k}}A^0_{{l}}}  = & \, 
-i \Big(Z^A_{{i 2}} \Big(\Big(4 {\lambda_2}  - {\lambda_5} \Big)Z^A_{{j 1}} \Big(Z^A_{{k 1}} Z^A_{{l 2}}  + Z^A_{{k 2}} Z^A_{{l 1}} \Big) + Z^A_{{j 2}} \Big(12 \Big(2 {\lambda_4}  + {\lambda_3}\Big)Z^A_{{k 2}} Z^A_{{l 2}}  \nonumber \\ 
 &+ \Big(4 {\lambda_2}  - {\lambda_5} \Big)Z^A_{{k 1}} Z^A_{{l 1}} \Big)\Big)+Z^A_{{i 1}} \Big(\Big(4 {\lambda_2}  - {\lambda_5} \Big)Z^A_{{j 2}} \Big(Z^A_{{k 1}} Z^A_{{l 2}}  + Z^A_{{k 2}} Z^A_{{l 1}} \Big) \nonumber \\ 
 &+ Z^A_{{j 1}} \Big(24 {\lambda}_{1} Z^A_{{k 1}} Z^A_{{l 1}}  + \Big(4 {\lambda_2}  - {\lambda_5} \Big)Z^A_{{k 2}} Z^A_{{l 2}} \Big)\Big)\Big) 
\\ 
\Gamma_{A^0_{{i}}A^0_{{j}}h_{{k}}h_{{l}}}  = & \, 
-\frac{i}{2} \Big(Z^A_{{i 2}} \Big(-2 \sqrt{2} {\lambda_5} Z^A_{{j 1}} \Big(Z_{{k 1}}^{H} Z_{{l 2}}^{H}  + Z_{{k 2}}^{H} Z_{{l 1}}^{H} \Big) + 2 Z^A_{{j 2}} \Big(4 \Big(2 {\lambda_4}  + {\lambda_3}\Big)Z_{{k 3}}^{H} Z_{{l 3}}^{H}  \nonumber \\ 
 &+ \Big(4 {\lambda_2}  - {\lambda_5} \Big)Z_{{k 1}}^{H} Z_{{l 1}}^{H}  + 8 {\lambda_4} Z_{{k 2}}^{H} Z_{{l 2}}^{H} \Big)\Big)+Z^A_{{i 1}} \Big(-2 \sqrt{2} {\lambda_5} Z^A_{{j 2}} \Big(Z_{{k 1}}^{H} Z_{{l 2}}^{H}  + Z_{{k 2}}^{H} Z_{{l 1}}^{H} \Big)\nonumber \\ 
 &+2 Z^A_{{j 1}} \Big(8 {\lambda}_{1} Z_{{k 1}}^{H} Z_{{l 1}}^{H}  + Z_{{k 2}}^{H} \Big(4 {\lambda_2} Z_{{l 2}}^{H}  + \sqrt{2} {\lambda_5} Z_{{l 3}}^{H} \Big) \nonumber \\ 
 & + Z_{{k 3}}^{H} \Big(\Big(4 {\lambda_2}  - {\lambda_5} \Big)Z_{{l 3}}^{H}  + \sqrt{2} {\lambda_5} Z_{{l 2}}^{H} \Big)\Big)\Big)\Big) 
\\ 
\Gamma_{A^0_{{i}}A^0_{{j}}H^+_{{k}}H^+_{{l}}}  = & \, 
\frac{i}{2} \Big(Z^A_{{i 1}} \Big(\sqrt{2} {\lambda_5} Z^A_{{j 2}} \Big(Z_{{k 1}}^{+} \Big(Z_{{l 2}}^{+} + Z_{{l 3}}^{+}\Big) + Z_{{k 2}}^{+} Z_{{l 1}}^{+}  + Z_{{k 3}}^{+} Z_{{l 1}}^{+} \Big)\nonumber \\ 
 &-2 Z^A_{{j 1}} \Big(8 {\lambda}_{1} Z_{{k 1}}^{+} Z_{{l 1}}^{+}  + Z_{{k 2}}^{+} \Big(4 {\lambda_2} Z_{{l 2}}^{+}  + {\lambda_5} Z_{{l 3}}^{+} \Big) + Z_{{k 3}}^{+} \Big(4 {\lambda_2} Z_{{l 3}}^{+}  + {\lambda_5} Z_{{l 2}}^{+} \Big)\Big)\Big)\nonumber \\ 
 &+Z^A_{{i 2}} \Big(\sqrt{2} {\lambda_5} Z^A_{{j 1}} \Big(Z_{{k 1}}^{+} \Big(Z_{{l 2}}^{+} + Z_{{l 3}}^{+}\Big) + Z_{{k 2}}^{+} Z_{{l 1}}^{+}  + Z_{{k 3}}^{+} Z_{{l 1}}^{+} \Big)\nonumber \\ 
 &-2 Z^A_{{j 2}} \Big(4 \Big(2 {\lambda_4}  + {\lambda_3}\Big)\Big(Z_{{k 2}}^{+} Z_{{l 2}}^{+}  + Z_{{k 3}}^{+} Z_{{l 3}}^{+} \Big) + \Big(4 {\lambda_2}  + {\lambda_5}\Big)Z_{{k 1}}^{+} Z_{{l 1}}^{+} \Big)\Big)\Big) 
\\ 
\Gamma_{A^0_{{i}}A^0_{{j}}{H}^{--}{H}^{--}}  = & \, 
-i \Big(4 \Big(2 {\lambda_4}  + 3 {\lambda_3} \Big)Z^A_{{i 2}} Z^A_{{j 2}}  + \Big(4 {\lambda_2}  + {\lambda_5}\Big)Z^A_{{i 1}} Z^A_{{j 1}} \Big) 
\\ 
\Gamma_{A^0_{{i}}h_{{j}}H^+_{{k}}H^+_{{l}}}  = & \, 
\frac{1}{2} \Big(\sqrt{2} Z^A_{{i 2}} \Big(4 {\lambda_3} Z_{{j 2}}^{H} \Big(- Z_{{k 2}}^{+} Z_{{l 3}}^{+}  + Z_{{k 3}}^{+} Z_{{l 2}}^{+} \Big) + {\lambda_5} Z_{{j 1}}^{H} \Big(Z_{{k 1}}^{+} \Big(- Z_{{l 3}}^{+}  + Z_{{l 2}}^{+}\Big) \nonumber \\ 
 &- Z_{{k 2}}^{+} Z_{{l 1}}^{+}  + Z_{{k 3}}^{+} Z_{{l 1}}^{+} \Big)\Big)+{\lambda_5} Z^A_{{i 1}} \Big(2 Z_{{j 1}}^{H} \Big(- Z_{{k 2}}^{+} Z_{{l 3}}^{+}  + Z_{{k 3}}^{+} Z_{{l 2}}^{+} \Big) \nonumber \\ 
 &+ \sqrt{2} Z_{{j 3}}^{H} \Big(Z_{{k 1}}^{+} \Big(- Z_{{l 2}}^{+}  + Z_{{l 3}}^{+}\Big) + Z_{{k 2}}^{+} Z_{{l 1}}^{+}  - Z_{{k 3}}^{+} Z_{{l 1}}^{+} \Big)\Big)\Big) 
\\ 
\Gamma_{A^0_{{i}}H^+_{{j}}H^+_{{k}}{H}^{--}}  = & \, 
4 \sqrt{2} {\lambda_3} Z^A_{{i 2}} \Big(  Z_{{j 3}}^{+} Z_{{k 3}}^{+} - Z_{{j 2}}^{+} Z_{{k 2}}^{+}\Big) + {\lambda_5} Z^A_{{i 1}} \Big(Z_{{j 1}}^{+} \Big(  Z_{{k 3}}^{+} - Z_{{k 2}}^{+}\Big) - Z_{{j 2}}^{+} Z_{{k 1}}^{+}  + Z_{{j 3}}^{+} Z_{{k 1}}^{+} \Big) 
\\ 
\Gamma_{h_{{i}}h_{{j}}h_{{k}}h_{{l}}}  = & \, 
\frac{i}{2} \Big(Z_{{i 1}}^{H} \Big(Z_{{j 2}}^{H} \Big(2 \sqrt{2} {\lambda_5} Z_{{k 3}}^{H} Z_{{l 1}}^{H}  -8 {\lambda_2} Z_{{k 2}}^{H} Z_{{l 1}}^{H}  + Z_{{k 1}}^{H} \Big(2 \sqrt{2} {\lambda_5} Z_{{l 3}}^{H}  -8 {\lambda_2} Z_{{l 2}}^{H} \Big)\Big)\nonumber \\ 
 &+2 Z_{{j 3}}^{H} \Big(\Big(-4 {\lambda_2}  + {\lambda_5}\Big)Z_{{k 3}}^{H} Z_{{l 1}}^{H}  + \sqrt{2} {\lambda_5} Z_{{k 2}}^{H} Z_{{l 1}}^{H}  + Z_{{k 1}}^{H} \Big(\Big(-4 {\lambda_2}  + {\lambda_5}\Big)Z_{{l 3}}^{H}  \nonumber \\ 
 &+ \sqrt{2} {\lambda_5} Z_{{l 2}}^{H} \Big)\Big)+2 Z_{{j 1}}^{H} \Big(-24 {\lambda}_{1} Z_{{k 1}}^{H} Z_{{l 1}}^{H}  + Z_{{k 2}}^{H} \Big(-4 {\lambda_2} Z_{{l 2}}^{H}  + \sqrt{2} {\lambda_5} Z_{{l 3}}^{H} \Big) \nonumber \\ 
 &+ Z_{{k 3}}^{H} \Big(\Big(-4 {\lambda_2}  + {\lambda_5}\Big)Z_{{l 3}}^{H}  + \sqrt{2} {\lambda_5} Z_{{l 2}}^{H} \Big)\Big)\Big)\nonumber \\ 
 &+Z_{{i 2}}^{H} \Big(-8 Z_{{j 2}}^{H} \Big(2 {\lambda_4} Z_{{k 3}}^{H} Z_{{l 3}}^{H}  + 6 \Big({\lambda_3} + {\lambda_4}\Big)Z_{{k 2}}^{H} Z_{{l 2}}^{H}  + {\lambda_2} Z_{{k 1}}^{H} Z_{{l 1}}^{H} \Big)\nonumber \\ 
 &+Z_{{j 1}}^{H} \Big(2 \sqrt{2} {\lambda_5} Z_{{k 3}}^{H} Z_{{l 1}}^{H}  -8 {\lambda_2} Z_{{k 2}}^{H} Z_{{l 1}}^{H}  + Z_{{k 1}}^{H} \Big(2 \sqrt{2} {\lambda_5} Z_{{l 3}}^{H}  -8 {\lambda_2} Z_{{l 2}}^{H} \Big)\Big)\nonumber \\ 
 &+2 Z_{{j 3}}^{H} \Big(-8 {\lambda_4} \Big(Z_{{k 2}}^{H} Z_{{l 3}}^{H}  + Z_{{k 3}}^{H} Z_{{l 2}}^{H} \Big) + \sqrt{2} {\lambda_5} Z_{{k 1}}^{H} Z_{{l 1}}^{H} \Big)\Big)\nonumber \\ 
 &+Z_{{i 3}}^{H} \Big(2 Z_{{j 1}}^{H} \Big(\Big(-4 {\lambda_2}  + {\lambda_5}\Big)Z_{{k 3}}^{H} Z_{{l 1}}^{H}  + \sqrt{2} {\lambda_5} Z_{{k 2}}^{H} Z_{{l 1}}^{H}  + Z_{{k 1}}^{H} \Big(\Big(-4 {\lambda_2}  + {\lambda_5}\Big)Z_{{l 3}}^{H}  \nonumber \\ 
 &+ \sqrt{2} {\lambda_5} Z_{{l 2}}^{H} \Big)\Big)+2 Z_{{j 2}}^{H} \Big(-8 {\lambda_4} \Big(Z_{{k 2}}^{H} Z_{{l 3}}^{H}  + Z_{{k 3}}^{H} Z_{{l 2}}^{H} \Big) + \sqrt{2} {\lambda_5} Z_{{k 1}}^{H} Z_{{l 1}}^{H} \Big)\nonumber \\ 
 &-2 Z_{{j 3}}^{H} \Big(4 \Big(2 {\lambda_4} Z_{{k 2}}^{H} Z_{{l 2}}^{H}  + 3 \Big(2 {\lambda_4}  + {\lambda_3}\Big)Z_{{k 3}}^{H} Z_{{l 3}}^{H} \Big) + \Big(4 {\lambda_2}  - {\lambda_5} \Big)Z_{{k 1}}^{H} Z_{{l 1}}^{H} \Big)\Big)\Big) 
\\ 
\Gamma_{h_{{i}}h_{{j}}H^+_{{k}}H^+_{{l}}}  = & \, 
\frac{i}{2} \Big(Z_{{i 1}}^{H} \Big(\sqrt{2} {\lambda_5} Z_{{j 3}}^{H} \Big(Z_{{k 1}}^{+} \Big(Z_{{l 2}}^{+} + Z_{{l 3}}^{+}\Big) + Z_{{k 2}}^{+} Z_{{l 1}}^{+}  + Z_{{k 3}}^{+} Z_{{l 1}}^{+} \Big)\nonumber \\ 
 &-2 Z_{{j 1}}^{H} \Big(8 {\lambda}_{1} Z_{{k 1}}^{+} Z_{{l 1}}^{+}  + Z_{{k 2}}^{+} \Big(4 {\lambda_2} Z_{{l 2}}^{+}  - {\lambda_5} Z_{{l 3}}^{+} \Big) + Z_{{k 3}}^{+} \Big(4 {\lambda_2} Z_{{l 3}}^{+}  - {\lambda_5} Z_{{l 2}}^{+} \Big)\Big)\Big)\nonumber \\ 
 &-4 Z_{{i 2}}^{H} \Big(2 Z_{{j 2}}^{H} \Big(2 \Big({\lambda_3} + {\lambda_4}\Big)\Big(Z_{{k 2}}^{+} Z_{{l 2}}^{+}  + Z_{{k 3}}^{+} Z_{{l 3}}^{+} \Big) + {\lambda_2} Z_{{k 1}}^{+} Z_{{l 1}}^{+} \Big) \nonumber \\ 
 &- \sqrt{2} {\lambda_3} Z_{{j 3}}^{H} \Big(Z_{{k 2}}^{+} Z_{{l 3}}^{+}  + Z_{{k 3}}^{+} Z_{{l 2}}^{+} \Big)\Big)+Z_{{i 3}}^{H} \Big(-2 Z_{{j 3}}^{H} \Big(4 \Big(2 {\lambda_4}  + {\lambda_3}\Big)\Big(Z_{{k 2}}^{+} Z_{{l 2}}^{+}  + Z_{{k 3}}^{+} Z_{{l 3}}^{+} \Big) \nonumber \\ 
 &+ \Big(4 {\lambda_2}  + {\lambda_5}\Big)Z_{{k 1}}^{+} Z_{{l 1}}^{+} \Big)+\sqrt{2} \Big(4 {\lambda_3} Z_{{j 2}}^{H} \Big(Z_{{k 2}}^{+} Z_{{l 3}}^{+}  + Z_{{k 3}}^{+} Z_{{l 2}}^{+} \Big) \nonumber \\ 
 &+ {\lambda_5} Z_{{j 1}}^{H} \Big(Z_{{k 1}}^{+} \Big(Z_{{l 2}}^{+} + Z_{{l 3}}^{+}\Big) + Z_{{k 2}}^{+} Z_{{l 1}}^{+}  + Z_{{k 3}}^{+} Z_{{l 1}}^{+} \Big)\Big)\Big)\Big) 
\\ 
\Gamma_{h_{{i}}h_{{j}}{H}^{--}{H}^{--}}  = & \, 
-i \Big(4 \Big(2 {\lambda_4}  + 3 {\lambda_3} \Big)Z_{{i 3}}^{H} Z_{{j 3}}^{H}  + \Big(4 {\lambda_2}  + {\lambda_5}\Big)Z_{{i 1}}^{H} Z_{{j 1}}^{H}  + 8 {\lambda_4} Z_{{i 2}}^{H} Z_{{j 2}}^{H} \Big) 
\\ 
\Gamma_{h_{{i}}H^+_{{j}}H^+_{{k}}{H}^{--}}  = & \, 
-i \Big(-4 \sqrt{2} {\lambda_3} Z_{{i 3}}^{H} \Big(Z_{{j 2}}^{+} Z_{{k 2}}^{+}  + Z_{{j 3}}^{+} Z_{{k 3}}^{+} \Big)- {\lambda_5} Z_{{i 1}}^{H} \Big(Z_{{j 1}}^{+} \Big(Z_{{k 2}}^{+} + Z_{{k 3}}^{+}\Big) + Z_{{j 2}}^{+} Z_{{k 1}}^{+}  \nonumber \\ 
 &+ Z_{{j 3}}^{+} Z_{{k 1}}^{+} \Big)+2 Z_{{i 2}}^{H} \Big(2 {\lambda_3} \Big(Z_{{j 2}}^{+} Z_{{k 3}}^{+}  + Z_{{j 3}}^{+} Z_{{k 2}}^{+} \Big) + {\lambda_5} Z_{{j 1}}^{+} Z_{{k 1}}^{+} \Big)\Big) 
\\ 
\Gamma_{H^+_{{i}}H^+_{{j}}H^+_{{k}}H^+_{{l}}}  = & \, 
-2 i \Big(Z_{{i 3}}^{+} \Big(2 \Big(4 \Big({\lambda_3} + {\lambda_4}\Big)Z_{{j 3}}^{+} Z_{{k 3}}^{+} Z_{{l 3}}^{+}  + {\lambda_2} Z_{{j 1}}^{+} \Big(Z_{{k 1}}^{+} Z_{{l 3}}^{+}  + Z_{{k 3}}^{+} Z_{{l 1}}^{+} \Big)\Big) \nonumber \\ 
 &+ Z_{{j 2}}^{+} \Big(4 {\lambda_4} \Big(Z_{{k 2}}^{+} Z_{{l 3}}^{+}  + Z_{{k 3}}^{+} Z_{{l 2}}^{+} \Big) - {\lambda_5} Z_{{k 1}}^{+} Z_{{l 1}}^{+} \Big)\Big)+Z_{{i 2}}^{+} \Big(2 \Big(4 \Big({\lambda_3} + {\lambda_4}\Big)Z_{{j 2}}^{+} Z_{{k 2}}^{+} Z_{{l 2}}^{+}  \nonumber \\ 
 &+ {\lambda_2} Z_{{j 1}}^{+} \Big(Z_{{k 1}}^{+} Z_{{l 2}}^{+}  + Z_{{k 2}}^{+} Z_{{l 1}}^{+} \Big)\Big) + Z_{{j 3}}^{+} \Big(4 {\lambda_4} \Big(Z_{{k 2}}^{+} Z_{{l 3}}^{+}  + Z_{{k 3}}^{+} Z_{{l 2}}^{+} \Big) - {\lambda_5} Z_{{k 1}}^{+} Z_{{l 1}}^{+} \Big)\nonumber \\ 
 &+Z_{{i 1}}^{+} \Big(2 {\lambda_2} \Big(Z_{{j 2}}^{+} \Big(Z_{{k 1}}^{+} Z_{{l 2}}^{+}  + Z_{{k 2}}^{+} Z_{{l 1}}^{+} \Big) \Big)+ Z_{{j 3}}^{+} \Big(Z_{{k 1}}^{+} Z_{{l 3}}^{+}  + Z_{{k 3}}^{+} Z_{{l 1}}^{+} \Big)\Big)\nonumber \\ 
 &+Z_{{j 1}}^{+} \Big(8 {\lambda}_{1} Z_{{k 1}}^{+} Z_{{l 1}}^{+}  - {\lambda_5} \Big(Z_{{k 2}}^{+} Z_{{l 3}}^{+}  + Z_{{k 3}}^{+} Z_{{l 2}}^{+} \Big)\Big)\Big)\Big) 
\\ 
\Gamma_{H^+_{{i}}{H}^{--}H^+_{{k}}{H}^{--}}  = & \, 
-i \Big(4 \Big(2 {\lambda_4}  + {\lambda_3}\Big)\Big(Z_{{i 2}}^{+} Z_{{k 2}}^{+}  + Z_{{i 3}}^{+} Z_{{k 3}}^{+} \Big) + \Big(4 {\lambda_2}  - {\lambda_5} \Big)Z_{{i 1}}^{+} Z_{{k 1}}^{+} \Big) 
\\ 
\Gamma_{{H}^{--}{H}^{--}{H}^{--}{H}^{--}}  = & \, 
-8 i \Big(2 {\lambda_4}  + {\lambda_3}\Big) 
\end{align} 

\subsection*{SSV}
\begin{align} 
\Gamma_{A^0_{{i}}h_{{j}}Z_{{\mu}}}  = & \, 
-\frac{1}{2} \Big(g_1 \sin\Theta_W   + g_2 \cos\Theta_W  \Big)\Big(2 Z^A_{{i 2}} Z_{{j 3}}^{H}  + Z^A_{{i 1}} Z_{{j 1}}^{H} \Big) 
\\ 
\Gamma_{A^0_{{i}}H^+_{{j}}W^+_{{\mu}}}  = & \, 
\frac{1}{2} g_2 \Big(\sqrt{2} Z^A_{{i 2}} Z_{{j 3}}^{+}  + Z^A_{{i 1}} Z_{{j 1}}^{+} \Big) 
\\ 
\Gamma_{A^0_{{i}}H^+_{{j}}W^+_{{\mu}}}  = & \, 
\frac{1}{2} g_2 \Big(\sqrt{2} Z^A_{{i 2}} Z_{{j 3}}^{+}  + Z^A_{{i 1}} Z_{{j 1}}^{+} \Big) 
\\ 
\Gamma_{h_{{i}}H^+_{{j}}W^+_{{\mu}}}  = & \, 
\frac{i}{2} g_2 \Big(2 Z_{{i 2}}^{H} Z_{{j 2}}^{+}  + \sqrt{2} Z_{{i 3}}^{H} Z_{{j 3}}^{+}  + Z_{{i 1}}^{H} Z_{{j 1}}^{+} \Big) 
\\ 
\Gamma_{h_{{i}}H^+_{{j}}W^+_{{\mu}}}  = & \, 
-\frac{i}{2} g_2 \Big(2 Z_{{i 2}}^{H} Z_{{j 2}}^{+}  + \sqrt{2} Z_{{i 3}}^{H} Z_{{j 3}}^{+}  + Z_{{i 1}}^{H} Z_{{j 1}}^{+} \Big) 
\\ 
\Gamma_{H^+_{{i}}H^+_{{j}}\gamma_{{\mu}}}  = & \, 
-\frac{i}{2} \Big(2 \Big(g_1 \cos\Theta_W  Z_{{i 3}}^{+} Z_{{j 3}}^{+}  + g_2 \sin\Theta_W  Z_{{i 2}}^{+} Z_{{j 2}}^{+} \Big) + \Big(g_1 \cos\Theta_W   + g_2 \sin\Theta_W  \Big)Z_{{i 1}}^{+} Z_{{j 1}}^{+} \Big) 
\\ 
\Gamma_{H^+_{{i}}{H}^{--}W^+_{{\mu}}}  = & \, 
-i g_2 Z_{{i 3}}^{+}  
\\ 
\Gamma_{H^+_{{i}}H^+_{{j}}Z_{{\mu}}}  = & \, 
-\frac{i}{2} \Big(-2 g_1 \sin\Theta_W  Z_{{i 3}}^{+} Z_{{j 3}}^{+}  + 2 g_2 \cos\Theta_W  Z_{{i 2}}^{+} Z_{{j 2}}^{+}  + \Big(- g_1 \sin\Theta_W   + g_2 \cos\Theta_W  \Big)Z_{{i 1}}^{+} Z_{{j 1}}^{+} \Big) 
\\ 
\Gamma_{{H}^{--}{H}^{--}\gamma_{{\mu}}}  = & \, 
i \Big(g_1 \cos\Theta_W   + g_2 \sin\Theta_W  \Big) 
\\ 
\Gamma_{{H}^{--}{H}^{--}Z_{{\mu}}}  = & \, 
i \Big(- g_1 \sin\Theta_W   + g_2 \cos\Theta_W  \Big) 
\\ 
\Gamma_{H^+_{{i}}{H}^{--}W^+_{{\mu}}}  = & \, 
i g_2 Z_{{i 3}}^{+}  
\end{align} 

\subsection*{SSVV}
\begin{align} 
\Gamma_{A^0_{{i}}A^0_{{j}}W^+_{{\mu}}W^+_{{\nu}}}  = & \, 
\frac{i}{2} g_{2}^{2} \Big(2 Z^A_{{i 2}} Z^A_{{j 2}}  + Z^A_{{i 1}} Z^A_{{j 1}} \Big) 
\\ 
\Gamma_{A^0_{{i}}A^0_{{j}}Z_{{\mu}}Z_{{\nu}}}  = & \, 
\frac{i}{2} \Big(g_1 \sin\Theta_W   + g_2 \cos\Theta_W  \Big)^{2} \Big(4 Z^A_{{i 2}} Z^A_{{j 2}}  + Z^A_{{i 1}} Z^A_{{j 1}} \Big) 
\\ 
\Gamma_{A^0_{{i}}H^+_{{j}}W^+_{{\mu}}\gamma_{{\nu}}}  = & \, 
\frac{1}{2} g_2 \Big(g_1 \cos\Theta_W  Z^A_{{i 1}} Z_{{j 1}}^{+}  + \sqrt{2} \Big(2 g_1 \cos\Theta_W   - g_2 \sin\Theta_W  \Big)Z^A_{{i 2}} Z_{{j 3}}^{+} \Big) 
\\ 
\Gamma_{A^0_{{i}}H^+_{{j}}W^+_{{\mu}}Z_{{\nu}}}  = & \, 
-\frac{1}{2} g_2 \Big(g_1 \sin\Theta_W  Z^A_{{i 1}} Z_{{j 1}}^{+}  + \sqrt{2} \Big(2 g_1 \sin\Theta_W   + g_2 \cos\Theta_W  \Big)Z^A_{{i 2}} Z_{{j 3}}^{+} \Big) 
\\ 
\Gamma_{A^0_{{i}}H^+_{{j}}\gamma_{{\mu}}W^+_{{\nu}}}  = & \, 
\frac{1}{2} g_2 \Big(- g_1 \cos\Theta_W  Z^A_{{i 1}} Z_{{j 1}}^{+}  + \sqrt{2} \Big(-2 g_1 \cos\Theta_W   + g_2 \sin\Theta_W  \Big)Z^A_{{i 2}} Z_{{j 3}}^{+} \Big) 
\\ 
\Gamma_{A^0_{{i}}{H}^{--}W^+_{{\mu}}W^+_{{\nu}}}  = & \, 
- \sqrt{2} g_{2}^{2} Z^A_{{i 2}}  
\\ 
\Gamma_{A^0_{{i}}H^+_{{j}}W^+_{{\mu}}Z_{{\nu}}}  = & \, 
\frac{1}{2} g_2 \Big(g_1 \sin\Theta_W  Z^A_{{i 1}} Z_{{j 1}}^{+}  + \sqrt{2} \Big(2 g_1 \sin\Theta_W   + g_2 \cos\Theta_W  \Big)Z^A_{{i 2}} Z_{{j 3}}^{+} \Big) 
\\ 
\Gamma_{A^0_{{i}}{H}^{--}W^+_{{\mu}}W^+_{{\nu}}}  = & \, 
\sqrt{2} g_{2}^{2} Z^A_{{i 2}}  
\\ 
\Gamma_{h_{{i}}h_{{j}}W^+_{{\mu}}W^+_{{\nu}}}  = & \, 
\frac{i}{2} g_{2}^{2} \Big(2 Z_{{i 3}}^{H} Z_{{j 3}}^{H}  + 4 Z_{{i 2}}^{H} Z_{{j 2}}^{H}  + Z_{{i 1}}^{H} Z_{{j 1}}^{H} \Big) 
\\ 
\Gamma_{h_{{i}}h_{{j}}Z_{{\mu}}Z_{{\nu}}}  = & \, 
\frac{i}{2} \Big(g_1 \sin\Theta_W   + g_2 \cos\Theta_W  \Big)^{2} \Big(4 Z_{{i 3}}^{H} Z_{{j 3}}^{H}  + Z_{{i 1}}^{H} Z_{{j 1}}^{H} \Big) 
\\ 
\Gamma_{h_{{i}}H^+_{{j}}W^+_{{\mu}}\gamma_{{\nu}}}  = & \, 
\frac{i}{2} g_2 \Big(2 g_2 \sin\Theta_W  Z_{{i 2}}^{H} Z_{{j 2}}^{+}  + g_1 \cos\Theta_W  Z_{{i 1}}^{H} Z_{{j 1}}^{+}  + \sqrt{2} \Big(2 g_1 \cos\Theta_W   - g_2 \sin\Theta_W  \Big)Z_{{i 3}}^{H} Z_{{j 3}}^{+} \Big) 
\\ 
\Gamma_{h_{{i}}H^+_{{j}}W^+_{{\mu}}Z_{{\nu}}}  = & \, 
-\frac{i}{2} g_2 \Big(-2 g_2 \cos\Theta_W  Z_{{i 2}}^{H} Z_{{j 2}}^{+}  + g_1 \sin\Theta_W  Z_{{i 1}}^{H} Z_{{j 1}}^{+}  \nonumber \\ 
 &+ \sqrt{2} \Big(2 g_1 \sin\Theta_W   + g_2 \cos\Theta_W  \Big)Z_{{i 3}}^{H} Z_{{j 3}}^{+} \Big) 
\\ 
\Gamma_{h_{{i}}H^+_{{j}}\gamma_{{\mu}}W^+_{{\nu}}}  = & \, 
\frac{i}{2} g_2 \Big(2 g_2 \sin\Theta_W  Z_{{i 2}}^{H} Z_{{j 2}}^{+}  + g_1 \cos\Theta_W  Z_{{i 1}}^{H} Z_{{j 1}}^{+}  + \sqrt{2} \Big(2 g_1 \cos\Theta_W   - g_2 \sin\Theta_W  \Big)Z_{{i 3}}^{H} Z_{{j 3}}^{+} \Big) 
\\ 
\Gamma_{h_{{i}}{H}^{--}W^+_{{\mu}}W^+_{{\nu}}}  = & \, 
i \sqrt{2} g_{2}^{2} Z_{{i 3}}^{H}  
\\ 
\Gamma_{h_{{i}}H^+_{{j}}W^+_{{\mu}}Z_{{\nu}}}  = & \, 
-\frac{i}{2} g_2 \Big(-2 g_2 \cos\Theta_W  Z_{{i 2}}^{H} Z_{{j 2}}^{+}  + g_1 \sin\Theta_W  Z_{{i 1}}^{H} Z_{{j 1}}^{+}  \nonumber \\ 
 &+ \sqrt{2} \Big(2 g_1 \sin\Theta_W   + g_2 \cos\Theta_W  \Big)Z_{{i 3}}^{H} Z_{{j 3}}^{+} \Big) 
\\ 
\Gamma_{h_{{i}}{H}^{--}W^+_{{\mu}}W^+_{{\nu}}}  = & \, 
i \sqrt{2} g_{2}^{2} Z_{{i 3}}^{H}  
\\ 
\Gamma_{H^+_{{i}}H^+_{{j}}W^+_{{\mu}}W^+_{{\nu}}}  = & \, 
-2 i g_{2}^{2} Z_{{i 2}}^{+} Z_{{j 2}}^{+}  
\\ 
\Gamma_{H^+_{{i}}H^+_{{j}}\gamma_{{\mu}}\gamma_{{\nu}}}  = & \, 
\frac{i}{2} \Big(\Big(g_1 \cos\Theta_W   + g_2 \sin\Theta_W  \Big)^{2} Z_{{i 1}}^{+} Z_{{j 1}}^{+} \nonumber \\ 
 &+4 \Big(g_{1}^{2} \cos\Theta_{W }^{2} Z_{{i 3}}^{+} Z_{{j 3}}^{+}  + g_{2}^{2} \sin\Theta_{W }^{2} Z_{{i 2}}^{+} Z_{{j 2}}^{+} \Big)\Big) 
\\ 
\Gamma_{H^+_{{i}}{H}^{--}\gamma_{{\mu}}W^+_{{\nu}}}  = & \, 
i g_2 \Big(2 g_1 \cos\Theta_W   + g_2 \sin\Theta_W  \Big)Z_{{i 3}}^{+}  
\\ 
\Gamma_{H^+_{{i}}H^+_{{j}}\gamma_{{\mu}}Z_{{\nu}}}  = & \, 
-\frac{i}{4} \Big(\Big(-2 g_1 g_2 \cos2 \Theta_W    + \Big(- g_{2}^{2}  + g_{1}^{2}\Big)\sin2 \Theta_W   \Big)Z_{{i 1}}^{+} Z_{{j 1}}^{+} \nonumber \\ 
 &+4 \sin2 \Theta_W   \Big(g_{1}^{2} Z_{{i 3}}^{+} Z_{{j 3}}^{+}  - g_{2}^{2} Z_{{i 2}}^{+} Z_{{j 2}}^{+} \Big)\Big) 
\\ 
\Gamma_{H^+_{{i}}{H}^{--}W^+_{{\mu}}Z_{{\nu}}}  = & \, 
i g_2 \Big(-2 g_1 \sin\Theta_W   + g_2 \cos\Theta_W  \Big)Z_{{i 3}}^{+}  
\\ 
\Gamma_{H^+_{{i}}H^+_{{j}}W^+_{{\mu}}W^+_{{\nu}}}  = & \, 
\frac{i}{2} g_{2}^{2} \Big(2 Z_{{i 2}}^{+} Z_{{j 2}}^{+}  + 4 Z_{{i 3}}^{+} Z_{{j 3}}^{+}  + Z_{{i 1}}^{+} Z_{{j 1}}^{+} \Big) 
\\ 
\Gamma_{H^+_{{i}}H^+_{{j}}Z_{{\mu}}Z_{{\nu}}}  = & \, 
\frac{i}{2} \Big(\Big(- g_1 \sin\Theta_W   + g_2 \cos\Theta_W  \Big)^{2} Z_{{i 1}}^{+} Z_{{j 1}}^{+} \nonumber \\ 
 &+4 \Big(g_{1}^{2} \sin\Theta_{W }^{2} Z_{{i 3}}^{+} Z_{{j 3}}^{+}  + g_{2}^{2} \cos\Theta_{W }^{2} Z_{{i 2}}^{+} Z_{{j 2}}^{+} \Big)\Big) 
\\ 
\Gamma_{{H}^{--}{H}^{--}\gamma_{{\mu}}\gamma_{{\nu}}}  = & \, 
2 i \Big(g_1 \cos\Theta_W   + g_2 \sin\Theta_W  \Big)^{2}  
\\ 
\Gamma_{{H}^{--}{H}^{--}\gamma_{{\mu}}Z_{{\nu}}}  = & \, 
-i \Big(-2 g_1 g_2 \cos2 \Theta_W    + \Big(- g_{2}^{2}  + g_{1}^{2}\Big)\sin2 \Theta_W   \Big) 
\\ 
\Gamma_{H^+_{{i}}{H}^{--}W^+_{{\mu}}\gamma_{{\nu}}}  = & \, 
i g_2 \Big(2 g_1 \cos\Theta_W   + g_2 \sin\Theta_W  \Big)Z_{{i 3}}^{+}  
\\ 
\Gamma_{H^+_{{i}}H^+_{{j}}W^+_{{\mu}}W^+_{{\nu}}}  = & \, 
-2 i g_{2}^{2} Z_{{i 2}}^{+} Z_{{j 2}}^{+}  
\\ 
\Gamma_{{H}^{--}{H}^{--}W^+_{{\mu}}W^+_{{\nu}}}  = & \, 
i g_{2}^{2}  
\\ 
\Gamma_{{H}^{--}{H}^{--}Z_{{\mu}}Z_{{\nu}}}  = & \, 
2 i \Big(- g_1 \sin\Theta_W   + g_2 \cos\Theta_W  \Big)^{2}  
\\ 
\Gamma_{H^+_{{i}}{H}^{--}W^+_{{\mu}}Z_{{\nu}}}  = & \, 
i g_2 \Big(-2 g_1 \sin\Theta_W   + g_2 \cos\Theta_W  \Big)Z_{{i 3}}^{+}  
\end{align} 

\subsection*{SVV}
\begin{align} 
\Gamma_{h_{{i}}W^+_{{\sigma}}W^+_{{\mu}}}  = & \, 
\frac{i}{2} g_{2}^{2} \Big(2 \sqrt{2} v_\chi Z_{{i 3}}^{H}  + 4 v_\chi Z_{{i 2}}^{H}  + v_\phi Z_{{i 1}}^{H} \Big) 
\\ 
\Gamma_{h_{{i}}Z_{{\sigma}}Z_{{\mu}}}  = & \, 
\frac{i}{2} \Big(g_1 \sin\Theta_W   + g_2 \cos\Theta_W  \Big)^{2} \Big(4 \sqrt{2} v_\chi Z_{{i 3}}^{H}  + v_\phi Z_{{i 1}}^{H} \Big) 
\\ 
\Gamma_{H^+_{{i}}W^+_{{\sigma}}\gamma_{{\mu}}}  = & \, 
\frac{i}{2} g_2 \Big(2 v_\chi \Big(\Big(2 g_1 \cos\Theta_W   - g_2 \sin\Theta_W  \Big)Z_{{i 3}}^{+}  + g_2 \sin\Theta_W  Z_{{i 2}}^{+} \Big) + g_1 v_\phi \cos\Theta_W  Z_{{i 1}}^{+} \Big) 
\\ 
\Gamma_{H^+_{{i}}W^+_{{\sigma}}Z_{{\mu}}}  = & \, 
-\frac{i}{2} g_2 \Big(2 v_\chi \Big(\Big(2 g_1 \sin\Theta_W   + g_2 \cos\Theta_W  \Big)Z_{{i 3}}^{+}  - g_2 \cos\Theta_W  Z_{{i 2}}^{+} \Big) + g_1 v_\phi \sin\Theta_W  Z_{{i 1}}^{+} \Big) 
\\ 
\Gamma_{H^+_{{i}}\gamma_{{\sigma}}W^+_{{\mu}}}  = & \, 
\frac{i}{2} g_2 \Big(2 v_\chi \Big(\Big(2 g_1 \cos\Theta_W   - g_2 \sin\Theta_W  \Big)Z_{{i 3}}^{+}  + g_2 \sin\Theta_W  Z_{{i 2}}^{+} \Big) + g_1 v_\phi \cos\Theta_W  Z_{{i 1}}^{+} \Big) 
\\ 
\Gamma_{{H}^{--}W^+_{{\sigma}}W^+_{{\mu}}}  = & \, 
2 i g_{2}^{2} v_\chi  
\\ 
\Gamma_{H^+_{{i}}W^+_{{\sigma}}Z_{{\mu}}}  = & \, 
-\frac{i}{2} g_2 \Big(2 v_\chi \Big(\Big(2 g_1 \sin\Theta_W   + g_2 \cos\Theta_W  \Big)Z_{{i 3}}^{+}  - g_2 \cos\Theta_W  Z_{{i 2}}^{+} \Big) + g_1 v_\phi \sin\Theta_W  Z_{{i 1}}^{+} \Big) 
\\ 
\Gamma_{{H}^{--}W^+_{{\sigma}}W^+_{{\mu}}}  = & \, 
2 i g_{2}^{2} v_\chi  
\end{align}

\subsection*{FFS}
\begin{align} 
& \Gamma^L_{\bar{d}_{{i \alpha}}d_{{j \beta}}A^0_{{k}}}  =   \,
- \frac{1}{\sqrt{2}} \delta_{\alpha \beta} \sum_{b=1}^{3}\sum_{a=1}^{3}U_{R,{i a}}^{d} Y_{d,{a b}}  U_{L,{j b}}^{d}  Z^A_{{k 1}} \hspace{.8cm}
 \Gamma^R_{\bar{d}_{{i \alpha}}d_{{j \beta}}A^0_{{k}}}  =   \,\frac{1}{\sqrt{2}} \delta_{\alpha \beta} \sum_{b=1}^{3}\sum_{a=1}^{3}U_{R,{j a}}^{d} Y_{d,{a b}}  U_{L,{i b}}^{d}  Z^A_{{k 1}}  
\\
&\Gamma^L_{\bar{e}_{{i}}e_{{j}}A^0_{{k}}}  =   \,
- \frac{1}{\sqrt{2}} \sum_{b=1}^{3}\sum_{a=1}^{3}U_{R,{i a}}^{e} Y_{e,{a b}}  U_{L,{j b}}^{e}  Z^A_{{k 1}} \hspace{1cm}
 \Gamma^R_{\bar{e}_{{i}}e_{{j}}A^0_{{k}}}  =   \,\frac{1}{\sqrt{2}} \sum_{b=1}^{3}\sum_{a=1}^{3}U_{R,{j a}}^{e} Y_{e,{a b}}  U_{L,{i b}}^{e}  Z^A_{{k 1}}  
\\ 
& \Gamma^L_{\bar{u}_{{i \alpha}}u_{{j \beta}}A^0_{{k}}}  =   \,
\frac{1}{\sqrt{2}} \delta_{\alpha \beta} \sum_{b=1}^{3}\sum_{a=1}^{3}U_{R,{i a}}^{u} Y_{u,{a b}}  U_{L,{j b}}^{u}  Z^A_{{k 1}} \hspace{.7cm}
 \Gamma^R_{\bar{u}_{{i \alpha}}u_{{j \beta}}A^0_{{k}}}  =   \,- \frac{1}{\sqrt{2}} \delta_{\alpha \beta} \sum_{b=1}^{3}\sum_{a=1}^{3}U_{R,{j a}}^{u} Y_{u,{a b}}  U_{L,{i b}}^{u}  Z^A_{{k 1}}  
\\
&\Gamma^L_{\bar{d}_{{i \alpha}}d_{{j \beta}}h_{{k}}}  =   \,
-i \frac{1}{\sqrt{2}} \delta_{\alpha \beta} \sum_{b=1}^{3}\sum_{a=1}^{3}U_{R,{i a}}^{d} Y_{d,{a b}}  U_{L,{j b}}^{d}  Z_{{k 1}}^{H} \hspace{.5cm}
 \Gamma^R_{\bar{d}_{{i \alpha}}d_{{j \beta}}h_{{k}}}  =   \,-i \frac{1}{\sqrt{2}} \delta_{\alpha \beta} \sum_{b=1}^{3}\sum_{a=1}^{3}U_{R,{j a}}^{d} Y_{d,{a b}}  U_{L,{i b}}^{d}  Z_{{k 1}}^{H}  
\\ 
& \Gamma^L_{\bar{u}_{{i \alpha}}d_{{j \beta}}H^+_{{k}}}  =   \,
i \delta_{\alpha \beta} \sum_{b=1}^{3}\sum_{a=1}^{3}U_{R,{i a}}^{u} Y_{u,{a b}}  U_{L,{j b}}^{d}  Z_{{k 1}}^{+} \hspace{1cm}
 \Gamma^R_{\bar{u}_{{i \alpha}}d_{{j \beta}}H^+_{{k}}}  =   \,-i \delta_{\alpha \beta} \sum_{b=1}^{3}\sum_{a=1}^{3}U_{R,{j a}}^{d} Y_{d,{a b}}  U_{L,{i b}}^{u}  Z_{{k 1}}^{+}  
\\
&\Gamma^L_{\bar{e}_{{i}}e_{{j}}h_{{k}}}  =   \,
-i \frac{1}{\sqrt{2}} \sum_{b=1}^{3}\sum_{a=1}^{3}U_{R,{i a}}^{e} Y_{e,{a b}}  U_{L,{j b}}^{e}  Z_{{k 1}}^{H} \hspace{1cm}
 \Gamma^R_{\bar{e}_{{i}}e_{{j}}h_{{k}}}  =   \,-i \frac{1}{\sqrt{2}} \sum_{b=1}^{3}\sum_{a=1}^{3}U_{R,{j a}}^{e} Y_{e,{a b}}  U_{L,{i b}}^{e}  Z_{{k 1}}^{H}  
\\ 
& \Gamma^L_{\bar{\nu}_{{i}}e_{{j}}H^+_{{k}}}  =  \,
0  \hspace{1cm}
 \Gamma^R_{\bar{\nu}_{{i}}e_{{j}}H^+_{{k}}}  =   \,-i \sum_{a=1}^{3}U_{R,{j a}}^{e} Y_{e,{a i}}  Z_{{k 1}}^{+}  
\\ 
& \Gamma^L_{\bar{u}_{{i \alpha}}u_{{j \beta}}h_{{k}}}  =   \,
-i \frac{1}{\sqrt{2}} \delta_{\alpha \beta} \sum_{b=1}^{3}\sum_{a=1}^{3}U_{R,{i a}}^{u} Y_{u,{a b}}  U_{L,{j b}}^{u}  Z_{{k 1}}^{H}  \hspace{.5cm}
 \Gamma^R_{\bar{u}_{{i \alpha}}u_{{j \beta}}h_{{k}}}  =   \,-i \frac{1}{\sqrt{2}} \delta_{\alpha \beta} \sum_{b=1}^{3}\sum_{a=1}^{3}U_{R,{j a}}^{u} Y_{u,{a b}}  U_{L,{i b}}^{u}  Z_{{k 1}}^{H}  
\\ 
&\Gamma^L_{\bar{d}_{{i \alpha}}u_{{j \beta}}H^+_{{k}}}  =   \,
-i \delta_{\alpha \beta} \sum_{b=1}^{3}\sum_{a=1}^{3}U_{R,{i a}}^{d} Y_{d,{a b}}  U_{L,{j b}}^{u}  Z_{{k 1}}^{+} \hspace{1cm}
 \Gamma^R_{\bar{d}_{{i \alpha}}u_{{j \beta}}H^+_{{k}}}  =   \,i \delta_{\alpha \beta} \sum_{b=1}^{3}\sum_{a=1}^{3}U_{R,{j a}}^{u} Y_{u,{a b}}  U_{L,{i b}}^{d}  Z_{{k 1}}^{+}  
\\ 
&\Gamma^L_{\bar{e}_{{i}}\nu_{{j}}H^+_{{k}}}  =   \,
-i \sum_{a=1}^{3}U_{R,{i a}}^{e} Y_{e,{a j}}  Z_{{k 1}}^{+} \,,
 \Gamma^R_{\bar{e}_{{i}}\nu_{{j}}H^+_{{k}}}  =   \,0 
\end{align}

\subsection*{GGS}
\begin{align} 
\Gamma_{\bar{\eta^+}\eta^+A^0_{{k}}}  = & \, 
\frac{1}{4} g_{2}^{2} \xi_{W^+} \Big(\sqrt{2} v_\chi Z^A_{{k 2}}  + v_\phi Z^A_{{k 1}} \Big) 
\\ 
\Gamma_{\bar{\eta^-}\eta^-A^0_{{k}}}  = & \, 
-\frac{1}{4} g_{2}^{2} \xi_{W^+} \Big(\sqrt{2} v_\chi Z^A_{{k 2}}  + v_\phi Z^A_{{k 1}} \Big) 
\\ 
\Gamma_{\bar{\eta^Z}\eta^{\gamma}h_{{k}}}  = & \, 
\frac{i}{8} \xi_{Z} \Big(2 g_1 g_2 \cos2 \Theta_W    + \Big(- g_{2}^{2}  + g_{1}^{2}\Big)\sin2 \Theta_W   \Big)\Big(4 \sqrt{2} v_\chi Z_{{k 3}}^{H}  + v_\phi Z_{{k 1}}^{H} \Big) 
\\ 
\Gamma_{\bar{\eta^+}\eta^{\gamma}H^+_{{k}}}  = & \, 
-\frac{i}{4} g_2 \xi_{W^+} \Big(2 v_\chi \Big(2 g_2 \sin\Theta_W  Z_{{k 2}}^{+}  + g_1 \cos\Theta_W  Z_{{k 3}}^{+} \Big) + v_\phi \Big(g_1 \cos\Theta_W   + g_2 \sin\Theta_W  \Big)Z_{{k 1}}^{+} \Big) 
\\ 
\Gamma_{\bar{\eta^-}\eta^{\gamma}H^+_{{k}}}  = & \, 
-\frac{i}{4} g_2 \xi_{W^+} \Big(2 v_\chi \Big(2 g_2 \sin\Theta_W  Z_{{k 2}}^{+}  + g_1 \cos\Theta_W  Z_{{k 3}}^{+} \Big) + v_\phi \Big(g_1 \cos\Theta_W   + g_2 \sin\Theta_W  \Big)Z_{{k 1}}^{+} \Big) 
\\ 
\Gamma_{\bar{\eta^+}\eta^+h_{{k}}}  = & \, 
-\frac{i}{4} g_{2}^{2} \xi_{W^+} \Big(4 v_\chi Z_{{k 2}}^{H}  + \sqrt{2} v_\chi Z_{{k 3}}^{H}  + v_\phi Z_{{k 1}}^{H} \Big) 
\\ 
\Gamma_{\bar{\eta^-}\eta^+{H}^{--}}  = & \, 
-\frac{i}{2} g_{2}^{2} v_\chi \xi_{W^+}  
\\ 
\Gamma_{\bar{\eta^Z}\eta^+H^+_{{k}}}  = & \, 
\frac{i}{4} g_2 \xi_{Z} \Big(g_1 \sin\Theta_W   + g_2 \cos\Theta_W  \Big)\Big(4 v_\chi Z_{{k 3}}^{+}  + v_\phi Z_{{k 1}}^{+} \Big) 
\\ 
\Gamma_{\bar{\eta^-}\eta^-h_{{k}}}  = & \, 
-\frac{i}{4} g_{2}^{2} \xi_{W^+} \Big(4 v_\chi Z_{{k 2}}^{H}  + \sqrt{2} v_\chi Z_{{k 3}}^{H}  + v_\phi Z_{{k 1}}^{H} \Big) 
\\ 
\Gamma_{\bar{\eta^Z}\eta^-H^+_{{k}}}  = & \, 
\frac{i}{4} g_2 \xi_{Z} \Big(g_1 \sin\Theta_W   + g_2 \cos\Theta_W  \Big)\Big(4 v_\chi Z_{{k 3}}^{+}  + v_\phi Z_{{k 1}}^{+} \Big) 
\\ 
\Gamma_{\bar{\eta^+}\eta^-{H}^{--}}  = & \, 
-\frac{i}{2} g_{2}^{2} v_\chi \xi_{W^+}  
\\ 
\Gamma_{\bar{\eta^Z}\eta^Zh_{{k}}}  = & \, 
-\frac{i}{4} \xi_{Z} \Big(g_1 \sin\Theta_W   + g_2 \cos\Theta_W  \Big)^{2} \Big(4 \sqrt{2} v_\chi Z_{{k 3}}^{H}  + v_\phi Z_{{k 1}}^{H} \Big) 
\\ 
\Gamma_{\bar{\eta^+}\eta^ZH^+_{{k}}}  = & \, 
-\frac{i}{4} g_2 \xi_{W^+} \Big(-2 g_1 v_\chi \sin\Theta_W  Z_{{k 3}}^{+}  + 4 g_2 v_\chi \cos\Theta_W  Z_{{k 2}}^{+}  + v_\phi \Big(- g_1 \sin\Theta_W   + g_2 \cos\Theta_W  \Big)Z_{{k 1}}^{+} \Big) 
\\ 
\Gamma_{\bar{\eta^-}\eta^ZH^+_{{k}}}  = & \, 
-\frac{i}{4} g_2 \xi_{W^+} \Big(-2 g_1 v_\chi \sin\Theta_W  Z_{{k 3}}^{+}  + 4 g_2 v_\chi \cos\Theta_W  Z_{{k 2}}^{+}  + v_\phi \Big(- g_1 \sin\Theta_W   + g_2 \cos\Theta_W  \Big)Z_{{k 1}}^{+} \Big) 
\end{align} 

\section{One-loop corrections}
\label{app:loopcorrections}
Here we give the expressions for the one-loop  tadpoles and self-energies. For that, we refer to the vertices listed in the last subsection. Since the corrections are applied to the mass matrices but not the 
mass eigenstates, external gauge eigenstates must be used. Therefore, the rotation matrices of these states must be replaced by the identity matrix. We label the corresponding fields as $\check{x}$. \\
In addition, we introduce the following abbreviations:
\begin{align}
X(a,b) =& X(p^2,a,b) \hspace{1cm} X=\{B_0,B_1,G_0,F_0\} \\ 
LR =& L \leftrightarrow R 
\end{align}

\subsection*{Tadpoles}
\begin{align} 
\delta t^{(1)}_{h} = & \, +{A_0\Big(m^2_{\eta^+}\Big)} {\Gamma_{\check{h}_{{i}},\bar{\eta^+},\eta^+}} +{A_0\Big(m^2_{\eta^-}\Big)} {\Gamma_{\check{h}_{{i}},\bar{\eta^-},\eta^-}} +{A_0\Big(m^2_{\eta^Z}\Big)} {\Gamma_{\check{h}_{{i}},\bar{\eta^Z},\eta^Z}} \nonumber \\ 
 &- {A_0\Big(m^2_{{H}^{--}}\Big)} {\Gamma_{\check{h}_{{i}},{H}^{{++}},{H}^{--}}} +4 {\Gamma_{\check{h}_{{i}},W^-,W^+}} \Big(-\frac{m^2_W}{2}    + {A_0\Big(m^2_W\Big)}\Big)+2 {\Gamma_{\check{h}_{{i}},Z,Z}} \Big(-\frac{m^2_{Z}}{2}    + {A_0\Big(m^2_{Z}\Big)}\Big)\nonumber \\ 
 &-\frac{1}{2} \sum_{a=1}^{2}{A_0\Big(m^2_{A^0_{{a}}}\Big)} {\Gamma_{\check{h}_{{i}},A^0_{{a}},A^0_{{a}}}}  - \sum_{a=1}^{3}{A_0\Big(m^2_{H^+_{{a}}}\Big)} {\Gamma_{\check{h}_{{i}},H^-_{{a}},H^+_{{a}}}}  -\frac{1}{2} \sum_{a=1}^{3}{A_0\Big(m^2_{h_{{a}}}\Big)} {\Gamma_{\check{h}_{{i}},h_{{a}},h_{{a}}}}  \nonumber \\ 
 &+6 \sum_{a=1}^{3}{A_0\Big(m^2_{d_{{a}}}\Big)} m_{d_{{a}}} \Big({\Gamma^L_{\check{h}_{{i}},\bar{d}_{{a}},d_{{a}}}} + {\Gamma^R_{\check{h}_{{i}},\bar{d}_{{a}},d_{{a}}}}\Big)
 +2 \sum_{a=1}^{3}{A_0\Big(m^2_{e_{{a}}}\Big)} m_{e_{{a}}} \Big({\Gamma^L_{\check{h}_{{i}},\bar{e}_{{a}},e_{{a}}}} + {\Gamma^R_{\check{h}_{{i}},\bar{e}_{{a}},e_{{a}}}}\Big) \nonumber \\ 
 &+6 \sum_{a=1}^{3}{A_0\Big(m^2_{u_{{a}}}\Big)} m_{u_{{a}}} \Big({\Gamma^L_{\check{h}_{{i}},\bar{u}_{{a}},u_{{a}}}} + {\Gamma^R_{\check{h}_{{i}},\bar{u}_{{a}},u_{{a}}}}\Big)  
\end{align}

\subsection*{Self-energies}
\begin{enumerate} 
\item {\bf CP-even Higgs}

\begin{align} 
& \Pi^H_{i,j}(p^2)\hspace{.3cm} = +2 \Big({B_0\Big(m^2_{Z},m^2_{Z}\Big)}-\frac{1}{2} \Big){\Gamma^*_{\check{h}_{{j}},Z,Z}} {\Gamma_{\check{h}_{{i}},Z,Z}} +4 \Big( {B_0\Big(m^2_W,m^2_W\Big)}-\frac{1}{2} \Big){\Gamma^*_{\check{h}_{{j}},W^-,W^+}} {\Gamma_{\check{h}_{{i}},W^-,W^+}} \nonumber \\ 
 &+{B_0\Big(m^2_{{H}^{--}},m^2_{{H}^{--}}\Big)} {\Gamma^*_{\check{h}_{{j}},{H}^{{++}},{H}^{--}}} {\Gamma_{\check{h}_{{i}},{H}^{{++}},{H}^{--}}} - {B_0\Big(m^2_{\eta^+},m^2_{\eta^+}\Big)} {\Gamma_{\check{h}_{{i}},\bar{\eta^+},\eta^+}} {\Gamma_{\check{h}_{{j}},\bar{\eta^+},\eta^+}} \nonumber \\ 
 &- {B_0\Big(m^2_{\eta^-},m^2_{\eta^-}\Big)} {\Gamma_{\check{h}_{{i}},\bar{\eta^-},\eta^-}} {\Gamma_{\check{h}_{{j}},\bar{\eta^-},\eta^-}} - {B_0\Big(m^2_{\eta^Z},m^2_{\eta^Z}\Big)} {\Gamma_{\check{h}_{{i}},\bar{\eta^Z},\eta^Z}} {\Gamma_{\check{h}_{{j}},\bar{\eta^Z},\eta^Z}} \nonumber \\ 
 &- {A_0\Big(m^2_{{H}^{--}}\Big)} {\Gamma_{\check{h}_{{i}},\check{h}_{{j}},{H}^{{++}},{H}^{--}}} +4 {\Gamma_{\check{h}_{{i}},\check{h}_{{j}},W^-,W^+}} \Big({A_0\Big(m^2_W\Big)}-\frac{m^2_W}{2}  \Big)+2 {\Gamma_{\check{h}_{{i}},\check{h}_{{j}},Z,Z}} \Big({A_0\Big(m^2_{Z}\Big)}-\frac{m^2_{Z}}{2}   \Big)\nonumber \\ 
 &-\frac{1}{2} \sum_{a=1}^{2}{A_0\Big(m^2_{A^0_{{a}}}\Big)} {\Gamma_{\check{h}_{{i}},\check{h}_{{j}},A^0_{{a}},A^0_{{a}}}}  +\frac{1}{2} \sum_{a=1}^{2}\sum_{b=1}^{2}{B_0\Big(m^2_{A^0_{{a}}},m^2_{A^0_{{b}}}\Big)} {\Gamma^*_{\check{h}_{{j}},A^0_{{a}},A^0_{{b}}}} {\Gamma_{\check{h}_{{i}},A^0_{{a}},A^0_{{b}}}}  \nonumber \\ 
 &- \sum_{a=1}^{3}{A_0\Big(m^2_{H^+_{{a}}}\Big)} {\Gamma_{\check{h}_{{i}},\check{h}_{{j}},H^-_{{a}},H^+_{{a}}}}  -\frac{1}{2} \sum_{a=1}^{3}{A_0\Big(m^2_{h_{{a}}}\Big)} {\Gamma_{\check{h}_{{i}},\check{h}_{{j}},h_{{a}},h_{{a}}}} +\frac{1}{2} \sum_{a,b=1}^{3}{B_0\Big(m^2_{h_{{a}}},m^2_{h_{{b}}}\Big)} {\Gamma^*_{\check{h}_{{j}},h_{{a}},h_{{b}}}} {\Gamma_{\check{h}_{{i}},h_{{a}},h_{{b}}}} \nonumber \\ 
 &+\sum_{a=1}^{3}\sum_{b=1}^{2}{B_0\Big(m^2_{h_{{a}}},m^2_{A^0_{{b}}}\Big)} {\Gamma^*_{\check{h}_{{j}},h_{{a}},A^0_{{b}}}} {\Gamma_{\check{h}_{{i}},h_{{a}},A^0_{{b}}}} + \sum_{a,b=1}^{3}{B_0\Big(m^2_{H^+_{{a}}},m^2_{H^+_{{b}}}\Big)} {\Gamma^*_{\check{h}_{{j}},H^-_{{a}},H^+_{{b}}}} {\Gamma_{\check{h}_{{i}},H^-_{{a}},H^+_{{b}}}} \nonumber \\ 
 &+3 \sum_{a,b=1}^{3}\left[{G_0\Big(m^2_{d_{{a}}},m^2_{d_{{b}}}\Big)} {\Gamma^{L*}_{\check{h}_{{j}},\bar{d}_{{a}},d_{{b}}}} {\Gamma^L_{\check{h}_{{i}},\bar{d}_{{a}},d_{{b}}}}  
 -2 m_{d_{{a}}} {B_0\Big(m^2_{d_{{a}}},m^2_{d_{{b}}}\Big)} m_{d_{{b}}} {\Gamma^{L*}_{\check{h}_{{j}},\bar{d}_{{a}},d_{{b}}}} {\Gamma^R_{\check{h}_{{i}},\bar{d}_{{a}},d_{{b}}}}    + LR \right] 
 \nonumber \\ 
 &+ \sum_{a,b=1}^{3}\left[{G_0\Big(m^2_{e_{{a}}},m^2_{e_{{b}}}\Big)} {\Gamma^{L*}_{\check{h}_{{j}},\bar{e}_{{a}},e_{{b}}}} {\Gamma^L_{\check{h}_{{i}},\bar{e}_{{a}},e_{{b}}}}  -2 m_{e_{{a}}} {B_0\Big(m^2_{e_{{a}}},m^2_{e_{{b}}}\Big)} m_{e_{{b}}} {\Gamma^{L*}_{\check{h}_{{j}},\bar{e}_{{a}},e_{{b}}}} {\Gamma^R_{\check{h}_{{i}},\bar{e}_{{a}},e_{{b}}}}   + LR  \right]\nonumber \\ 
 &+3 \sum_{a,b=1}^{3}\left[{G_0\Big(m^2_{u_{{a}}},m^2_{u_{{b}}}\Big)} {\Gamma^{L*}_{\check{h}_{{j}},\bar{u}_{{a}},u_{{b}}}} {\Gamma^L_{\check{h}_{{i}},\bar{u}_{{a}},u_{{b}}}}   
 -2 m_{u_{{a}}} {B_0\Big(m^2_{u_{{a}}},m^2_{u_{{b}}}\Big)} m_{u_{{b}}} {\Gamma^{L*}_{\check{h}_{{j}},\bar{u}_{{a}},u_{{b}}}} {\Gamma^R_{\check{h}_{{i}},\bar{u}_{{a}},u_{{b}}}}   + LR \right]\nonumber \\ 
 &+\sum_{b=1}^{2}{\Gamma^*_{\check{h}_{{j}},Z,A^0_{{b}}}} {\Gamma_{\check{h}_{{i}},Z,A^0_{{b}}}} {F_0\Big(m^2_{A^0_{{b}}},m^2_{Z}\Big)} +2 \sum_{b=1}^{3}{\Gamma^*_{\check{h}_{{j}},W^-,H^+_{{b}}}} {\Gamma_{\check{h}_{{i}},W^-,H^+_{{b}}}} {F_0\Big(m^2_{H^+_{{b}}},m^2_W\Big)}   
\end{align} 
\item {\bf CP-odd Higgs}

\begin{align} 
& \Pi^A_{i,j}(p^2) \hspace{.3cm} = - {B_0\Big(m^2_{\eta^+},m^2_{\eta^+}\Big)} {\Gamma_{\check{A}^0_{{i}},\bar{\eta^+},\eta^+}} {\Gamma_{\check{A}^0_{{j}},\bar{\eta^+},\eta^+}} - {B_0\Big(m^2_{\eta^-},m^2_{\eta^-}\Big)} {\Gamma_{\check{A}^0_{{i}},\bar{\eta^-},\eta^-}} {\Gamma_{\check{A}^0_{{j}},\bar{\eta^-},\eta^-}} \nonumber \\ 
 &- {A_0\Big(m^2_{{H}^{--}}\Big)} {\Gamma_{\check{A}^0_{{i}},\check{A}^0_{{j}},{H}^{{++}},{H}^{--}}} +4 {\Gamma_{\check{A}^0_{{i}},\check{A}^0_{{j}},W^-,W^+}} \Big(-\frac{m^2_W}{2}    + {A_0\Big(m^2_W\Big)}\Big)\nonumber \\ 
 &+2 {\Gamma_{\check{A}^0_{{i}},\check{A}^0_{{j}},Z,Z}} \Big(-\frac{m^2_{Z}}{2}    + {A_0\Big(m^2_{Z}\Big)}\Big)-\frac{1}{2} \sum_{a=1}^{2}{A_0\Big(m^2_{A^0_{{a}}}\Big)} {\Gamma_{\check{A}^0_{{i}},\check{A}^0_{{j}},A^0_{{a}},A^0_{{a}}}}  \nonumber \\ 
 &+\frac{1}{2} \sum_{a=1}^{2}\sum_{b=1}^{2}{B_0\Big(m^2_{A^0_{{a}}},m^2_{A^0_{{b}}}\Big)} {\Gamma^*_{\check{A}^0_{{j}},A^0_{{a}},A^0_{{b}}}} {\Gamma_{\check{A}^0_{{i}},A^0_{{a}},A^0_{{b}}}}  - \sum_{a=1}^{3}{A_0\Big(m^2_{H^+_{{a}}}\Big)} {\Gamma_{\check{A}^0_{{i}},\check{A}^0_{{j}},H^-_{{a}},H^+_{{a}}}}  \nonumber \\ 
 &-\frac{1}{2} \sum_{a=1}^{3}{A_0\Big(m^2_{h_{{a}}}\Big)} {\Gamma_{\check{A}^0_{{i}},\check{A}^0_{{j}},h_{{a}},h_{{a}}}}  +\sum_{a=1}^{3}\sum_{b=1}^{2}{B_0\Big(m^2_{h_{{a}}},m^2_{A^0_{{b}}}\Big)} {\Gamma^*_{\check{A}^0_{{j}},h_{{a}},A^0_{{b}}}} {\Gamma_{\check{A}^0_{{i}},h_{{a}},A^0_{{b}}}} \nonumber \\ 
 &+ \sum_{a,b=1}^{3}{B_0\Big(m^2_{H^+_{{a}}},m^2_{H^+_{{b}}}\Big)} {\Gamma^*_{\check{A}^0_{{j}},H^-_{{a}},H^+_{{b}}}} {\Gamma_{\check{A}^0_{{i}},H^-_{{a}},H^+_{{b}}}} +\frac{1}{2} \sum_{a,b=1}^{3}{B_0\Big(m^2_{h_{{a}}},m^2_{h_{{b}}}\Big)} {\Gamma^*_{\check{A}^0_{{j}},h_{{a}},h_{{b}}}} {\Gamma_{\check{A}^0_{{i}},h_{{a}},h_{{b}}}}  \nonumber \\ 
 &-3 \sum_{a,b=1}^{3}\left[2 m_{d_{{a}}} {B_0\Big(m^2_{d_{{a}}},m^2_{d_{{b}}}\Big)} m_{d_{{b}}} {\Gamma^{L*}_{\check{A}^0_{{j}},\bar{d}_{{a}},d_{{b}}}} {\Gamma^R_{\check{A}^0_{{i}},\bar{d}_{{a}},d_{{b}}}}   -{G_0\Big(m^2_{d_{{a}}},m^2_{d_{{b}}}\Big)} {\Gamma^{L*}_{\check{A}^0_{{j}},\bar{d}_{{a}},d_{{b}}}} {\Gamma^L_{\check{A}^0_{{i}},\bar{d}_{{a}},d_{{b}}}}   + LR \right] \nonumber \\ 
 &-\sum_{a,b=1}^{3} \left[2 m_{e_{{a}}} {B_0\Big(m^2_{e_{{a}}},m^2_{e_{{b}}}\Big)} m_{e_{{b}}} {\Gamma^{L*}_{\check{A}^0_{{j}},\bar{e}_{{a}},e_{{b}}}} {\Gamma^R_{\check{A}^0_{{i}},\bar{e}_{{a}},e_{{b}}}}  -{G_0\Big(m^2_{e_{{a}}},m^2_{e_{{b}}}\Big)} {\Gamma^{L*}_{\check{A}^0_{{j}},\bar{e}_{{a}},e_{{b}}}} {\Gamma^L_{\check{A}^0_{{i}},\bar{e}_{{a}},e_{{b}}}}   + LR \right]\nonumber \\ 
 &-3 \sum_{a,b=1}^{3}\left[2m_{u_{{a}}} {B_0\Big(m^2_{u_{{a}}},m^2_{u_{{b}}}\Big)} m_{u_{{b}}} {\Gamma^{L*}_{\check{A}^0_{{j}},\bar{u}_{{a}},u_{{b}}}} {\Gamma^R_{\check{A}^0_{{i}},\bar{u}_{{a}},u_{{b}}}}    - \sum_{a,b=1}^{3}{G_0\Big(m^2_{u_{{a}}},m^2_{u_{{b}}}\Big)} {\Gamma^{L*}_{\check{A}^0_{{j}},\bar{u}_{{a}},u_{{b}}}} {\Gamma^L_{\check{A}^0_{{i}},\bar{u}_{{a}},u_{{b}}}}   +LR\right] \nonumber \\ 
 &+\sum_{b=1}^{3}{\Gamma^*_{\check{A}^0_{{j}},Z,h_{{b}}}} {\Gamma_{\check{A}^0_{{i}},Z,h_{{b}}}} {F_0\Big(m^2_{h_{{b}}},m^2_{Z}\Big)} +2 \sum_{b=1}^{3}{\Gamma^*_{\check{A}^0_{{j}},W^-,H^+_{{b}}}} {\Gamma_{\check{A}^0_{{i}},W^-,H^+_{{b}}}} {F_0\Big(m^2_{H^+_{{b}}},m^2_W\Big)}   
\end{align} 
\item {\bf Charged Higgs}

\begin{align} 
& \Pi^{H^+}_{i,j}(p^2) \hspace{.3cm}= 4 \Big({B_0\Big(0,m^2_W\Big)} -\frac{1}{2} \Big){\Gamma^*_{\check{H}^-_{{j}},W^+,\gamma}} {\Gamma_{\check{H}^-_{{i}},W^+,\gamma}} +4 \Big({B_0\Big(m^2_W,m^2_{Z}\Big)}-\frac{1}{2}\Big){\Gamma^*_{\check{H}^-_{{j}},Z,W^+}} {\Gamma_{\check{H}^-_{{i}},Z,W^+}} \nonumber \\ 
 &- {B_0\Big(m^2_{\eta^Z},m^2_{\eta^-}\Big)} {\Gamma_{\check{H}^-_{{i}},\bar{\eta^-},\eta^Z}} {\Gamma_{\check{H}^+_{{j}},\eta^-,\bar{\eta^Z}}} - {B_0\Big(m^2_{\eta^+},m^2_{\eta^Z}\Big)} {\Gamma_{\check{H}^-_{{i}},\bar{\eta^Z},\eta^+}} {\Gamma_{\check{H}^+_{{j}},\eta^Z,\bar{\eta^+}}} \nonumber \\ 
 &- {A_0\Big(m^2_{{H}^{--}}\Big)} {\Gamma_{\check{H}^+_{{i}},\check{H}^-_{{j}},{H}^{{++}},{H}^{--}}} +{\Gamma^*_{\check{H}^-_{{j}},{H}^{{++}},W^-}} {\Gamma_{\check{H}^-_{{i}},{H}^{{++}},W^-}} {F_0\Big(m^2_{{H}^{--}},m^2_W\Big)} \nonumber \\ 
 &+4 {\Gamma_{\check{H}^+_{{i}},\check{H}^-_{{j}},W^-,W^+}} \Big(-\frac{m^2_W}{2}    + {A_0\Big(m^2_W\Big)}\Big)+2 {\Gamma_{\check{H}^+_{{i}},\check{H}^-_{{j}},Z,Z}} \Big(-\frac{m^2_{Z}}{2}    + {A_0\Big(m^2_{Z}\Big)}\Big)\nonumber \\ 
 &-\frac{1}{2} \sum_{a=1}^{2}{A_0\Big(m^2_{A^0_{{a}}}\Big)} {\Gamma_{\check{H}^+_{{i}},\check{H}^-_{{j}},A^0_{{a}},A^0_{{a}}}}  - \sum_{a=1}^{3}{A_0\Big(m^2_{H^+_{{a}}}\Big)} {\Gamma_{\check{H}^+_{{i}},\check{H}^-_{{j}},H^-_{{a}},H^+_{{a}}}}  \nonumber \\ 
 &-\frac{1}{2} \sum_{a=1}^{3}{A_0\Big(m^2_{h_{{a}}}\Big)} {\Gamma_{\check{H}^+_{{i}},\check{H}^-_{{j}},h_{{a}},h_{{a}}}}  +\sum_{a=1}^{3}\sum_{b=1}^{2}{B_0\Big(m^2_{H^+_{{a}}},m^2_{A^0_{{b}}}\Big)} {\Gamma^*_{\check{H}^-_{{j}},H^+_{{a}},A^0_{{b}}}} {\Gamma_{\check{H}^-_{{i}},H^+_{{a}},A^0_{{b}}}} \nonumber \\ 
 &+ \sum_{a,b=1}^{3}{B_0\Big(m^2_{H^+_{{a}}},m^2_{h_{{b}}}\Big)} {\Gamma^*_{\check{H}^-_{{j}},H^+_{{a}},h_{{b}}}} {\Gamma_{\check{H}^-_{{i}},H^+_{{a}},h_{{b}}}} +\sum_{b=1}^{3}{\Gamma^*_{\check{H}^-_{{j}},Z,H^+_{{b}}}} {\Gamma_{\check{H}^-_{{i}},Z,H^+_{{b}}}} {F_0\Big(m^2_{H^+_{{b}}},m^2_{Z}\Big)}  \nonumber \\ 
 &-3 \sum_{a,b=1}^{3}\left[2 m_{d_{{a}}}{B_0\Big(m^2_{d_{{a}}},m^2_{u_{{b}}}\Big)} m_{u_{{b}}} {\Gamma^{L*}_{\check{H}^-_{{j}},\bar{d}_{{a}},u_{{b}}}} {\Gamma^R_{\check{H}^-_{{i}},\bar{d}_{{a}},u_{{b}}}}  -{G_0\Big(m^2_{d_{{a}}},m^2_{u_{{b}}}\Big)} {\Gamma^{L*}_{\check{H}^-_{{j}},\bar{d}_{{a}},u_{{b}}}} {\Gamma^L_{\check{H}^-_{{i}},\bar{d}_{{a}},u_{{b}}}}    + LR\right] \nonumber \\ 
 &-\sum_{a,b=1}^{3} \left[2 m_{e_{{a}}}{B_0\Big(m^2_{e_{{a}}},m^2_{\nu_{{b}}}\Big)} m_{\nu_{{b}}} {\Gamma^{L*}_{\check{H}^-_{{j}},\bar{e}_{{a}},\nu_{{b}}}} {\Gamma^R_{\check{H}^-_{{i}},\bar{e}_{{a}},\nu_{{b}}}} -{G_0\Big(m^2_{e_{{a}}},m^2_{\nu_{{b}}}\Big)} {\Gamma^{L*}_{\check{H}^-_{{j}},\bar{e}_{{a}},\nu_{{b}}}} {\Gamma^L_{\check{H}^-_{{i}},\bar{e}_{{a}},\nu_{{b}}}}   + LR \right]\nonumber \\ 
 &+\sum_{b=1}^{2}{\Gamma^*_{\check{H}^-_{{j}},W^+,A^0_{{b}}}} {\Gamma_{\check{H}^-_{{i}},W^+,A^0_{{b}}}} {F_0\Big(m^2_{A^0_{{b}}},m^2_W\Big)} +\sum_{b=1}^{3}{B_0\Big(m^2_{{H}^{--}},m^2_{H^+_{{b}}}\Big)} {\Gamma^*_{\check{H}^-_{{j}},{H}^{{++}},H^-_{{b}}}} {\Gamma_{\check{H}^-_{{i}},{H}^{{++}},H^-_{{b}}}} \nonumber \\ 
 &+\sum_{b=1}^{3}{\Gamma^*_{\check{H}^-_{{j}},W^+,h_{{b}}}} {\Gamma_{\check{H}^-_{{i}},W^+,h_{{b}}}} {F_0\Big(m^2_{h_{{b}}},m^2_W\Big)} +\sum_{b=1}^{3}{\Gamma^*_{\check{H}^-_{{j}},\gamma,H^+_{{b}}}} {\Gamma_{\check{H}^-_{{i}},\gamma,H^+_{{b}}}} {F_0\Big(m^2_{H^+_{{b}}},0\Big)}  
\end{align} 

\item {\bf Doubly-charged Higgs }
\begin{align} 
& \Pi^{H^{++}}(p^2) \hspace{.3cm}= +2 |{\Gamma_{{H}^{{++}},W^-,W^-}}|^2 \Big(-\frac{1}{2}   + {B_0\Big(m^2_W,m^2_W\Big)}\Big)- {B_0\Big(m^2_{\eta^-},m^2_{\eta^+}\Big)} {\Gamma_{{H}^{--},\eta^+,\bar{\eta^-}}} {\Gamma_{{H}^{{++}},\bar{\eta^+},\eta^-}} \nonumber \\ 
 &- {A_0\Big(m^2_{{H}^{--}}\Big)} {\Gamma_{{H}^{--},{H}^{{++}},{H}^{{++}},{H}^{--}}} +|{\Gamma_{{H}^{{++}},{H}^{--},\gamma}}|^2 {F_0\Big(m^2_{{H}^{--}},0\Big)} +|{\Gamma_{{H}^{{++}},{H}^{--},Z}}|^2 {F_0\Big(m^2_{{H}^{--}},m^2_{Z}\Big)} \nonumber \\ 
 &+4 {\Gamma_{{H}^{--},{H}^{{++}},W^-,W^+}} \Big(-\frac{m^2_W}{2}    + {A_0\Big(m^2_W\Big)}\Big)+2 {\Gamma_{{H}^{--},{H}^{{++}},Z,Z}} \Big(-\frac{m^2_{Z}}{2}    + {A_0\Big(m^2_{Z}\Big)}\Big)\nonumber \\ 
 &-\frac{1}{2} \sum_{a=1}^{2}{A_0\Big(m^2_{A^0_{{a}}}\Big)} {\Gamma_{{H}^{--},{H}^{{++}},A^0_{{a}},A^0_{{a}}}}  - \sum_{a=1}^{3}{A_0\Big(m^2_{H^+_{{a}}}\Big)} {\Gamma_{{H}^{--},{H}^{{++}},H^-_{{a}},H^+_{{a}}}}  \nonumber \\ 
 &-\frac{1}{2} \sum_{a=1}^{3}{A_0\Big(m^2_{h_{{a}}}\Big)} {\Gamma_{{H}^{--},{H}^{{++}},h_{{a}},h_{{a}}}}  +\frac{1}{2} \sum_{a,b=1}^{3}|{\Gamma_{{H}^{{++}},H^-_{{a}},H^-_{{b}}}}|^2 {B_0\Big(m^2_{H^+_{{a}}},m^2_{H^+_{{b}}}\Big)}  \nonumber \\ 
 &+\sum_{b=1}^{3}|{\Gamma_{{H}^{{++}},{H}^{--},h_{{b}}}}|^2 {B_0\Big(m^2_{{H}^{--}},m^2_{h_{{b}}}\Big)} +\sum_{b=1}^{3}|{\Gamma_{{H}^{{++}},W^-,H^-_{{b}}}}|^2 {F_0\Big(m^2_{H^+_{{b}}},m^2_W\Big)}  
\end{align} 
\end{enumerate}

\section{Counter-terms}
\label{app:CTs}
\allowdisplaybreaks

Here we present the counter-terms necessary to renormalise the scalar sector of the GM model on-shell. 
\begin{align}
\label{eq:CT1}
\delta\lambda_{1}= & -\frac{1}{8} v_\phi^{-3} \Big(- v_\phi \Pi^H_{11}  + \delta t_1\Big)\\
\delta\lambda_{2a}= & +\delta\lambda_{2b}+\frac{1}{8} v_\phi^{-2} v_\chi^{-1} \Big(-2 \sqrt{2} \delta t_3  -2 v_\phi \Pi^H_{12}  + 4 \delta t_2  + \sqrt{2} v_\phi \Pi^H_{13} \Big)\nonumber \\ 
 &+\frac{1}{2} v_\phi^{-2} \Big(8 v_\chi^{2}  + v_\phi^{2}\Big)^{-1} \Big(4 \sqrt{2} \delta t_3 v_\chi  -8 \delta t_2 v_\chi  + v_\phi \Big(4 v_\chi \Pi^{H^+}_{12}  -4 v_\chi \Pi^{H^+}_{13}  - v_\phi \Pi^{H^+}_{22}  + v_\phi \Pi^{H^+}_{33} \Big)\Big)\\
\delta\lambda_{3b}= & \frac{1}{8} v_\chi^{-2} \Big(8 v_\chi^{2}  + v_\phi^{2}\Big)^{-2} \Big(-1024 \delta\lambda_{4c} v_\chi^{6} +64 v_\chi^{4} \Big(-4 \delta\lambda_{4c} v_\phi^{2}  + \sqrt{2} \Pi^H_{23}  + \Pi^{++}\Big) \nonumber \\
& -4 v_\phi^{2} v_\chi \Big(2 \delta t_2  + \sqrt{2} \delta t_3  + v_\phi \Big(-3 \Pi^{H^+}_{13}  - \Pi^{H^+}_{12}  + \sqrt{2} \Pi^A_{12} \Big)\Big)-32 v_\chi^{3} \Big(\sqrt{2} \delta t_3  + v_\phi \Big(- \Pi^{H^+}_{13}  + \Pi^{H^+}_{12}\Big)\Big)\nonumber \\ 
 &+8 v_\phi v_\chi^{2} \Big(v \Big(-2 \Pi^{H^+}_{11}  + 2 \sqrt{2} \Pi^H_{23}  -2 \delta\lambda_{4c} v_\phi^{2}  + 2 \Pi^{H^{++} } - \Pi^{H^+}_{33}  + \Pi^A_{11} + \Pi^{H^+}_{22}\Big) + \delta t_1\Big) \nonumber\\
 & +v_\phi^{4} \Big(-2 \Big(\Pi^{H^+}_{23} + \Pi^{H^+}_{33}\Big) + \sqrt{2} \Pi^H_{23}  + \Pi^{H^{++} }+ \Pi^A_{22}\Big)\Big) \label{eq:app:couterterm_lam3b}\\
\delta\lambda_{3c}= & \frac{1}{8} v_\chi^{-2} \Big(8 v_\chi^{2}  + v_\phi^{2}\Big)^{-2} \Big(-1024 \delta\lambda_{4c} v_\chi^{6} -8 \delta t_2 v_\chi \Big(4 v_\chi^{2}  + v_\phi^{2}\Big)-4 \sqrt{2} \delta t_3 v_\chi \Big(4 v_\chi^{2}  + v_\phi^{2}\Big)\nonumber \\
& +4 v_\phi^{3} v_\chi \Big(\Pi^{H^+}_{12} + \Pi^{H^+}_{13}\Big)+v_\phi^{4} \Big(-2 \Pi^{H^+}_{23}  + \sqrt{2} \Pi^H_{23} \Big)\nonumber \\ 
 &-32 v_\chi^{4} \Big(2 \Pi^{H^+}_{23}  -2 \sqrt{2} \Pi^H_{23}  + 8 \delta\lambda_{4c} v_\phi^{2}  - \Pi^{H^+}_{22}  - \Pi^{H^+}_{33} \Big)+8 v_\phi v_\chi^{2} \Big(-2 \delta\lambda_{4c} v_\phi^{3}  \nonumber \\
 & + v_\phi \Big(-2 \Pi^{H^+}_{23}  + 2 \sqrt{2} \Pi^H_{23}  - \Pi^{H^+}_{11}  + \Pi^{H^+}_{22} + \Pi^{H^+}_{33}\Big) + \delta t_1\Big)\Big) \label{eq:app:couterterm_lam3c}\\
\delta\lambda_{4a}= & \frac{1}{16} v_\chi^{-2} \Big(8 v_\chi^{2}  + v_\phi^{2}\Big)^{-2} \Big(-1024 \Big(\delta\lambda_{3a} + \delta\lambda_{4c}\Big)v_\chi^{6} -64 v_\chi^{4} \Big(-2 \Pi^H_{22}  + 4 \Big(\delta\lambda_{3a} + \delta\lambda_{4c}\Big)v_\phi^{2}  - \sqrt{2} \Pi^H_{23} \Big) \nonumber \\
& -32 v_\chi^{3} \Big(4 \delta t_2  + v_\phi \Big(- \Pi^{H^+}_{12}  + \Pi^{H^+}_{13}\Big)\Big)+4 v_\phi^{2} v_\chi \Big(-6 \delta t_2  - \sqrt{2} \delta t_3  + v_\phi \Big(3 \Pi^{H^+}_{12}  + \Pi^{H^+}_{13}\Big)\Big)\nonumber \\ 
 &+v_\phi^{4} \Big(2 \Pi^H_{22}  -2 \Big(\Pi^{H^+}_{22} + \Pi^{H^+}_{23}\Big) + \sqrt{2} \Pi^H_{23} \Big)+8 v_\phi v_\chi^{2} \Big(2 \delta t_1  + v_\phi \Big(-2 \Pi^{H^+}_{11}  + 2 \sqrt{2} \Pi^H_{23} \nonumber \\
 & -2 \Big(\delta\lambda_{3a} + \delta\lambda_{4c}\Big)v_\phi^{2}   + 4 \Pi^H_{22}  - \Pi^{H^+}_{22}  + \Pi^{H^+}_{33}\Big)\Big)\Big)\\
\delta\lambda_{4b}= & \frac{1}{16} v_\chi^{-2} \Big(8 v_\chi^{2}  + v_\phi^{2}\Big)^{-2} \Big(1024 \delta\lambda_{4c} v_\chi^{6} -64 v_\chi^{4} \Big(-4 \delta\lambda_{4c} v_\phi^{2}  - \Pi^H_{33}  + \sqrt{2} \Pi^H_{23}  + \Pi^{++}\Big)\nonumber \\ 
& +32 v_\phi v_\chi^{3} \Big(- \Pi^{H^+}_{13}  + \Pi^{H^+}_{12}\Big)-4 v_\phi^{2} v_\chi \Big(-2 \delta t_2  + \sqrt{2} \delta t_3  + v_\phi \Big(-2 \sqrt{2} \Pi^A_{12}  + 3 \Pi^{H^+}_{13}  + \Pi^{H^+}_{12}\Big)\Big)\nonumber \\ 
 &+8 v_\phi^{2} v_\chi^{2} \Big(-2 \Pi^A_{11}  + 2 \Pi^{H^+}_{11}  + 2 \Pi^H_{33}  -2 \sqrt{2} \Pi^H_{23}  + 2 \delta\lambda_{4c} v_\phi^{2}  -2 \Pi^{H^{++} } - \Pi^{H^+}_{22}  + \Pi^{H^+}_{33}\Big)\nonumber \\ 
& +v_\phi^{4} \Big(2 \Big(- \Pi^A_{22}  + \Pi^{H^+}_{23} + \Pi^{H^+}_{33}\Big) - \sqrt{2} \Pi^H_{23}  - \Pi^{H^{++} } + \Pi^H_{33}\Big)\Big)\\
\delta\lambda_{5a}= & v_\phi^{-1} v_\chi^{-1} \Big(8 v_\chi^{2}  + v_\phi^{2}\Big)^{-2} \Big(256 \delta\lambda_{2b} v_\phi v_\chi^{5} +v_\phi^{3} \Big(2 \delta t_2  - v_\phi \Pi^H_{12} \Big)+8 v_\phi v_\chi^{2} \Big(2 \delta t_2  - \sqrt{2} \delta t_3  \nonumber \\ 
& + v_\phi \Big(-2 \Pi^H_{12}  - \Pi^{H^+}_{13}  + \sqrt{2} \Pi^A_{12}  + \Pi^{H^+}_{12}\Big)\Big)-64 v_\chi^{4} \Big(- \Pi^{H^+}_{12}  + \Pi^H_{12} + \Pi^{H^+}_{13}\Big)\nonumber \\ 
 &+2 v_\phi^{3} v_\chi \Big(2 \delta\lambda_{2b} v_\phi^{2}  - \Pi^A_{22}  - \Pi^{H^+}_{22}  + \Pi^{H^+}_{33}\Big)+16 v_\chi^{3} \Big(v \Big(4 \delta\lambda_{2b} v_\phi^{2}  - \Pi^A_{11}  - \Pi^{H^+}_{22}  + \Pi^{H^+}_{33}\Big) + \delta t_1\Big)\Big)\\
\delta\lambda_{5b}= & v_\phi^{-1} v_\chi^{-1} \Big(8 v_\chi^{2}  + v_\phi^{2}\Big)^{-2} \Big(256 \delta\lambda_{2b} v_\phi v_\chi^{5} +v_\phi^{3} \Big(2 \delta t_2  - v_\phi \Pi^H_{12} \Big)-32 v_\chi^{4} \Big(2 \Pi^H_{12}  - \Pi^{H^+}_{12}  + \Pi^{H^+}_{13}\Big)\nonumber \\ 
& +4 v_\phi v_\chi^{2} \Big(2 \delta t_2  - \sqrt{2} \delta t_3  + v_\phi \Big(3 \Pi^{H^+}_{12}  -4 \Pi^H_{12}  + \Pi^{H^+}_{13}\Big)\Big)+2 v_\phi^{3} v_\chi \Big(2 \delta\lambda_{2b} v_\phi^{2}  - \Pi^{H^+}_{22}  - \Pi^{H^+}_{23} \Big)\nonumber \\ 
 &+8 v_\chi^{3} \Big(2 \delta t_1  + v_\phi \Big(-2 \Pi^{H^+}_{11}  + 8 \delta\lambda_{2b} v_\phi^{2}  - \Pi^{H^+}_{22}  + \Pi^{H^+}_{33}\Big)\Big)\Big)\\
\delta M_{1a}= & 2 v_\phi^{-1} \Big(8 v_\chi^{2}  + v_\phi^{2}\Big)^{-2} \Big(-256 \delta\lambda_{2b} v_\phi v_\chi^{5} +v_\phi^{3} \Big( v_\phi \Pi^H_{12} -4 \delta t_2 \Big)+8 v_\phi v_\chi^{2} \Big(\sqrt{2} \delta t_3 -2 \delta t_2  \nonumber \\ 
 &+ v_\phi \Big(2 \Pi^H_{12}  -3 \Pi^{H^+}_{12}  - \Pi^{H^+}_{13} \Big)\Big)+64 v_\chi^{4} \Big(- \Pi^{H^+}_{12}  + \Pi^H_{12} + \Pi^{H^+}_{13}\Big)\nonumber \\ 
 &+4 v_\phi^{3} v_\chi \Big(- \delta\lambda_{2b} v_\phi^{2}  + \Pi^{H^+}_{22} + \Pi^{H^+}_{23}\Big)-16 v_\chi^{3} \Big(2 \delta t_1  + v_\phi \Big(-2 \Pi^{H^+}_{11}  + 4 \delta\lambda_{2b} v_\phi^{2}  - \Pi^{H^+}_{22}  + \Pi^{H^+}_{33}\Big)\Big)\Big)\\
\delta M_{1b}= & v_\phi^{-1} \Big(8 v_\chi^{2}  + v_\phi^{2}\Big)^{-2} \Big(-512 \delta\lambda_{2b} v_\phi v_\chi^{5} -4 \delta t_2 \Big(4 v_\phi v_\chi^{2}  + v_\phi^{3}\Big)+2 v_\phi^{3} \Big(- \sqrt{2} \delta t_3  + v_\phi \Pi^H_{12} \Big)\nonumber \\ 
& +64 v_\chi^{4} \Big(2 \Pi^H_{12}  - \Pi^{H^+}_{12}  + \Pi^{H^+}_{13}\Big) -8 v_\phi v_\chi^{2} \Big(- \sqrt{2} \delta t_3  + v_\phi \Big(2 \sqrt{2} \Pi^A_{12}  + 3 \Pi^{H^+}_{12}  -4 \Pi^H_{12}  + \Pi^{H^+}_{13}\Big)\Big)\nonumber \\ 
 &+4 v_\phi^{3} v_\chi \Big(-2 \delta\lambda_{2b} v_\phi^{2}  + \Pi^A_{22} + \Pi^{H^+}_{22} + \Pi^{H^+}_{23}\Big)-16 v_\chi^{3} \Big(4 \delta t_1  + v_\phi \Big(-2 \Pi^A_{11}  -2 \Pi^{H^+}_{11}  \nonumber \\ 
 &+ 8 \delta\lambda_{2b} v_\phi^{2}  - \Pi^{H^+}_{22}  + \Pi^{H^+}_{33}\Big)\Big)\Big)\\
\delta M_{2}= & \frac{1}{12} v_\chi^{-2} \Big(8 v_\chi^{2}  + v_\phi^{2}\Big)^{-2} \Big(1024 \delta\lambda_{4c} v_\chi^{7} +v_\phi^{4} \Big(-2 \delta t_2  + v_\phi \Pi^H_{12} \Big)-64 v_\chi^{5} \Big(4 \Big(- \delta\lambda_{4c}  + \delta\lambda_{2b}\Big)v_\phi^{2}  \nonumber \\ 
&+ \sqrt{2} \Pi^H_{23} \Big) +4 v_\phi^{2} v_\chi^{2} \Big(-2 \delta t_2  + \sqrt{2} \delta t_3  + v_\phi \Big(-3 \Pi^{H^+}_{12}  + 4 \Pi^H_{12}  - \Pi^{H^+}_{13} \Big)\Big)\nonumber \\ 
 &+32 v_\phi v_\chi^{4} \Big(2 \Pi^H_{12}  - \Pi^{H^+}_{12}  + \Pi^{H^+}_{13}\Big)+v_\phi^{4} v_\chi \Big(2 \Big(\Pi^{H^+}_{22} + \Pi^{H^+}_{23}\Big) -4 \delta\lambda_{2b} v_\phi^{2}  - \sqrt{2} \Pi^H_{23} \Big)\nonumber \\ 
& -8 v_\phi v_\chi^{3} \Big(2 \delta t_1  + v_\phi \Big(-2 \Pi^{H^+}_{11}  + 2 \sqrt{2} \Pi^H_{23}  -2 \delta\lambda_{4c} v_\phi^{2}  + 8 \delta\lambda_{2b} v_\phi^{2}  - \Pi^{H^+}_{22}  + \Pi^{H^+}_{33}\Big)\Big)\Big)\\
\delta\mu_2^2= & \frac{1}{2} v_\phi^{-1} \Big(8 v_\chi^{2}  + v_\phi^{2}\Big)^{-2} \Big(-768 \delta\lambda_{2b} v_\phi v_\chi^{6} +v_\phi^{4} \Big(3 \delta t_1  - v_\phi \Pi^H_{11} \Big)\nonumber \\ 
&- v_\phi^{3} v_\chi \Big(2 \sqrt{2} \delta t_3  -2 v_\phi \Pi^H_{12}  + 8 \delta t_2  + \sqrt{2} v_\phi \Pi^H_{13} \Big) +64 v_\chi^{5} \Big(-2 \Pi^{H^+}_{12}  + 2 \Pi^H_{12}  + 2 \Pi^{H^+}_{13}  - \sqrt{2} \Pi^H_{13} \Big)\nonumber \\ 
 &-16 v_\phi v_\chi^{3} \Big(2 \delta t_2  - \sqrt{2} \delta t_3  + v_\phi \Big(-2 \Pi^H_{12}  + 3 \Pi^{H^+}_{12}  + \sqrt{2} \Pi^A_{12}  + \sqrt{2} \Pi^H_{13}  + \Pi^{H^+}_{13}\Big)\Big)\nonumber \\ 
& +4 v_\phi^{2} v_\chi^{2} \Big(12 \delta t_1  + v_\phi \Big(2 \Big(\Pi^{H^+}_{22} + \Pi^{H^+}_{23}\Big) -3 \delta\lambda_{2b} v_\phi^{2}  -4 \Pi^H_{11}  + \Pi^A_{22}\Big)\Big)\nonumber \\ 
 &+32 v_\chi^{4} \Big(3 \delta t_1  + v_\phi \Big(-2 \Pi^H_{11}  + 2 \Pi^{H^+}_{11}  -6 \delta\lambda_{2b} v_\phi^{2}  - \Pi^{H^+}_{33}  + \Pi^A_{11} + \Pi^{H^+}_{22}\Big)\Big)\Big)\\
\delta \mu_\eta^2= & \frac{1}{4} \Big(8 v_\chi^{2}  + v_\phi^{2}\Big)^{-2} \Big(1024 \delta\lambda_{4c} v_\chi^{6} -8 \delta\lambda_{2b} \Big(8 v_\phi v_\chi^{2}  + v_\phi^{3}\Big)^{2} +64 v_\chi^{4} \Big(-2 \Pi^H_{22}  -3 \sqrt{2} \Pi^H_{23}  + 4 \delta\lambda_{4c} v_\phi^{2} \Big)\nonumber \\ 
& +96 v_\chi^{3} \Big(4 \delta t_2  + v_\phi \Big(- \Pi^{H^+}_{12}  + \Pi^{H^+}_{13}\Big)\Big) +12 v_\phi^{2} v_\chi \Big(6 \delta t_2  + \sqrt{2} \delta t_3  - v_\phi \Big(3 \Pi^{H^+}_{12}  + \Pi^{H^+}_{13}\Big)\Big)\nonumber \\ 
& +v_\phi^{4} \Big(-2 \Pi^H_{22}  -3 \sqrt{2} \Pi^H_{23}  + 6 \Big(\Pi^{H^+}_{22} + \Pi^{H^+}_{23}\Big)\Big)+8 v_\phi v_\chi^{2} \Big(-6 \delta t_1  + v_\phi \Big(2 \delta\lambda_{4c} v_\phi^{2}  + 3 \Pi^{H^+}_{22}  \nonumber \\ 
& -3 \Pi^{H^+}_{33}  -4 \Pi^H_{22}  + 6 \Pi^{H^+}_{11}  -6 \sqrt{2} \Pi^H_{23} \Big)\Big)\Big)\\
\delta \mu_\chi^2= & \frac{1}{4} \Big(16 \delta\lambda_{4c} v_\chi^{2} +v_\chi^{-1} \Big(2 \sqrt{2} \delta t_3  + 2 v_\phi \Pi^H_{12}  -4 \delta t_2  - \sqrt{2} v_\phi \Pi^H_{13} \Big)-2 \Big(4 \delta\lambda_{2b} v_\phi^{2}  + \sqrt{2} \Pi^H_{23}  + \Pi^H_{33}\Big)\nonumber \\ 
 &+\Big(8 v_\chi^{2}  + v_\phi^{2}\Big)^{-1} \Big(2 v_\phi^{2} \Big(2 \Pi^{H^+}_{11}  - \Pi^{H^+}_{33}  + \Pi^A_{11} + \Pi^{H^+}_{22}\Big) -6 \delta t_1 v_\phi  + 8 v_\chi \Big(4 \delta t_2  + \sqrt{2} \delta t_3  - v_\phi \Pi^{H^+}_{12}  \nonumber \\ 
 &+ v_\phi \Pi^{H^+}_{13} \Big)\Big)+2 v_\phi^{2} \Big(8 v_\chi^{2}  + v_\phi^{2}\Big)^{-2} \Big(8 \Big(\sqrt{2} \delta t_3  + \delta t_2\Big)v_\chi  + v_\phi \Big(3 \delta t_1  -4 v_\chi \Big(2 \Big(\Pi^{H^+}_{12} + \Pi^{H^+}_{13}\Big) \nonumber \\ 
&+ \sqrt{2} \Pi^A_{12} \Big)  + v_\phi \Big(-2 \Pi^{H^+}_{11}  + 2 \Pi^{H^+}_{23}  - \Pi^A_{11}  + \Pi^A_{22} + \Pi^{H^+}_{22} + \Pi^{H^+}_{33}\Big)\Big)\Big)\Big) \label{eq:CT2}
\end{align}

\section{Tree-level unitarity conditions with \SARAH}
\label{app:scattering}
The tree-level unitarity conditions can be obtained with \SARAH as follows
\begin{lstlisting}
(* loading SARAH and the model *)
<< SARAH.m
Start["Georgi-Machacek"];

(* extracting all scalars *)
AllScalars = Transpose[Select[Particles[GaugeES], #[[4]] == S &]][[1]];

(* Creating all possible 2-tuples with the scalars and the complex conjugates ones *)
AllScalars = Join[AllScalars, conj /@ AllScalars];
pairs = Intersection[Map[Sort[#] &, Tuples[AllScalars, {2}], {1}]];

(* calculating the scattering matrix *)
ScatterMatrix = 
  Table[ 
    (* symmetry factors *)
    If[pairs[[i2, 2]] === pairs[[i2, 1]], 1/Sqrt[2], 1] 
    If[pairs[[i1, 2]] === pairs[[i1, 1]], 1/Sqrt[2], 1] 
    (* derivation *)
    D[
      D[
        D[
          D[
            LagSSSS[GaugeES],  (* the scalar potential in SARAH *)
          pairs[[i1, 1]]], 
        pairs[[i1, 2]]], 
      pairs[[i2, 1]]], 
    pairs[[i2, 2]]], 
  {i1, 1, Length[pairs]}, {i2, 1,Length[pairs]}];
   
(* Taking the eigenvalues *)   
Eigenvalues[ScatterMatrix]  
\end{lstlisting}
This method can also be used for any other model implemented in \SARAH. Only if coloured scalars are present in the model, 
one needs to take care of the colour factor in addition.

\end{appendix}

\bibliography{GM}
\bibliographystyle{ArXiv}

\end{document}